\documentclass[%
reprint,
superscriptaddress,
 amsmath,amssymb,
 aps,
pra,
]{revtex4-2}

\usepackage{xifthen}  
\usepackage{graphicx}
\usepackage{dcolumn}
\usepackage{bm}
\usepackage{derivative}
\usepackage{lipsum}

\usepackage{xspace}
\usepackage{hyperref}        
\usepackage{siunitx}
\usepackage{layout}
\usepackage[normalem]{ulem}

\def\REDLINE{0} 
\if\REDLINE1
    \newcommand{\rladd}[1]{\textcolor{red}{#1}}
    \newcommand{\rlremove}[1]{\st{#1}}
\else
    \newcommand{\rladd}[1]{#1}
    \newcommand{\rlremove}[1]{\unskip}
\fi

\usepackage{soul}
\usepackage[dvipsnames]{xcolor}
\definecolor{myred}{rgb}{1, 0.123, 0.404}
\definecolor{mypink}{rgb}{1, 0.2, 0.745}
\definecolor{mycyan}{rgb}{0.090, 0.667, 0.553}
\definecolor{freshgreen}{rgb}{0.051, 0.796, 0.733}
\definecolor{mygreen}{rgb}{0.184, 0.792, 0.0}
\definecolor{myviolet}{rgb}{0.256, 0.207, 1.00}
\definecolor{myorange}{rgb}{0.9, 0.44, 0.0}
\definecolor{mypink2}{rgb}{0.898, 0.0, 0.451}
\definecolor{mypurple}{rgb}{0.659, 0.251, 1.000}

\def\COLORIZE{0} 
\if\COLORIZE1
    
    \newcommand{\pink}[1]{\textcolor{mypink}{#1}}
    \newcommand{\myorange}[1]{\textcolor{myorange}{#1}}
    \newcommand{\fgreen}[1]{\textcolor{freshgreen}{#1}}
\else
    
    \newcommand{\pink}[1]{#1}
    \newcommand{\myorange}[1]{#1}
    \newcommand{\fgreen}[1]{#1}
\fi

\usepackage[textsize=small]{todonotes}

\newcommand{\molhhc}{H$_2^+$}
\newcommand{\molco}{CO}
\newcommand{\molocs}{OCS}
\newcommand{\molbrmet}{CH$_3$Br}

\newcommand{\myunit}[2][]{%
  \ifthenelse{\isempty{#1}}%
    {\ensuremath{#2}}
    {\ensuremath{{#1}\:{#2}}}
}

\newcommand{\betazero}[1]{\ensuremath{ \beta_{#1}^{(0)} }}
\newcommand{\betaone}[1]{\ensuremath{ \beta_{#1}^{(1)} }}
\newcommand{\omgzero}[2]{\ensuremath{ \Omega_{#1}^{(0)} (#2) }}
\newcommand{\omgone}[2]{\ensuremath{ \Omega_{#1}^{(1)} (#2) }}
\newcommand{\retazero}[2]{\ensuremath{ \mathcal R_{#1}^{(0)} (#2) }}
\newcommand{\retaone}[2]{\ensuremath{ \mathcal R_{#1}^{(1)} (#2) }}
\newcommand{\oetazero}[2]{\ensuremath{ \mathcal O_{#1}^{(0)} (#2) }}

\newcommand{\phizero}[2]{\ensuremath{ \phi_{#1}^{(0)} (#2) }}
\newcommand{\phione}[2]{\ensuremath{ \phi_{#1}^{(1)} (#2) }}
\newcommand{\wfnzero}[1]{\ensuremath{ \psi^{(0)} (#1) }}
\newcommand{\wfnone}[1]{\ensuremath{ \psi^{(1)} (#1) }}

\let\oldtheequation\theequation
\makeatletter
\def\tagform@#1{\maketag@@@{\ignorespaces#1\unskip\@@italiccorr}}
\renewcommand{\theequation}{(\oldtheequation)}
\makeatother
\begin{document}

\preprint{APS/123-QED}

\setstcolor{red}

\title{Orbital distortion and parabolic channel effects transforming minima \\
in molecular ionization probabilities into maxima 
}

\author{Imam S. Wahyutama}
\email{iwahyutama@ucmerced.edu}
\affiliation{Department of Physics and Astronomy, Louisiana State University, Baton Rouge, Louisiana 70803, USA}

\author{Denawakage D. Jayasinghe}
\affiliation{Department of Chemistry, Louisiana State University, Baton Rouge, Louisiana 70803, USA}

\author{Fran\c{c}ois Mauger}
\affiliation{Department of Physics and Astronomy, Louisiana State University, Baton Rouge, Louisiana 70803, USA}

\author{Kenneth Lopata}
\affiliation{Department of Chemistry, Louisiana State University, Baton Rouge, Louisiana 70803, USA}
\affiliation{Center for Computation and Technology, Louisiana State University, Baton Rouge, Louisiana 70808, USA}

\author{Kenneth J. Schafer}
\affiliation{Department of Physics and Astronomy, Louisiana State University, Baton Rouge, Louisiana 70803, USA}

\date{\today}

\begin{abstract}
In the tunneling regime and at sufficiently low field amplitudes, the shape of orientation-dependent molecular ionization rate curves usually resembles the shape of the ionized orbital. As the ionizing field strength increases, the shape of the ionization rate can deviate from \pink{this pattern}. The oft-cited explanation is that the increasing contribution of excited states relative to the ground \pink{state modifies the distribution}. In this paper, we show that orbital distortion and parabolic channel effects, which are independent of excited-state effects, can also significantly modify the angular dependence of the yields of widely studied molecules where excited state effects are negligible. For example, \rladd{as the ionizing field increases} we find that
\rladd{(i) in \molco{}, orbital distortion is responsible for the switch of the position of the global maximum in the orientation-dependent ionization rate to the opposite orientation, and (ii)} in \molbrmet{}, the interplay between orbital distortion and parabolic channel effects transforms a local minimum \rlremove{in the orientation-dependent ionization rate} to a local maximum \rlremove{as the ionizing field strength increases}. To simulate orbital distortion and parabolic channel effects, we use the one-electron weak-field asymptotic theory including the first-order correction (OE-WFAT(1)) in the integral representation. Since OE-WFAT(1) incurs expensive computations when the number of orientation angles is large, we also reformulate the original OE-WFAT(1) \myorange{algorithm} into a partial-wave expansion form, which greatly enhances the efficiency of the method.
\end{abstract}

\maketitle

\section{Introduction}

Molecular tunnel ionization is sensitive to the shape of the ionized \myorange{orbital}, a property that has been exploited in several methods of strong-field physics, such as molecular orbital tomography \cite{mot-2004, mot-multiel-2006, mot-attoimaging-2010, mot-generalized-2011, mot-hhs-molimage-2014, mot-h2o-2023, slimp2-2024}, laser-induced electron diffraction \cite{lied-science-2008, lied-review-2010, lied-self-image-2010, lied-nh3-umbrella-2021}, and the three-step model of high-harmonic generation \cite{sfi-plasma-perspective-1993, super-intense-1993, ati-1993}. \pink{Using ionization as a probe of} ultrafast dynamics also relies on the correlation between ionization probability and the shape of the instantaneous \myorange{ionized orbital}. Furthermore, a robust understanding of molecular tunnel ionization, being the first step in strong-field processes, is essential in analyzing experimental data. \pink{These factors highlight} the need for a comprehensive theory of molecular tunnel ionization that accurately captures the field and orientation dependences of the ionization probability and how these two dependences are intertwined. \pink{To address this need}, several successful adiabatic tunneling theories have been formulated.
These include the Ammosov-Delone-Krainov theory \cite{adk-1986} and more sophisticated theories designed for adiabatic tunneling simulations in single-electron \cite{moadk-2002, moadk-kjeldsen-2005, ft-tunnel-atom-2010, wfat_theory1-2011, wfat-structurefactors-2013, ir_oewfat-2016, ft-tunnel-molecule-2016} as well as correlated many-electron molecular models \cite{tr_mewfat-2014, tr_mewfat_app2-2014, tr_mewfat_app-2017, wahyutama-dft-mewfat-2022, wahyutama-mewfat-fundamental-2025}.

At the heart of the aforementioned applications of molecular tunnel ionization is the intuition that the orientation dependence of ionization yield reflects the shape of the ionized orbital \cite{channel-ati-2013, CO-orbdistort-2015, CO-permdipole-2017}, which is often the highest occupied molecular orbital (HOMO). This mapping, however, is generally only reliable at low ionizing field strengths since, as the field strength increases, the shape of the ionization yield curve deviates from that of the HOMO \cite{OCS-CS2-tddft-2011, CO-OCS-tddft-2014}. \myorange{Often, excited state effects, that is, when the ionized orbital is lower than the HOMO, \pink{are cited as} the main contribution to this deviation.}
In this work, we will show that there are factors other than excited state effect that can \pink{lead to such a field-dependent deviation}:
orbital distortion and higher parabolic channels. \pink{This deviation} can be so significant that, for example, a minimum in the orientation-dependent ionization rate turns into a maximum \pink{as the field strength increases}\cite{channel-ati-2013}.

\myorange{\pink{In this paper we use ``orbital distortion'' to refer} to the modification of the field-free orbitals due to the ionizing field. Using field-distorted orbitals in tunneling ionization calculations leads to a more accurate ionization rate compared to those obtained with field-free orbitals \cite{orbdistort-hhg-2014, CO-orbdistort-2015, oewfat1-h2_c-2015}. Likewise, as the field strength increases, several final parabolic states of the outgoing electron may contribute to the tunneling rate. We refer to these as parabolic channel effects.}
\myorange{In the present work, we investigate the first-order orbital distortion and parabolic channel effects in tunnel ionization from \molco{}, \molocs{}, and \molbrmet{}, and show that the two effects can be prominent in situations where excited the state contribution is otherwise negligible, stressing the importance of taking these two effects into account.
Since the first-order distortion is generally much more significant than the higher orders within the range of the applicability of our method described below \cite{high-order-stark-1981}, we will sometimes omit the order specification when mentioning orbital distortion.
From an applications viewpoint, correctly incorporating these effects into tunneling ionization theories \cite{mot-2004, mot-multiel-2006, mot-attoimaging-2010, mot-generalized-2011, mot-hhs-molimage-2014, mot-h2o-2023, slimp2-2024, lied-science-2008, lied-review-2010, lied-self-image-2010, lied-nh3-umbrella-2021, sfi-plasma-perspective-1993, super-intense-1993, ati-1993} will extend their capability to higher field intensities.}

\myorange{To calculate orientation-dependent ionization rates in the present work, we use the one-electron weak-field asymptotic theory including the first-order correction (OE-WFAT(1)) in the \textit{integral representation} (IR) \cite{ir_oewfat-2016}.} OE-WFAT(1) is a molecular tunnel ionization theory for static fields that enables a formal, nonheuristic treatment of orbital distortion \cite{wfat1_atom_ref-2013, oewfat1-h2_c-2015}. 
\myorange{Note that there are two representations of OE-WFAT(1): the \textit{tail representation} (TR) \cite{wfat1_atom_ref-2013, oewfat1-h2_c-2015}, and the integral representation (IR) \cite{ir_oewfat-2016}. We use the IR in the present work because it is numerically more stable than the TR \rladd{when using Gaussian basis to represent the orbitals}.} 

\pink{Orbital distortion may manifest more prominently within certain orientation regions than others, and these regions cannot be determined \textit{a priori}. Therefore, the capability to efficiently calculate tunnel ionization probabilities across the entire range of molecular orientation angles is indispensable.}
However, when there are a large number of orientation angles to calculate, the original formulation of OE-WFAT(1) in the IR requires extremely long computations due to \pink{certain computationally expensive two-electron integrals that occur}. For this reason, we reformulate OE-WFAT(1) in the IR \cite{ir_oewfat-2016} into a partial-wave expansion form \cite{wfat_partial_wave1-2017, ir_oewfat_grid-2018}. The partial-wave reformulation of OE-WFAT(1) enables efficient and equally accurate computation of ionization rates by allowing storage of computationally expensive but orientation-independent quantities in memory or on disk, which can later be reused for the calculation of the rates at arbitrary orientations. This reformulation results in a speed-up factor that increases linearly with the number of orientation angles.
Our OE-WFAT(1) algorithm is implemented as part of the WFAT module in the development version of \textsc{NWChem} \cite{nwchem-github-dev-2025}. Another program called \textsc{PyStructureFactor} \cite{pystructurefactor-2023} also provides WFAT functionality at the zeroth-order level of approximation (OE-WFAT(0)).

This paper is organized as follows: In \autoref{sec:intro-oewfat1}, the formulation of OE-WFAT(1) in the IR is summarized. In \autoref{sec:orb_distort_parabolic}, orbital distortion and parabolic channel effects are defined in the context of WFAT. Readers who are only interested in the result may want to focus on this section and only refer to the other theory sections as needed. In \autoref{sec:derive_q}, we present the reformulation of OE-WFAT(1) in partial-wave decomposition form. In \autoref{sec:degenerate}, we discuss how to efficiently treat degeneracies. \autoref{sec:validation} presents the validation of our partial-wave expansion formulation of OE-WFAT(1). In \autoref{sec:co_ocs}, we discuss \pink{the importance of orbital distortion in tunnel ionization from CO and OCS molecules for reproducing the orientation dependence observed in reference data}. Finally, \autoref{sec:ch3br} presents an example of a case where a minimum in the orientation-dependent rate transforms to a maximum as the field increases.  
The prospective applications and improvement of OE-WFAT(1) are presented in \autoref{sec:conclusion}.

\section{Theory}

\subsection{Overview of OE-WFAT(1) \label{sec:intro-oewfat1}}

\begin{figure}[!tbp]
    \includegraphics[width=0.9\columnwidth]{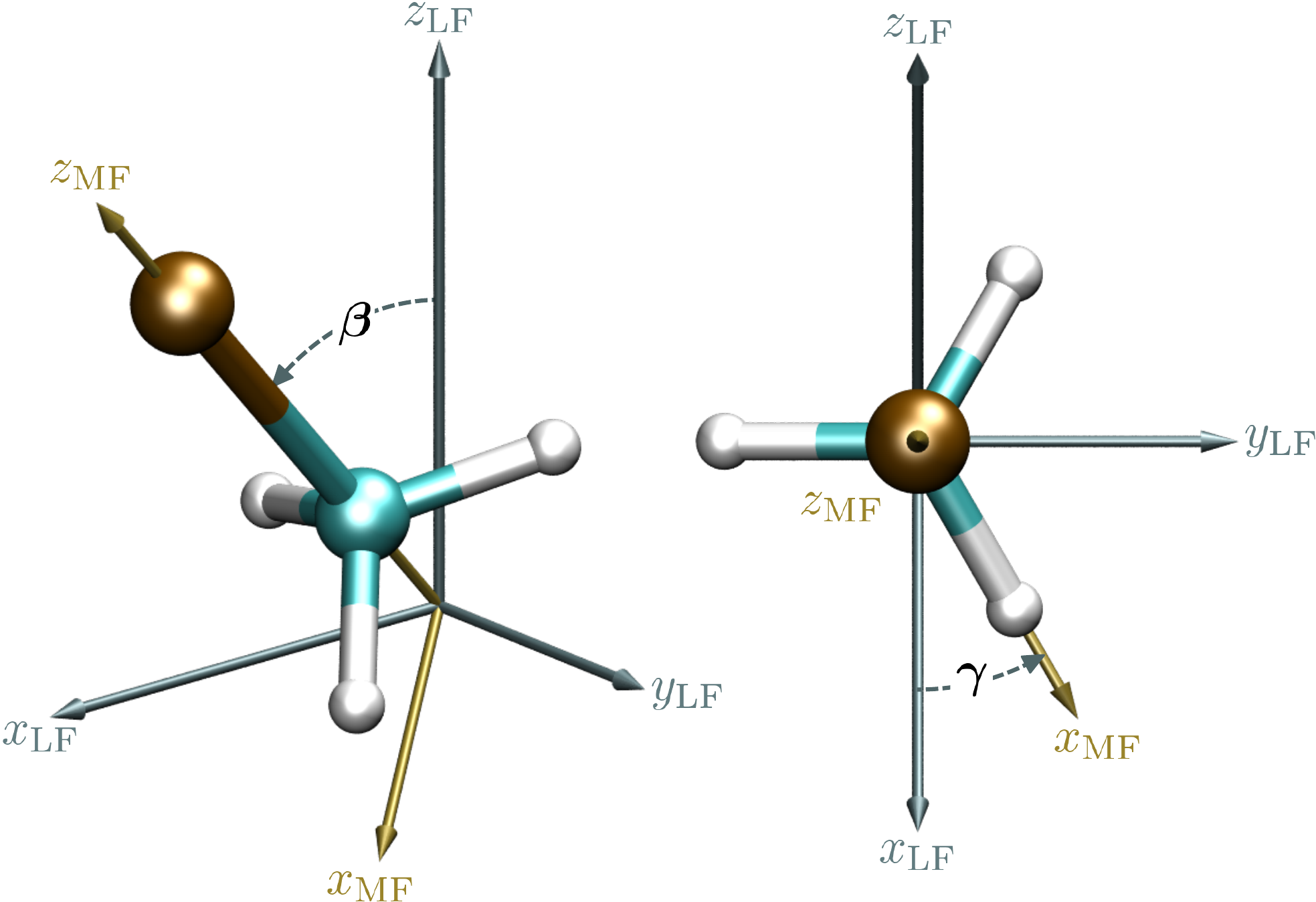}
    \caption{An illustration of Euler orientation angles $(\beta,\gamma)$. 
    \myorange{$\beta$ is the angle between $z_\text{MF}$ and $z_\text{LF}$, while $\gamma$ rotates $x_\text{MF}$ (and $y_\text{MF}$) around $z_\text{MF}$.}
    On the right is the view along $z_\text{MF}$ toward the origin. 
    \myorange{The third Euler angle (not shown) rotates the MF around $z_\textrm{LF}$, and so, is unnecessary. This angle has been set to \ang{0} so that $z_\text{MF}$ stays in the $x_\text{LF}z_\text{LF}$ plane.}
    The field points in the positive $z_\text{LF}$.}
    \label{fig:angle_def}
\end{figure}

In this section, we will summarize the mathematical steps that lead us to the OE-WFAT ionization formula including first-order \pink{corrections in the field. The} detailed derivation can be found in Ref. \cite{ir_oewfat-2016}. 
We start by considering a molecule \myorange{at a certain orientation in space} exposed to a static electric field $\mathbf F = F\hat z, \, F \geq 0$, where $\hat z$ is the positive $z$ direction in the lab frame (LF) and $F$ lies in a range of field strengths in which single-electron ionization dominates over higher-order ionization. The external field in this situation usually affects the outermost electron more strongly than the remaining ones. The other electrons can therefore be treated as a screening potential so that the potential felt by the outermost electron is expressed as an effective potential $V(\mathbf r)$\pink{, where the screening effect is modeled through two-electron interaction terms \cite{ir_oewfat-2016}}. The \myorange{time-independent} Schr\"odinger equation in the LF that describes such a system is
\begin{gather}
    \left(
    -\frac{1}{2}\nabla^2 + V(\mathbf r) + Fz
    \right)
    \psi(\mathbf r)
    =
    E \psi(\mathbf r)
    \label{eq:schroedinger_eq}.
\end{gather}
Unless otherwise specified, we use atomic units throughout.
\myorange{The goal is to obtain the expression for $\psi(\mathbf{r})$ in the asymptotic region, which will then be used to derive} the ionization rate formula as a function of the orientation angle between the external field and the molecule-bound axes, i.e. the molecular frame (MF). The relationship between the two coordinate systems is governed by the rotation matrix $\hat R \equiv \hat R(\beta,\gamma)$, where $(\beta,\gamma)$ are the Euler angles that characterize the orientation, so that the position vectors of a point in the MF and LF are connected through $\mathbf r_\textrm{MF} = \hat R \, \mathbf r_\textrm{LF}$. The elements of $\hat R$ are
\begin{gather}
    \hat R =
    \begin{pmatrix}
     \cos\beta \cos\gamma & \sin\gamma & -\sin\beta \cos\gamma\\
    -\cos\beta \sin\gamma & \cos\gamma &  \sin\beta \sin\gamma\\
                \sin\beta &          0 &             \cos\beta
    \end{pmatrix}.
\end{gather}
\myorange{To illustrate the action of $\hat R$ on a molecule, we present an example of a \molbrmet{} molecule that has been oriented by $(\beta,\gamma)$ with respect to the LF in \autoref{fig:angle_def}. Since the field is uniform, the third Euler angle is irrelevant and can be set to an arbitrary value, $\ang{0}$ in our simulations.} 

WFAT is a rigorously derived tunnel ionization theory based on the expansion of the ionized wave function in parabolic channel basis in the asymptotic region. WFAT has been used in some works to help analyze experimental data \cite{wfat-application-laser-elstru-2015, wfat-application-atto-cm-2015, dissoc-tunnel-no-2019, wfat-application-hhs-2024}. In general, for OE-WFAT, the solution of \autoref{eq:schroedinger_eq} is expanded as follows \cite{wfat1_atom_ref-2013, ir_oewfat-2016},
\begin{gather}
    \psi(\mathbf r) = \eta^{-1/2} \sum_\nu F_\nu(\eta) \Phi_\nu(\xi, \varphi),
    \label{eq:wfn_expand}
\end{gather}
where 
\begin{gather}
    \Phi_\nu(\xi,\varphi) = \phi_\nu(\xi) \frac{e^{im\varphi}}{\sqrt{2\pi}},
\end{gather}
and $(\xi,\eta,\varphi)$ are coordinates in the parabolic coordinate system \cite{wfat_theory1-2011}. $\Phi_\nu(\xi,\varphi)$ are called parabolic channel functions and are characterized by the set of parabolic quantum numbers
\begin{gather}
    \nu \equiv (n_\xi,m), \hspace{3mm} n_\xi = 0,1,\ldots, \hspace{3mm} m = 0, \pm 1, \ldots.
\end{gather}
$\phi_\nu(\xi)$ is the solution of the following eigenproblem 
\begin{gather}
    \left(
    \odv*[fun]{\xi\odv*{}{\xi}}{\xi}
    - \frac{m^2}{4\xi}
    + Z
    + \frac{E\xi}{2}
    - \frac{F\xi^2}{4}
    - \beta_\nu
    \right)
    \phi_\nu(\xi)
    = 0,
    \label{eq:diff_eq_phi}
\end{gather}
where $Z$ is the total charge of the cation. By substituting Eq. \eqref{eq:wfn_expand} into Eq. \eqref{eq:schroedinger_eq} and using Eq. \eqref{eq:diff_eq_phi}, one may obtain the equation satisfied by $F_\nu(\eta)$. 

In the asymptotic region where $\eta \to \infty$ and also where the ionized electron propagates, $\psi(\mathbf{r})$ may be expressed as
\begin{gather}
    \psi(\mathbf r) \big|_{\eta \to \infty} = 
    \eta^{-1/2} 
    \sum_\nu f_\nu \, p_\nu(\eta) 
    \Phi_\nu(\xi, \varphi),
    \label{eq:wfn_expand_as}
\end{gather}
where $p_\nu(\eta) = f_\nu^{-1} F_\nu(\eta) \big|_{\eta \to \infty}$. The expression for $F_\nu(\eta) \big|_{\eta \to \infty}$, and hence $p_\nu(\eta)$, can be found in Refs. 
\cite{wfat_theory1-2011, wfat1_atom_ref-2013, oewfat1-h2_c-2015}.
\fgreen{The total ionization rate (or probability) $\Gamma$ is then calculated as the total flux of electrons leaving a surface of constant $\eta$ as $\eta$ approaches infinity \cite{wfat1_atom_ref-2013}. This results in the formula $\Gamma \approx \sum_\nu \Gamma_\nu$, where $\Gamma_\nu = |f_\nu|^2$.
}

In OE-WFAT(1), the partial ionization probability $\Gamma_\nu$ is expanded into powers of $F$ and then truncated up to the first power, which gives 
\begin{align}
    \Gamma_\nu \approx& \,\, \Gamma_\nu^{(1)}
    =
    W_\nu(F) e^{-2\varkappa \mu_z}
    \bigg(
    |g_\nu|^2
    \Big(
    1
    + A_\nu F \ln \frac{F}{4\varkappa^2}
    \nonumber \\
    & \,
    + \tilde B_\nu F 
    + 2F o_\nu
    \Big)
    + \, 2F \, \textrm{Re}\left[g_\nu^* h_\nu\right]
    \bigg),
    \label{eq:wfat1_rate}
\end{align}
where $\varkappa = \sqrt{2|E^{(0)}|}$, $E^{(0)}$ is the field-free orbital energy, $\mu_z$ is the $z$ component of the dipole vector in the LF, that is, 
\begin{equation}
    \mu_z \equiv [\boldsymbol\mu_\textrm{LF}]_3 = \left[\hat R^T (\beta,\gamma) \boldsymbol \mu_\textrm{MF} \right]_3,
    \label{eq:rotate_dipole}
\end{equation}
where  $\mu_\textrm{MF}$ is a 3-element vector whose element is equal to $- \big\langle \psi^{(0)} \big| x_s \big| \psi^{(0)} \big\rangle$ with $x_s$ being Cartesian coordinates in the MF. \pink{$W_\nu(F)$ is called field factor and is given by}
\begin{equation}
    W_\nu(F) = \frac{\varkappa}{2}     
    \left(\frac{4\varkappa^2}{F}\right)^{2Z/\varkappa-2n_\xi-|m|-1}
    \exp\left( -\frac{2\varkappa^3}{3F} \right)
    \label{eq:field_factor}
\end{equation}
The expressions for $A_\nu$, $\tilde B_\nu$, and $o_\nu$ can be found in Appendix \ref{app:define_quantities}.

\pink{The ionization rate formula \autoref{eq:wfat1_rate} depends on the ionized orbital through the \textit{asymptotic coefficients}, $g_\nu$ and $h_\nu$.} \myorange{Their integral formulas in the MF are given by \cite{ir_oewfat-2016}}
\begin{gather}
    g_\nu = 
    \int d\mathbf r^3  \,
    \Omega_{\bar\nu}^{(0)}(\hat R^{-1} \mathbf r) \, V_c(\mathbf r) \, \psi^{(0)}(\mathbf r) ,
    \label{eq:define_gnu}
    \\
    h_\nu =
    \int d\mathbf r^3 
    \left( 
    \Omega_{\bar\nu}^{(0)}(\hat R^{-1} \mathbf r) \, \psi^{(1)}(\mathbf r) 
    +
    \Omega_{\bar\nu}^{(1)}(\hat R^{-1} \mathbf r) \, \psi^{(0)}(\mathbf r) 
    \right) \nonumber \\
    \times V_c(\mathbf r).
    \label{eq:define_hnu}
\end{gather}
\pink{Here, \wfnzero{\mathbf r} and \wfnone{\mathbf r} represent the field-free orbital and its first-order perturbation correction, respectively, so that the field-dressed orbital is approximated as $\psi(\mathbf r) \approx \wfnzero{\mathbf r} + F \wfnone{\mathbf r}$. The terms \omgzero{\nu}{\mathbf r} and \omgone{\nu}{\mathbf r} are the function that determines the final state of the ionized electron and its first-order perturbation correction, respectively. The potential term $V_c(\mathbf r) = V(\mathbf r) + Zr^{-1}$ appears as a result of the Lippmann-Schwinger integral reformulation of WFAT [20]. This potential term is absent in the TR formulation.}
Equations \eqref{eq:define_gnu}--\eqref{eq:define_hnu} constitute the most computationally expensive steps due to the required numerical integrations using \pink{spatial} grid points and the evaluation of two-electron potential terms inside $V(\mathbf r)$ at each of the grid points. \myorange{We note that $g_\nu$ and $h_\nu$ depend on the choice of origin through $V_c(\mathbf{ r)}$ and $\Omega_\nu^{(0,1)}(\mathbf{r})$. This origin dependence, however, is canceled by the origin dependence of $\exp(-2\varkappa\mu_z)$ when calculating $\Gamma_\nu$ (see \autoref{eq:wfat1_rate}).} 

The first-order correction (or distortion) of the field-free orbital may be expressed as
\begin{gather}
    \psi^{(1)} (\mathbf r) = \sum_{s=1}^3 \hat R_{s3} \, \psi_s^{(1)} (\mathbf r), 
    \label{eq:define_psi1}
\end{gather}
where 
\begin{gather}
    \psi_s^{(1)}(\mathbf r) 
    = \sum_{i; v_i \notin \mathcal D[n]} v_i(\mathbf r) \frac{\langle v_i |x_s| v_n \rangle}{\varepsilon_n - \varepsilon_i}.
    \label{eq:define_psi_s}
\end{gather}
$v_i$ and $\varepsilon_i$ are the eigenvectors and the corresponding eigenvalues of the unperturbed Hamiltonian, respectively. $n$ is the index of the unperturbed ionized wave function, i.e. $\psi^{(0)} \equiv v_n$ and $E^{(0)} \equiv \varepsilon_n$. $\mathcal D[n]$ is the degenerate subspace spanned by all unperturbed eigenstates with energy $\varepsilon_n$, which, for the case of non-degenerate state, consists of just $v_n$, and $x_s$ with $s=1,2,3$ are the three Cartesian coordinates in the MF. The addition of $F \psi^{(1)}(\mathbf{r})$ into $\psi^{(0)}(\mathbf{r})$ will distort the shape of the latter based on the field strength and its direction relative to the molecule, and will thus affect the orientation dependence of ionization rate. The effects of such a distortion to the orbital are one of the main subjects of the present work.

The functions \omgzero{\nu}{\mathbf{r}} and \omgone{\nu}{\mathbf{r}} are given by \cite{ir_oewfat-2016}
\begin{align}
    \Omega_\nu^{(0)}(\mathbf r) =
    &\,
    -\frac{2 \varkappa^{\beta_\nu^{(0)}/\varkappa-1}}{\sqrt{2\pi\eta}}
    \mathcal R_\nu^{(0)}(\eta) \phi_\nu^{(0)}(\xi) \, e^{im\varphi} ,
    \label{eq:Omega0_explicit}
    \\
    \Omega_\nu^{(1)}(\mathbf r) =
    &\,
    -\frac{2 \varkappa^{\beta_\nu^{(0)}/\varkappa-1}}{\sqrt{2\pi\eta}}
    e^{im\varphi}
    \nonumber
    \\
    &\,
    \times \left(
    \mathcal R_\nu^{(1)}(\eta) \phi_\nu^{(0)}(\xi) \,
    + \mathcal R_\nu^{(0)}(\eta)  
    \phi_\nu^{(1)}(\xi)
    \right) ,
    \label{eq:Omega1_explicit}
\end{align}
where \betazero{\nu}, \retazero{\nu}{\eta}, \retaone{\nu}{\eta}, \phizero{\nu}{\xi}, and \phione{\nu}{\xi} can be found in Appendix \ref{app:define_quantities}. Rewriting \omgzero{\nu}{\mathbf{r}} and \omgone{\nu}{\mathbf{r}} as an expansion in terms of spherical harmonics constitutes the first step in deriving the partial-wave expansion method presented in \autoref{sec:derive_q}.

In Ref. \cite{ir_oewfat_grid-2018}, integrals of the form \autoref{eq:define_gnu} are computed by partitioning the integral into Becke cells \cite{becke_cell-1988}, a method which we shall also employ here. Each partition is an integral over the space defined by the cell centered around a nucleus. These individual partitioned integrals are then computed using a 3D quadrature before being summed over to yield the desired final integral. Using this integration method, one can of course directly compute the integrals in \autoref{eq:define_gnu} and \autoref{eq:define_hnu}.
However, such a direct method, dubbed the \textit{explicit method} hereafter, has a severe drawback related to \myorange{a computationally expensive integral}---we will explain more about this in the following.
The orientation dependence in Eqs. \eqref{eq:define_gnu} and \eqref{eq:define_hnu} is encoded in the orientation between the $\Omega_\nu(\mathbf r)$'s and the wave functions as well as $V_c(\mathbf r)$: \myorange{the $\Omega_\nu(\mathbf r)$'s are bound to $z_\text{LF}$ (which is where the field points) whereas the wave functions and $V_c(\mathbf r)$ are bound to the MF. While Eqs. \eqref{eq:define_gnu} and \eqref{eq:define_hnu} have been written in the MF, one can also recast them into their LF form. However, since Becke cells are molecule-bound, the algorithm to calculate these integrals in the LF is essentially identical to that in the MF.}
Performing the integral of \autoref{eq:define_hnu} in the MF for a large number of orientations, however, can be very expensive
because: 
1) $\Omega_\nu^{(1)}(\mathbf r)$ must be recomputed whenever the orientation is changed, and
2) the computation of $\mathcal R_\nu^{(1)}(\eta)$ contained in $\Omega_\nu^{(1)}(\mathbf r)$ involves yet another integral that must be computed numerically (see Eq. \eqref{eq:define_r1_eta}).
In  \autoref{sec:derive_q}, we shall derive a set of equations that allow for a highly efficient computation of \autoref{eq:define_hnu} in the MF for a large number of orientation angles, we will call this the \textit{partial wave method}.

We end this section by noting that we have expressed some quantities such as $\psi^{(1)}(\mathbf{r})$ (Eq. \eqref{eq:define_psi1}), $\beta_\nu^{(1)}$ (Eq. \eqref{eq:define_beta1}), $o_\nu$ (Eq. \eqref{eq:define_onu}), and the expansion coefficients in Eq. \eqref{eq:phi1_expand} in the decomposed form of $\sum_i d_i f_i$ where $f_i$ is orientation-independent while $d_i$ depends on the orientation. This is done because to implement the efficient computation of OE-WFAT(1) ionization rates in \autoref{sec:derive_q}, 
we need to defer incorporating the orientation angle into the calculation until the last step (which is the computation of the ionization rate) when all computationally expensive, orientation-independent quantities have been calculated and stored in memory or on disk. Throughout this paper, a quantity is cast in the above decomposed form if it involves or is needed in expensive calculations (for example, see Appendix \ref{app:define_quantities}).

\subsection{Orbital distortion and parabolic channel effects in tunnel ionization \label{sec:orb_distort_parabolic}}

Equation \eqref{eq:wfat1_rate} gives the OE-WFAT(1) rate formula for ionization to the parabolic channel $\nu$. We note that without the first-order correction, the OE-WFAT(0) partial rate reads \cite{wfat_theory1-2011, oewfat0-linearmol-2012, wfat-structurefactors-2013, wfat_partial_wave1-2017, ir_oewfat_grid-2018}
\begin{equation}
    \label{eq:wfat0_rate}
    \Gamma_\nu^{(0)} = W_\nu(F) \exp(-2 \varkappa \mu_z) |g_\nu|^2.
\end{equation}
Since the field and orientation dependences of $\Gamma_\nu^{(0)}$ are decoupled into 
$W_\nu(F)$ and $\exp(-2 \varkappa \mu_z) |g_\nu|^2$, respectively, in OE-WFAT(0), the angular shape of partial ionization rates \myorange{is independent of the field}. 
In the following, we present our study of the effects of the quantities that appear in going from $\Gamma_\nu^{(0)}$ (\autoref{eq:wfat0_rate}) to $\Gamma_\nu^{(1)}$ (\autoref{eq:wfat1_rate}) on the angular shape of ionization rate curves. This amounts to the first-order orbital distortion effect. These additional quantities depend on $\psi^{(1)}(\mathbf{r})$ and other quantities related to the first-order corrections in field \pink{strength. Consequently, they cause the field and orientation dependences of the ionization rates to become coupled, unlike in OE-WFAT(0) where these dependences are decoupled.}

The total ionization rate is obtained by summing over all parabolic channels, $\Gamma \approx \sum_\nu \Gamma_\nu$ \cite{wfat1_atom_ref-2013}. One then needs to apply the appropriate order of approximation to each parabolic channel based on the order of $F$ in $W_\nu(F)$. For the lowest three parabolic channels \pink{in first order}, this gives \cite{wfat1_atom_ref-2013}
\begin{equation}
    \Gamma \approx \Gamma_{00}^{(1)} + \Gamma_{0,+1}^{(0)} + \Gamma_{0,-1}^{(0)}.
    \label{eq:total_ion_rate}
\end{equation}
In many cases, $\Gamma_{00}$ \pink{(approximated as $\Gamma_{00}^{(1)}$ in \autoref{eq:total_ion_rate})} is the dominant contributor to the total rate. The dependence of $\Gamma_{00}$ on $(\beta,\gamma)$ resembles the shape of $\psi^{(0)}(\mathbf{r})$ at low field strengths. \pink{This term is therefore responsible for the mapping between} $\psi^{(0)}(\mathbf{r})$ and the orientation-dependent yield curve. The parabolic channel effect investigated in our work amounts to the contributions from $\Gamma_{0,\pm1}^{(0)}$. 

For the analyses of molecular ionization rates employed in this paper, we define the \textit{normalized rate} $\tilde \Gamma$ as $\tilde \Gamma = \Gamma / W_{00}$.

\subsection{Partial-wave decomposition of OE-WFAT(1) \label{sec:derive_q}}

\myorange{In computing the MF integrals in Eqs. \eqref{eq:define_gnu}--\eqref{eq:define_hnu}, we need to apply $\hat R$ on \omgzero{\nu}{\mathbf{r}} and \omgone{\nu}{\mathbf{r}}. Therefore, we can expect to gain some computational efficiency if these functions are expanded in terms of the eigenfunctions of the generator of the rotation operator, namely, the spherical harmonics. This so-called partial-wave expansion method has been employed in Refs. \cite{wfat_partial_wave1-2017, ir_oewfat_grid-2018} to efficiently compute the integral in Eq. \eqref{eq:define_gnu}.}
In this method, $\Omega_\nu^{(0)}(\mathbf r)$ is expressed as
\begin{gather}
    \label{eq:Omega0_expand}
    \Omega_\nu^{(0)}(\mathbf r) = 
    \sum_{l=|m|}^\infty 
    R_l^\nu(r) \,
    Y_{lm}(\theta,\varphi),
\end{gather}
where the radial function $R_l^\nu(r)$ is found to be
\begin{gather}
    R_l^\nu(r) =
    \omega_l^\nu
    (\varkappa r)^l
    e^{-\varkappa r}
    M(l+1-Z/\varkappa, 2l+2, 2\varkappa r), 
    \label{eq:define_rnu}
\end{gather}
with
\begin{align}
    \omega_l^\nu
    =& \,\,
    (-1)^{l+(|m|-m)/2+1} \,
    2^{l+3/2} \,
    \varkappa^{\betazero{\nu}/\varkappa}   \nonumber \\
    &\,\,
    \times
    \sqrt{(2l+1) (l+m)! (l-m)! (|m|+n_\xi)! n_\xi!}   \nonumber \\
    &\,\,
    \times
    \frac{l!}{(2l+1)!} 
    \sum_{k=0}^{\textrm{min}(n_\xi,l-|m|)}
    \frac
    {1}
    {k! (l-k)! (|m|+k)!}   \nonumber \\
    &\,\, 
    \times
    \frac{\Gamma(l+v-k)}{(l-|m|-k) !(n_\xi-k)!},
    \label{eq:define_omega_nu}
\end{align}
$\Gamma(z)$ is the gamma function, and $v = 1 + n_\xi - Z/\varkappa$.
\myorange{By applying $\hat R$ on \omgzero{\nu}{\mathbf{r}} in \autoref{eq:Omega0_expand} and using the result in \autoref{eq:define_gnu}, one obtains the partial-wave expansion of $g_\nu$,}
\begin{align}
    g_\nu(\beta,\gamma) = 
    & \,
    \sum_{l=|m|}^\infty \sum_{m'=-l}^l
    d_{mm'}^l(\beta) e^{-im'\gamma} I_{lm'}^\nu,
    \label{eq:gnu_partial_wave}
\end{align}
where $d_{mm'}^l(\beta)$ is the Wigner function \cite{varshalovich_angular_momentum-1988} and
\begin{gather}
    I_{lm'}^\nu =
    \int d\mathbf r^3 \,
    R_\nu^l(r) Y_{lm'}^*(\theta,\varphi) V_c(\mathbf r) \, \psi^{(0)}(\mathbf r).
    \label{eq:define-i}
\end{gather}
\myorange{In the partial-wave method, $g_\nu$ is calculated through \autoref{eq:gnu_partial_wave}, while in the explicit method, it is obtained from \autoref{eq:define_gnu}.}

To apply the partial-wave expansion for $h_\nu$ in \autoref{eq:define_hnu}, we need the partial-wave expansion for \omgzero{\nu}{\mathbf{r}}, given in Eqs. \eqref{eq:Omega0_expand}--\eqref{eq:define_omega_nu}, and \omgone{\nu}{\mathbf{r}},
\begin{gather}
    \label{eq:Omega1_expand}
    \Omega_\nu^{(1)}(\mathbf r) = 
    \sum_{l=|m|}^\infty 
    Q_l^\nu(r) \,
    Y_{lm}(\theta,\varphi).
\end{gather}
\pink{In \autoref{eq:Omega1_expand}, the objective is to determine $Q_l^\nu(r)$, which we derive in the present work. The partial-wave sum in \autoref{eq:Omega1_expand} must reproduce \omgone{\nu}{\mathbf{r}} given in \autoref{eq:Omega1_explicit}.}
Appendix \ref{app:derive_q} describes the derivation to solve for $Q_l^\nu(r)$. Here, we will only invoke the final result, namely
\begin{gather}
    Q_l^\nu(r) = Q_1^{\nu l}(r) + \mu_z Q_2^{\nu l}(r),
    \label{eq:q_split}
\end{gather}
where    
\begin{align}    
    Q_1^{\nu l}(r) 
    =
    & \,
    d_1^{\nu l} g_l^\nu(r)
    + 
    \frac{2^{2l+2} \varkappa \, \Gamma(l+1-Z/\varkappa)}{(2l+1)!} 
    \int_0^r dr' r'^2   \nonumber
    \\
    & \, 
    \times 
    \left(
    k_{l|m|} r' R_{l-1}^\nu(r')
    +
    k_{l+1,|m|} r' R_{l+1}^\nu(r') 
    \right)
    \nonumber
    \\
    & \, 
    \times
    \left(
    g_l^\nu(r) h_l^\nu(r')
    -
    g_l^\nu(r') h_l^\nu(r)
    \right),
    \label{eq:define_q1}
    \\
    Q_2^{\nu l}(r) 
    =
    & \,
    d_2^{\nu l} g_l^\nu(r)
    + 
    \frac{2^{2l+2} \varkappa \, \Gamma(l+1-Z/\varkappa)}{(2l+1)!}
    \int_0^r dr' r'^2 \nonumber
    \\
    & \, 
    \times
    R_l^\nu(r')
    \left(
    g_l^\nu(r) h_l^\nu(r')
    -
    g_l^\nu(r') h_l^\nu(r)
    \right).
    \label{eq:define_q2}
\end{align}
Here, $k_{l|m|} = \sqrt{(l^2-m^2)(4l^2-1)^{-1}}$, $g_l^\nu(r)$ and $h_l^\nu(r)$ are defined in \autoref{eq:def_g} and \autoref{eq:def_h}, respectively, and 
\begin{widetext}
\begin{align}
    d_r^{\nu l}
    =
    & \,
    (-1)^{l+(m-|m|)/2+1} \,
    2^{l+3/2} \,
    \varkappa^{\beta_\nu^{(0)}/\varkappa}
    \sqrt{(2l+1) (l+m)! (l-m)!} \frac{l!}{(2l+1)!} \nonumber
    \\
    & \,
    \times
    \left[
    \sum_{i=n_\xi-2; i \neq n_\xi}^{n_\xi+2} 
    \sqrt{(i+|m|)! \, i!} \, C_r^{im} \sum_{k=0}^{\textrm{min}(i,l-|m|)}
    \frac{\Gamma(l+v-k)}{k!(l-k)!(|m|+k)!(l-|m|-k)!(i-k)!}\, 
    + 
    \frac{\sqrt{(n_\xi+|m|)! n_\xi!}}{\varkappa^2 \Gamma(v+|m|)}
    \right. \nonumber
    \\
    & \,
    \left.
    \times
    \left\{
    \frac{(-1)^{|m|+1}}{\Gamma(v)}
    \left(
    e^1_r H_1^{\nu l}
    +
    e^2_r H_2^{\nu l}
    +
    e^3_r H_3^{\nu l}
    \right)
    - 
    \Gamma(1-v)
    \left(
    e^1_r \tilde H_1^{\nu l}
    +
    e^2_r \tilde H_2^{\nu l}
    +
    e^3_r \tilde H_3^{\nu l}
    \right)
    \right\}
    \right] ,
    \label{eq:define_d}
\end{align}
\end{widetext}
where 
$H_p^{\nu l}$ and $\tilde H_p^{\nu l}$ are given in \autoref{eq:define_h_app} and \autoref{eq:define_ht_app}, respectively;
$e^1_r = b^\nu_r \varkappa$; 
$e^2_r = -\delta_{r2}/2$;
$e^3_r = \delta_{r1}/(4\varkappa)$;
$\delta_{rr'}$ is the Kronecker delta function;
$b_r^\nu$ is given in Eqs. \eqref{eq:define_b1}--\eqref{eq:define_b2};
and
$C_r^{im}$ is given in Eq. \eqref{eq:phi1_expand_c}. 
\myorange{Equations \eqref{eq:q_split}--\eqref{eq:define_d} constitute the main mathematical result of this paper.}
Having obtained $Q_l^\nu(r)$, we can now go back to Eq. \eqref{eq:define_hnu} and follow the same steps \myorange{that lead to \autoref{eq:gnu_partial_wave}} to express $h_\nu$ as a series of its partial wave components,
\begin{align}
    h_\nu(\beta,\gamma) = 
    & \,
    \sum_{l=|m|}^\infty \sum_{m'=-l}^l
    \Bigg( K_1^{\nu lm'} + \mu_z K_2^{\nu lm'} + \nonumber
    \\
    & \,
    \sum_{s=1}^3 \hat R_{s3} J_s^{\nu lm'} \Bigg) 
    d_{mm'}^l(\beta) e^{-im'\gamma} ,
    \label{eq:hnu_partial_wave}
\end{align}
where
\begin{gather}
    J_s^{\nu lm'} =
    \int d\mathbf r^3 \,
    R_\nu^l(r) Y_{lm'}^*(\theta,\varphi) V_c(\mathbf r) \, \psi_s^{(1)}(\mathbf r),
    \label{eq:define-js}
    \\
    K_r^{\nu lm'} =
    \int d\mathbf r^3 \,
    Q_r^{\nu l}(r) Y_{lm'}^*(\theta,\varphi) V_c(\mathbf r) \, \psi^{(0)}(\mathbf r),
    \label{eq:define-kr}
\end{gather}
with $r = 1,2$ and $s = 1,2,3$. \myorange{Equations \eqref{eq:define_hnu} and \eqref{eq:hnu_partial_wave} give the formulas for $h_\nu$ in the explicit and partial-wave methods, respectively.} We have thus obtained all of the quantities needed by OE-WFAT(1) in the partial-wave formulation. 

A procedure to efficiently perform this series of calculations consists of the following steps. First, the summations in \autoref{eq:Omega0_expand} and \autoref{eq:Omega1_expand} have to be truncated at a certain maximum angular momentum $L_\textrm{max}$. Then, before the loop over orientation angles, calculate $o_r^\nu$, $I_{lm'}^\nu$, $J_s^{\nu lm'}$, and $K_r^{\nu lm'}$ for all values of their indices. These are the most computationally expensive quantities due to their dependence on integrals that can only be computed numerically. 
Once $o_r^\nu$, $I_{lm'}^\nu$, $J_s^{\nu lm'}$, and $K_r^{\nu lm'}$ have been obtained for all of their indices, inside the loop that goes through all orientation angles, one can finally calculate the rate via Eq. \eqref{eq:total_ion_rate}, \eqref{eq:wfat0_rate}, and \eqref{eq:wfat1_rate}. The speed-up factor relative to the explicit method realized by this procedure scales linearly with the number of orientation angles. The procedure above applies to non-degenerate $\psi^{(0)}$. For degenerate situations, some additional computations need to be inserted before the final step of calculating ionization rate. This will be explained in the next section.

\subsection{Degenerate eigenstates \label{sec:degenerate}}
The correction terms resulting from a perturbation expansion, e.g. \autoref{eq:define_psi_s}, depend on energy differences appearing in the denominator. In the presence of degeneracy, this necessitates a special treatment of the unperturbed eigenfunctions so that the perturbation series does not have any singularities. In this additional treatment, the unperturbed eigenfunctions are transformed such that the dipole matrix component along the applied field within the degenerate subspace is diagonal, that is, $D_{ij} = \langle v_i | \sum_{s=1}^3 \hat R_{s3} x_s | v_j \rangle \propto \delta_{ij}$ with $v_i,v_j \in \mathcal D[n]$.
Therefore, at every orientation angle, one has to diagonalize a $\text{dim}(\mathcal D[n])$-by-$\text{dim}(\mathcal D[n])$ matrix $\mathbf D$ whose elements are given by $D_{ij}$ as defined above. So, suppose that $\mathbf t_{n'}$ is a particular eigenvector of $\mathbf D$, then the corresponding \rlremove{wave} \rladd{eigen}function is given by
\begin{gather}
    \tilde v_{n'} (\mathbf r) = \sum_{i; v_i \in \mathcal D[n]} t_{in'} v_i(\mathbf r),
    \hspace{3mm} 
    n' \in [1, \text{dim}(\mathcal D[n])],
\end{gather}
where $t_{in'}$ are the $i$-th element of $\mathbf{t}_{n'}$. Since $g_\nu$ and $h_\nu$ are linear with respect to the unperturbed ionized wave function, their values associated with the ionization from $\tilde v_{n'}(\mathbf{r})$ can be calculated as a linear combination of their values for ionization from $v_i(\mathbf{r})$,
\begin{gather}
    \tilde g_{n'\nu} = \sum_{i; v_i \in \mathcal D[n]} t_{in'} g_{i\nu},
    \label{eq:new_gnu}
    \\
    \tilde h_{n'\nu} = \sum_{i; v_i \in \mathcal D[n]} t_{in'} h_{i\nu},
    \label{eq:new_hnu}
\end{gather}
where $g_{i\nu}$ is calculated through Eq. \eqref{eq:gnu_partial_wave} using $v_i(\mathbf r)$ in place of $\psi^{(0)}$. $h_{i\nu}$ is calculated similarly through Eq. \eqref{eq:hnu_partial_wave}.

\myorange{Since $\mu_z$, $\alpha_{zz}$, and the choice of origin, which are the properties of the ionized orbital, are required in OE-WFAT(1), transforming the ionized orbitals from $v_n$ to $\tilde v_{n'}$ must generally be followed by updating the aforementioned orbital properties. The new dipole moment, $\tilde \mu_z$, and polarizability component, $\tilde \alpha_{zz}$, are calculated using \autoref{eq:rotate_dipole} and \autoref{eq:rotate_polarizability}, respectively, except that $\psi^{(0)}$ is replaced by $\tilde v_{n'}$. If, however, the optimal origin is used \cite{wfat_partial_wave1-2017, ir_oewfat_grid-2018}, which is the case in the present work, $\tilde \mu_z$ and $\mu_z$ are identical and equal to the total dipole vector of the neutral (see Appendix \ref{app:origin}), so updating the dipole moment becomes unnecessary. Therefore, when the optimal origin is used, one only needs to update $\alpha_{zz}$ and the optimal origin itself.
However, since the choice of origin affects $g_{i\nu}$ and $h_{i\nu}$, updating the origin means that $\tilde g_{i\nu}$ and $\tilde h_{i\nu}$ cannot be computed through \autoref{eq:new_gnu} and \autoref{eq:new_hnu}---they will have to be recomputed at every orientation angle using the expensive \autoref{eq:define_gnu} and \autoref{eq:define_hnu}. To avoid this expensive step, in the remainder of this paper, we therefore assume that the change in origin is negligible for all orientation angles, a condition which is satisfied when the off-diagonal elements of $\mathbf{D}$ are much smaller than its diagonal elements. We will show that this approximation is well satisfied in \molbrmet{}.}

The total ionization rate including the first-order correction from this degenerate manifold is then calculated as 
$
    \Gamma = \sum_{n'; \tilde v_{n'} \in \mathcal D[n]} \Gamma(n'),
$
where $\Gamma(n')$ is the ionization rate from $\tilde v_{n'}$ obtained through Eq. \eqref{eq:total_ion_rate}.

\section{Results and Discussion}
\subsection{Validation of the partial wave expansion: noble gas elements and \molhhc{} \label{sec:validation}}

\begin{table*}[t]
   \caption{\label{tab:properties} 
   Ionization properties of noble gas atoms calculated using OE-WFAT(1) with partial wave expansion. 
   $\boldsymbol{\alpha}_\textrm{MF}$ is the molecular-frame polarizability tensor (see \autoref{eq:mf_polarizability}), $a_\nu = o_\nu + h_\nu/g_\nu$, and $B_\nu = \tilde B_\nu + 2 \Re{a_\nu}$. \myorange{The values of $a_{00}$ and $B_{00}$ inside the parentheses are taken from Ref. \cite{wfat1_atom_ref-2013}, which are obtained using TR OE-WFAT(1), and are used as the reference for our partial-wave method.} All values are in atomic units.
   }
   \begin{ruledtabular}
      \begin{tabular}{crrrrrrr}
         & & & & \\[-0.9em]
         Atom &
         $u_1$ &
         $u_2$ &
         $E^{(0)}$&
         $[\boldsymbol \alpha_\text{MF}]_{zz}$ &
         $a_{00}$ &
         $A_{00}$ &
         $B_{00}$ \\
         & & & & & & & \\[-0.9em]
         \hline
         & & & & & & & \\[-0.8em]
         Ne   &  $1.704$  &   $2.810$  &  $-0.793$  &  $ 0.152$  &  $-0.882$ ($-0.8$)  &  $ 0.246$  &  $ -2.786$ ($ -2.6$) \\
         Ar   &  $0.933$  &   $3.600$  &  $-0.579$  &  $ 1.323$  &  $-2.184$ ($-2.1$)  &  $ 0.158$  &  $ -7.752$ ($ -7.7$) \\
         Kr   &  $1.340$  &   $4.311$  &  $-0.515$  &  $ 2.098$  &  $-2.849$ ($-2.8$)  &  $ 0.042$  &  $-10.518$ ($-10.5$) \\
         Xe   &  $1.048$  &   $5.197$  &  $-0.446$  &  $ 3.080$  &  $-4.791$ ($-4.8$)  &  $-0.222$  &  $-16.459$ ($-16.4$) \\
      \end{tabular}
   \end{ruledtabular}
\end{table*}

We will first verify the validity of our partial-wave formulation of OE-WFAT(1)
using single-electron ionization rates from Ne, Ar, Kr, Xe, and \molhhc{}.
As a reference, we use the results for Ne, Ar, Kr, and Xe in Ref. \cite{wfat1_atom_ref-2013}, and \molhhc{} in Ref. \cite{oewfat1-h2_c-2015}, \pink{which employed OE-WFAT(1) in the TR}.

We model our noble gas targets using a single active electron approximation, where the effective potential takes the form of
\begin{gather}
    V(r) = - \frac{Z_\text{eff}(r)}{r}.
\end{gather}
The effective charge $Z_\text{eff}(r)$, which models the Coulomb screening by the inactive electrons, is given by \cite{wfat1_atom_ref-2013}
\begin{gather}
    Z_\text{eff}(r) = N - (N-1)\left[ 1 - \left(\frac{u_2}{u_1}(e^{u_1r}-1) + 1\right)^{-1} \right],
    \label{eq:zeff}
\end{gather}
where $u_1$ and $u_2$ are screening parameters that can be adjusted to tune the ionization potential. The unperturbed wave functions $\psi^{(0)}(\mathbf r)$ of the active valence electron in He, Ne, Ar, Kr, and Xe are described by the $2p_0$, $3p_0$, $4p_0$, and $5p_0$ states, respectively, where the subscript indicates the magnetic quantum number with the quantization axis being along $z_\text{LF}$. \fgreen{Table \ref{tab:properties} presents the values of the screening parameters, binding energy, and polarizability component of our noble gas targets.}
We discretize the Schr\"odinger equation for our noble gas targets and \molhhc using a basis formed by a direct product between finite element discrete variable representation (FEDVR) radial functions \cite{fedvr-grid-scatter-2000, fedvr-three-body-2004, fedvr-parallel-solver-2006, fedvr-quant_dyn-2011} and associated-Legendre DVR (ALDVR) angular functions (see Appendix \ref{app:dvr_basis}).

\begin{figure}[!tbp]
    \includegraphics[width=\columnwidth]{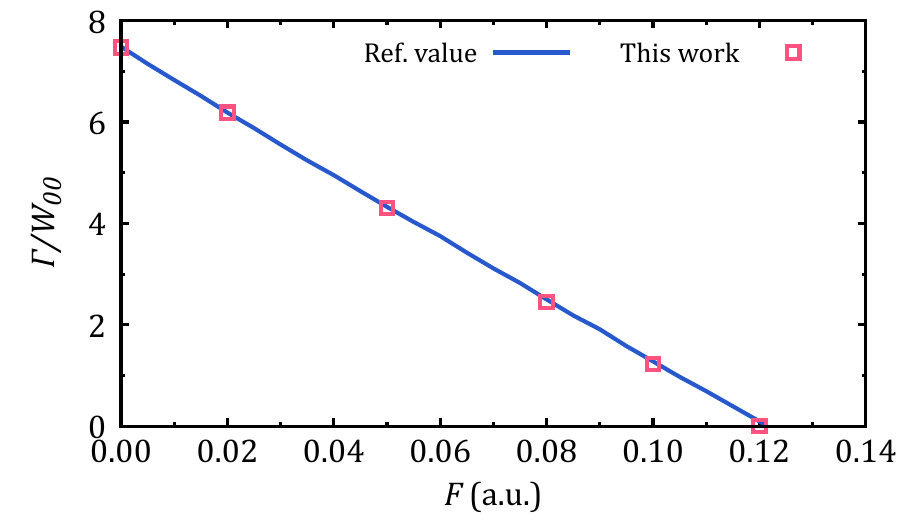}
    \caption{Normalized ionization rate from Ar for a range of field strengths. The reference value is extracted from Ref. \cite{wfat1_atom_ref-2013}.}
    \label{fig:ar-scanfield}
\end{figure}

Using OE-WFAT(1) with partial-wave expansion, we calculated several OE-WFAT(1) parameters in our noble gas targets taken at $(\beta,\gamma) = (\ang{0},\ang{0})$, and present them in \autoref{tab:properties}.
We can see that $a_{00}$ and $B_{00}$ \myorange{obtained in the present work} agree up to the first decimal place with those obtained using the TR OE-WFAT(1) (the values inside the parentheses) \cite{wfat1_atom_ref-2013}.
The normalized ionization rate of Ar at several field strengths is compared to the result of Ref. \cite{wfat1_atom_ref-2013} in Fig. \ref{fig:ar-scanfield}, where it is seen that the agreement is excellent. For our atomic targets, due to their spherical symmetry and the chosen parabolic channel of $\nu=(0,0)$, $L_\textrm{max}=2$ is sufficient---the partial-wave components of $I_{l0}^{00}$, $J_s^{(00)l0}$ for all $s$, and $K_r^{(00)l0}$ for all $r$ where $l > 2$ are identically zero.

\begin{figure}[!tbp]
    \includegraphics[width=\columnwidth]{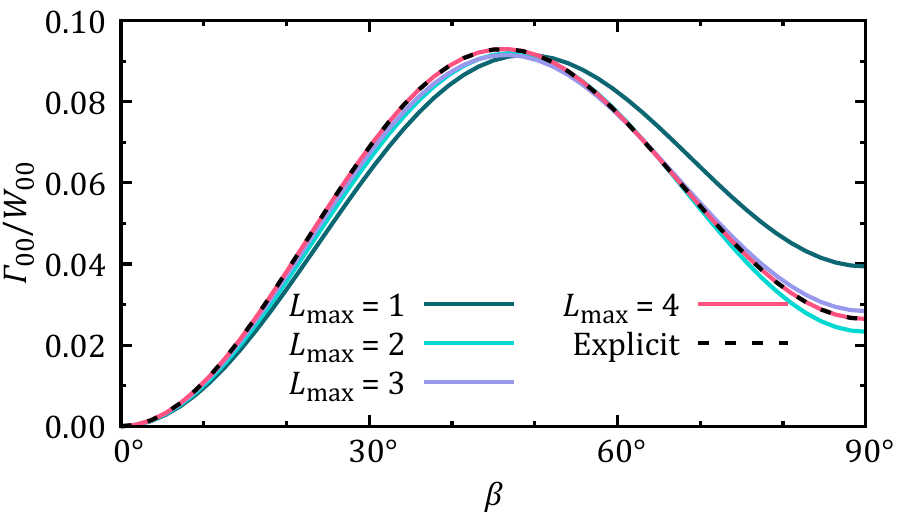}
    \caption{
    Angular momentum convergence of $\Gamma_{00}^{(1)}/W_{00}$ from the $2p\pi^+$ state of \molhhc{} calculated at $F=0.018$.}
    \label{fig:h2_c-scanlmax}
\end{figure}

Next, we examine \molhhc{}. We set the internuclear distance to \qty{1.058}{\AA} (\qty{2}{bohr}). Figure \ref{fig:h2_c-scanlmax} presents the convergence behavior of $\Gamma_{00}^{(1)}/W_{00}$ at $F=0.018$ from the $2p\pi^+$ eigenstate in terms of the maximum angular momentum of the partial-wave expansions of Eqs. \eqref{eq:Omega0_expand} and \eqref{eq:Omega1_expand}. Here, we can see that the rate converges at $L_\textrm{max} = 4$ even though the rate at $L_\text{max}=2$ already has a nice agreement with the \pink{rate obtained using the explicit formula \autoref{eq:define_gnu} and \autoref{eq:define_hnu} (denoted as ``Explicit'' in \autoref{fig:h2_c-scanlmax})}. The convergence is fast because the eigenfunctions of \molhhc{} deviate only slightly from eigenfunctions of hydrogen-like atoms.

\begin{figure}
    \centering
    \includegraphics[width=1\linewidth]{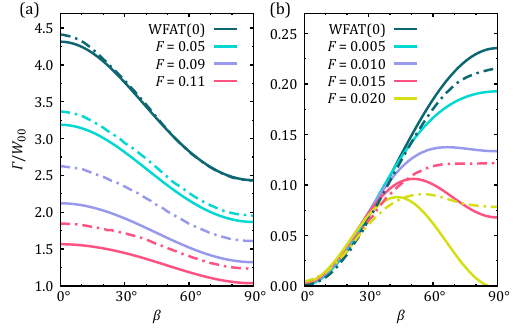}
    \caption{Normalized rates from the (a) $1s\sigma$ and (b) $2p\pi^+$ states of \molhhc{} for several field strengths obtained using the partial-wave formulation of OE-WFAT(1) (solid lines). \myorange{Whenever available, we also show the normalized rates calculated with the same field strength extracted from Ref. \cite{oewfat1-h2_c-2015} as the dot-dashed lines. $z_\text{MF}$ is along the molecular axis.}}
     \label{fig:h2_c-scanfield}
\end{figure}

We illustrate the effect of the first-order correction in the ionization from the $1s\sigma$ and $2p\pi^+$ eigenstates as the field strength increases in Fig. \ref{fig:h2_c-scanfield}(a) and \ref{fig:h2_c-scanfield}(b), respectively. For these results, we use $L_\textrm{max} = 12$. We note that the normalized rate at $F=0$, denoted as ``WFAT(0)'', is identical to the result without the first-order correction \myorange{(compare \autoref{eq:wfat1_rate} to \autoref{eq:wfat0_rate})}.
One may see that there are noticeable discrepancies between our results and those from Ref. \cite{oewfat1-h2_c-2015} in \autoref{fig:h2_c-scanfield}. We speculate that these discrepancies are due to: (1) the difference in the radial basis for discretizing the wave functions (Laguerre-based DVR in Ref. \cite{oewfat1-h2_c-2015} vs. FEDVR here), and (2) the difference in the WFAT representation (TR in Ref. \cite{oewfat1-h2_c-2015} vs. IR in the present work). In the WFAT formulation, ionization rates are sensitive to the behavior of the ionized wave functions in the asymptotic region. The two possible factors mentioned previously determine the behavior of the computation in this region---(1) Laguerre functions have an infinite range, while FEDVR functions are localized 
\footnote{We choose to use FEDVR instead of Laguerre-based DVR for the radial basis because the latter is sensitive to the choice of the scaling factor in the argument of the exponential prefactor appearing in Laguerre functions, which, if improperly chosen, can make the WFAT integrals (Eq. \eqref{eq:define-kr}, \eqref{eq:define-js}, and \eqref{eq:define-i}) diverge.}, 
and 
(2) the TR involves the evaluation of a limit $\eta \to \infty$
while the IR does not involve such a procedure. Despite this discrepancy, the trend of the first-order-corrected rates as the field increases in our results is the same as those in Ref. \cite{oewfat1-h2_c-2015}.

\subsection{Orbital distortion effect in \molco{} and \molocs{} \label{sec:co_ocs}}

Orientation-dependent ionization rates of \molco{} have been extensively studied in previous works---experimental rates can be found in Refs. \cite{CO-NO-twocolor-2011, CO-ionrate1-2012}, \pink{while theoretical rates obtained using a range of methods can be found in Refs. \cite{CO-dyncore-2013, wfat-structurefactors-2013, wfat-structurefactors-atomic-data-2015, CO_hacc-2015, CO-orbdistort-2015, CO-permdipole-2017, CO-permdipole-2017, CO_tdse_sae-2020, wahyutama-mewfat-fundamental-2025}. The rates in most of these works are calculated at a particular ionizing field value and agree with experimental data in which the maximum rate occurs when the field points from C to O \cite{CO-ionrate1-2012, CO-dyncore-2013}. A scan of field strength is performed in Ref. \cite{CO_hacc-2015}, which found that the ratio of the rate when the field points from C to O to that at the opposite orientation gets larger as the field gets stronger.}

Orbital distortion effects in CO have been investigated in Ref. \cite{CO-orbdistort-2015} \pink{using strong-field approximation and its modifications. In the present work, we investigate the same effect by performing a field scan, which goes beyond what was done in Ref. \cite{CO-orbdistort-2015}.}
\myorange{A field scan to study orbital distortion effects is impossible with OE-WFAT(0) because the field and orientation dependences are decoupled. This means that OE-WFAT(0) cannot reconcile the discrepancy with experiments taken at high intensities \cite{wfat-structurefactors-2013, wfat-structurefactors-atomic-data-2015}.} \pink{In fact, the rate calculated by OE-WFAT(0) disagrees with the published experimental rate \cite{wfat-structurefactors-2013, wfat-structurefactors-atomic-data-2015}.}
\pink{Here, we demonstrate that orbital distortion, as treated in OE-WFAT(1), is necessary to achieve qualitative agreement with these experiments, thereby illuminating the underlying physics that governs the experimental observations.}

\begin{figure*}[t]
    \centering
    \includegraphics[width=\linewidth]{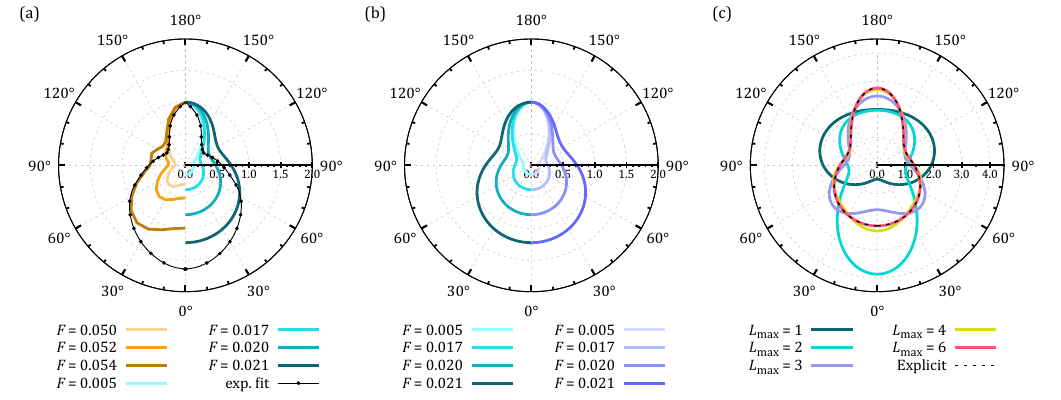}
    \caption{$\beta$-dependent ionization rates from \molco{} for several field strengths and $L_\text{max}$'s. \myorange{$z_\text{MF}$ is along the molecular axis, and} at $\beta=\ang{0}$, the field points from C to O.
     (a) Left: yields calculated using RT-TDDFT. Right: total rates $\Gamma$ calculated by OE-WFAT(1), normalized so that the rates at $\beta=\ang{180}$ are equal to unity. The fit of an experimental data extracted from Ref. \cite{CO-dyncore-2013} is also presented (line with circles).
     (b) Left: The same curves as in the right half of panel (a). Right: Same as the left half of the current panel except that in \autoref{eq:total_ion_rate}, we set $\Gamma_{0,\pm 1}^{(0)}=0$. The normalization factor of the rate in the left half for the same field is used in the right half.
     (c) Angular momentum convergence of $\Gamma_{00}^{(1)}/W_{00}$ calculated at $F=0.02$.}
     \label{fig:co-scanfield}
\end{figure*}

\myorange{\pink{To carry out calculations on \molco{}}, we set the C-O internuclear distance to} \qty{1.124}{\AA}, and then perform a density functional theory (DFT) calculation using the tuned range-separated (RS) functional LC-PBE0* with RS parameters of $\alpha_\textrm{RS}=0.27$ and $\gamma_\textrm{RS}=0.37$ and the cc-pvtz basis to obtain the HOMO that will be used as the ionized orbital. The HOMO has a binding energy of \qty{-0.522076}{H} and \rlremove{has} a $\sigma$ symmetry.
The total ionization rates of \molco{} calculated using OE-WFAT(1) (\autoref{eq:total_ion_rate}) at several field strengths with $L_\textrm{max}=15$ are shown as solid lines in the right half of \autoref{fig:co-scanfield}(a). 
\myorange{We can see that, as the field increases from $0.005$ to $0.021$, the orientation of the maximum shifts from $\beta=\ang{180}$ (the field pointing from O to C), which is also the orientation of the maximum of OE-WFAT(0)-calculated rates \cite{wfat-structurefactors-2013, wfat-structurefactors-atomic-data-2015}, to $\beta=\ang{0}$}.
At $F=0.021$, the OE-WFAT(1) rate resembles the experimental data (line with circles), although it cannot capture the slight elongation along $\beta=\ang{0}$. 
\rlremove{The field-dependent behavior revealed by OE-WFAT(1) is in agreement with the result in Ref. [55], which employs hybrid anti-symmetrized coupled channel basis (haCC) method, and our real-time time-dependent DFT (RT-TDDFT) simulations, shown in the left half of Fig. 5(a). We note that the haCC and RT-TDDFT results are independent of perturbation order. Our RT-TDDFT calculations employ the same functional as above and aug-cc-pvtz basis augmented with some additional diffuse basis functions.}

\rladd{We also calculate ionization rates from \molco{} using real-time time-dependent DFT (RT-TDDFT) following the procedure given in Ref. \cite{angle-sfi-tuned-rs-2016} and show the results in the left half of \autoref{fig:co-scanfield}(a). Our RT-TDDFT simulations on \molco{} employ the same functional as the DFT calculation above to obtain the HOMO, aug-cc-pvtz basis augmented with some additional diffuse basis functions to capture the ionized electron flux, a time step of \qty[mode=text]{0.25}{$\text{a.u.}$}, and a quasi-static field as described in Ref. \cite{ch3x_ionization-2017}. Our OE-WFAT(1) results are seen to be in good agreement with the RT-TDDFT results.}
The peak at $\beta=\ang{40}$ in the RT-TDDFT results, however, is absent in the OE-WFAT(1) results. \pink{Understanding this discrepancy requires further investigation beyond the scope of this paper.} 
\rladd{Our OE-WFAT(1) simulations also reproduce the field-dependent behavior of the rate from \molco{} revealed by hybrid anti-symmetrized coupled channel basis (haCC) method in Ref. \cite{CO_hacc-2015}. We note that both haCC and RT-TDDFT are independent of perturbation order.}

To understand parabolic channel effects in \molco{}, we calculated the total rates with the $\Gamma_{0,\pm1}^{(0)}$ contributions omitted from \autoref{eq:total_ion_rate}, shown in the right half of \autoref{fig:co-scanfield}(b). They look similar to the rates that include the $\Gamma_{0,\pm1}^{(0)}$ contributions (left half) with the similarity decreasing as $F$ increases. This indicates that parabolic channel effects in the strong-field ionization from \molco{} are negligible.

\pink{We note that the range of field strengths over which the maximum is observed to shift is different in our OE-WFAT(1) calculations and our RT-TDDFT simulations.} This range may also be different from the experimental range \cite{CO-dyncore-2013}.
Such a difference in the range of fields is caused by the limited accuracy of the ionization potential and the imperfect asymptotic tail modeled by Gaussian basis, \pink{as used in \textsc{NWChem}} \cite{nwchem-2010, nwchem2-2020}.
Consequently, WFAT-calculated rates from a wave function expanded in Gaussian bases \myorange{can be} used to understand the trend of the orientation-dependent ionization rate as a function of external field, \pink{but they are less reliable for predicting the magnitude at any particular field strength}.

Figure \ref{fig:co-scanfield}(c) shows the convergence behavior of the normalized ionization rates of \molco{} in terms of $L_\textrm{max}$. 
The curve reaches convergence at $L_\text{max}=6$.
Running OE-WFAT(1) simulations on $101$ different $\beta$ points to produce \autoref{fig:co-scanfield}(c), we observe a speed-up of about $68$ times when using our partial-wave method with $L_\textrm{max}=6$ compared to the explicit method.

Like \molco{}, the strong-field ionization properties of \molocs{} have been widely studied. In some works, alignment- instead of orientation-dependent ionization yields from \molocs{} are obtained where the maximum ionization occurs at $\beta=\ang{90}$ \cite{OCS-circpol-2011, OCS-CS2-tddft-2011, OCS-asymtop-2011, OCS-align-sfi-2016, OCS-single-double-2018, OCS-multiel-2020}. In the present work, we calculate orientation- rather than alignment-dependent ionization rates of \molocs{} since alignment conflates features in the ionization rates at $\beta$ and $\ang{180} - \beta$. This will suppress the weaker features \pink{between the two orientations}. Orbital distortion effects in OCS were studied in Ref. \cite{CO-orbdistort-2015}, but, just like CO, \pink{a field scan adds to the understanding of orbital distortion effects}.

\begin{figure*}[t]
    \centering
    \includegraphics[width=\linewidth]{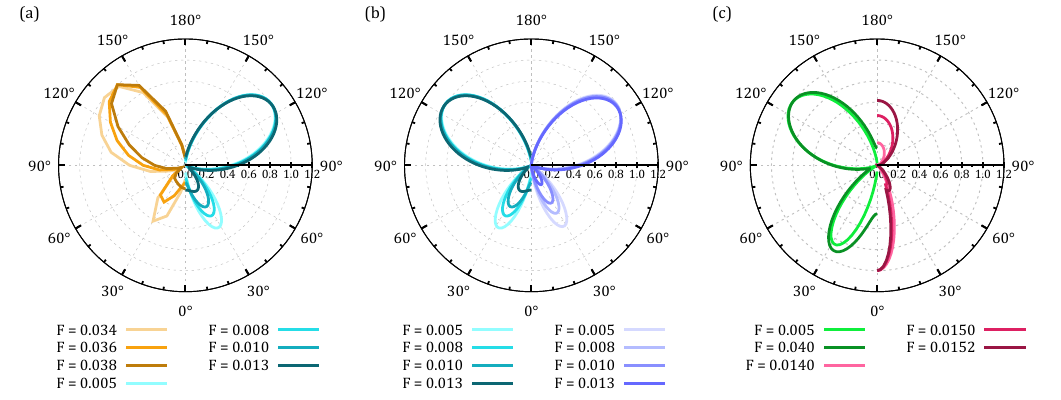}
    \caption{$\beta$-dependent ionization rates from \molocs{} for several field strengths. \myorange{$z_\text{MF}$ is along the molecular axis, and} at $\beta=\ang{0}$, the field points from O to S.
    (a) Left: yields calculated using RT-TDDFT. Right: total rates $\Gamma$ calculated by OE-WFAT(1). 
    (b) Left: the same curves as in the right half of panel (a). Right: same as the left half of the current panel except that in \autoref{eq:total_ion_rate}, we set $\Gamma_{0,\pm 1}^{(0)}=0$. The normalization factor of the rate in the left half for the same field is used in the right half.
    (c) Left: total rates $\Gamma$ calculated by OE-WFAT(0), that is, $\Gamma_{00}^{(1)}$ in \autoref{eq:total_ion_rate} is replaced with $\Gamma_{00}^{(0)}$. Right: total rates $\Gamma$ calculated by OE-WFAT(1) for ionization from HOMO-1. In panels (a) and (c), both halves are normalized so that the maximum of each rate is equal to unity.}
    \label{fig:ocs-scanfield}
\end{figure*}

OCS is a linear molecule with O--C and C--S bond lengths of \qty{1.153}{\AA} and \qty{1.562}{\AA}, respectively.
We choose the ionized state to be the HOMO obtained through a DFT calculation employing a tuned RS functional of LC-PBE0* using RS parameters of $\alpha_\textrm{RS}=0$ and $\gamma_\textrm{RS}=0.409$ and a basis set of cc-pvtz. The binding energy $E^{(0)}$ of the HOMO is \qty{-0.418839}{H}.
The HOMO of \molocs{} is doubly degenerate, nevertheless, the algorithm for degenerate states in \autoref{sec:degenerate} is unnecessary since $\mathbf{D}$ is already diagonal in \molocs{}'s HOMO degenerate subspace. The total rate is given by the sum of the rates from the two degenerate HOMOs (see \autoref{sec:degenerate}). This results in an axially symmetric total rate, i.e., $\Gamma$ is a function of $\beta$ only, so that the analyses can be confined to a slice at an arbitrary $\gamma$. For all OE-WFAT calculations on \molocs{} in the following, we use $L_\textrm{max}=15$.
\rlremove{The right half of Fig. 5(a) shows} \rladd{Figure \ref{fig:ocs-scanfield}(a) compares} the slices of orientation-dependent total rates of \molocs{} at several field strengths calculated \rladd{by OE-WFAT(1) }through \autoref{eq:total_ion_rate}\rladd{(right half) and RT-TDDFT (left half). The functional used in the RT-TDDFT simulations is the same as that in the DFT calculation for this molecule above, while the basis type, time step, and the form of the quasi-static field are the same as those in the \molco{} RT-TDDFT simulations.}
We can see that OE-WFAT(1) \rlremove{(right half)} have an excellent agreement with RT-TDDFT \rlremove{(left half)} in the behavior of the orientation-dependent rates as the field strength increases except for the \rlremove{stretching} \rladd{narrowing} of the lobes \rlremove{toward the parallel orientation} in the RT-TDDFT results, which may be due to higher-than-first-order effects. 

The right half of \autoref{fig:ocs-scanfield}(b) shows OE-WFAT(1)-calculated rates at several field strengths with the contributions from parabolic channels $\nu = (0,\pm 1)$ omitted in \autoref{eq:total_ion_rate}. For comparison, results for the same field strength shown in the right half of \autoref{fig:ocs-scanfield}(a) are shown in the left half of \autoref{fig:ocs-scanfield}(b). The overall shapes of the curves in both halves of \autoref{fig:ocs-scanfield}(b) are very similar, with the main difference being the rates at $\beta=\ang{0}$. This indicates that, like the case of \molco{}, orbital distortion plays an important role in strong-field ionization of \molocs{}, and that parabolic channel effects are negligible. In \molocs{}, orbital distortion modifies orientation-dependent rates by suppressing the rates around $\beta=\ang{0}$, that is, at orientations when the field has a component that points from O to S, relative to the rates in the opposite orientations.

In the left half of \autoref{fig:ocs-scanfield}(c), we demonstrate the effect of the omission of first-order corrections in WFAT, that is, when $\Gamma_{00}^{(1)}$ in \autoref{eq:total_ion_rate} is replaced by $\Gamma_{00}^{(0)}$. As can be seen, even though the two field strengths used are very different, the overall shapes are still similar. The effects of excited states are shown in the right half of \autoref{fig:ocs-scanfield}(c), which shows ionization rates from HOMO-1. It is evident that orbital distortion causes the lobe at $\beta=\ang{0}$ to become weaker relative to the lobe at $\beta=\ang{180}$ as the field strength increases. This means that as field-induced effects become stronger, one can expect to observe increasing contribution from the first excited state at orientations where the field has a component that points from S to O (around $\beta=\ang{180}$). This region of orientations is the opposite of the region where orbital distortion-induced modification is the most prominent, namely around $\beta=\ang{0}$.

We would like to point out that the orbital distortion-induced modifications to the ionization rates in \molco{} and \molocs{} studied here are qualitatively similar to the modifications resulting from the use of ME-WFAT(0) in our earlier publication \cite{wahyutama-mewfat-fundamental-2025}. However, the cause of the modifications is entirely different. In the case of Ref. \cite{wahyutama-mewfat-fundamental-2025}, it is the dipole moment that makes the rates in \molco{} and \molocs{} calculated by ME-WFAT(0) different from those calculated by OE-WFAT(0). \myorange{Orbital distortion in the present work and the many-electron treatment in Ref. \cite{wahyutama-mewfat-fundamental-2025} independently improve the rates in \molco{} and \molocs{} calculated by OE-WFAT(0). When combined together, 
we expect that the trend in the field dependence of ionization rates observed here should be preserved. In weakly correlated molecules, the main difference from the present results should be in the magnitude of the rates for the same field strength, and thus, also in the range of fields in which the field-dependent effects in the present work are observed.}

\subsection{Parabolic channel effect in \molbrmet{} \label{sec:ch3br}}

\begin{figure*}[t]
    \centering
    \includegraphics[width=\linewidth]{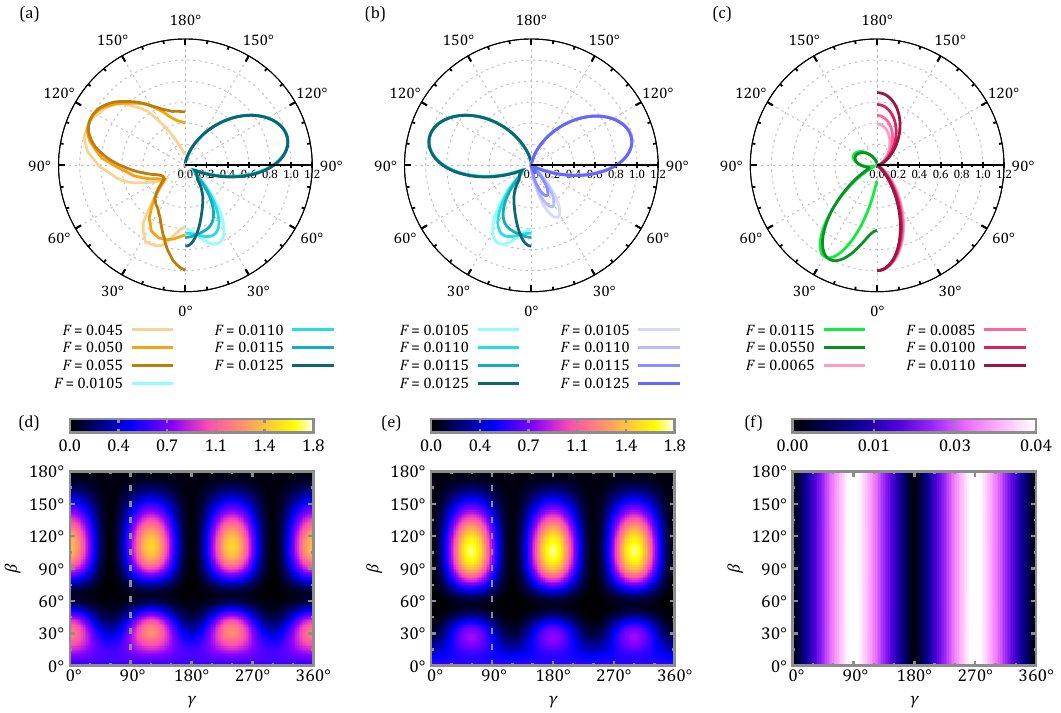}
    \caption{(a-e) Orientation-dependent ionization rates from \molbrmet{} for several field strengths. The definition of $(\beta,\gamma)$ for this molecule is shown in \autoref{fig:angle_def}.
     (a) Left: yields calculated using TDCI, extracted from Ref. \cite{ch3x_ionization-2017}. Right: total rates $\Gamma$ calculated by OE-WFAT(1). 
     (b) Left: the same curves as in the right half of panel (a). Right: same as the left half of the current panel except that in \autoref{eq:total_ion_rate}, we set $\Gamma_{0,\pm 1}^{(0)}=0$. The normalization factor of the rate in the left half for the same field is used in the right half.
     (c) Left: $\Gamma$ calculated using OE-WFAT(0), that is, $\Gamma_{00}^{(1)}$ in \autoref{eq:total_ion_rate} is replaced with $\Gamma_{00}^{(0)}$. Right: total rates $\Gamma$ calculated by OE-WFAT(1) for ionization from HOMO-1. In panels (a) and (c), both halves are normalized so that the maximum of each rate is equal to unity.
     (d) and (e) show the full orientation-dependent normalized rates at $F=0.0110$ from the two HOMOs of \molbrmet{} that diagonalize $\mathbf{D}$.
     (f) The average of $|\boldsymbol{\mu}_\text{MF} - \tilde{\boldsymbol{\mu}}_\text{MF}|$ in bohr for the two HOMOs.
     }
    \label{fig:ch3br}
\end{figure*}

Orientation-dependent ionization yields of \molbrmet{} have previously been calculated using time-dependent configuration interaction (TDCI) \cite{ch3x_ionization-2017, cap-reduce-cost-2024} and RT-TDDFT \cite{halomethane_ionization-2019}. \molbrmet{} belongs to the $C_{3v}$ point group. For our simulations, the C--Br and C--H bond lengths are \qty{1.943}{\AA} and \qty{1.075}{\AA}, respectively, and the HCBr angle is \ang{107.71}.
Since we will compare our OE-WFAT(1) simulations against TDCI results in Ref. \cite{ch3x_ionization-2017} where the orbitals are generated through a Hartree-Fock calculation, we use this method to obtain the HOMO of \molbrmet{}. The HOMO is doubly degenerate and has a binding energy of \qty{-0.401145}{H}. 
Both HOMOs have a nodal plane that contains the C--Br bond and a nodal surface located somewhere between C and Br.
When the ionizing intensity is low, these nodal planes determine the orientations of minimum ionization rate.
This can be seen from the yield curve for $F=0.045$ calculated by TDCI shown in the left half of \autoref{fig:ch3br}(a) (reproduced from Ref. \cite{ch3x_ionization-2017}).
It shows that local minima occur when the field is parallel or antiparallel to the C--Br bond, or when it makes an angle of $\beta = {\sim} \ang{60}$ with this bond.
As the field increases, the minimum at $\beta = \ang{0}$ turns to a local maximum when the field reaches $0.055$. In this section, we will reveal the mechanism responsible for this minimum-to-maximum transformation.

In our OE-WFAT(1) calculations, we use the algorithm for degenerate states in \autoref{sec:degenerate} since $\mathbf{D}$ is not diagonal in the subspace of the degenerate HOMOs of \molbrmet{}. 
For all OE-WFAT calculations on \molbrmet{} in the following, we use $L_\textrm{max}=15$.
Using OE-WFAT(1), we have calculated the total rate for several field strengths, their slices at $\gamma=\ang{90}$ are shown in the right half of \autoref{fig:ch3br}(a), where we see that the behavior of the minimum at $\beta = \ang{0}$ as field strength increases as computed by TDCI (left half of \autoref{fig:ch3br}(a)) is reproduced. Even though the ranges of field strength used for the left and right halves of \autoref{fig:ch3br}(a) are different, we can see that both show the same trend toward higher field strengths. 

Figure \ref{fig:ch3br}(b) elucidates the importance of parabolic channels $\nu= (0, \pm1)$ to the observed behavior of the rate at $\beta=\ang{0}$. The left half of \autoref{fig:ch3br}(b) shows the same data as those in the right half of \autoref{fig:ch3br}(a) whereas the right half shows the rates obtained by omitting $\Gamma_{0,\pm1}^{(0)}$ in \autoref{eq:total_ion_rate} using the same fields as the left half. Here we see that the rate at $\beta=\ang{0}$ is always a minimum for all field strengths.
From \autoref{fig:ch3br}(a) and \autoref{fig:ch3br}(b), we can therefore understand that the observed minimum-to-maximum transformation in the ionization rate at $\beta=\ang{0}$ is the result of the interplay between orbital distortion and parabolic channel contributions. As field strength increases, the former suppresses the growth of the rates in the $(0,0)$ parabolic channel around $\beta=\ang{0}$ so that at a certain field strength, the contributions from $(0,\pm1)$ parabolic channels start to dominate the ionization around this angle.

We rule out the possibility of excited state effects contributing to the minimum-to-maximum transformation of the rate at $\beta=\ang{0}$ by analyzing the orientation-dependent rate from HOMO-1, shown in right half of \autoref{fig:ch3br}(c). The behavior of the field dependence is the same as the case of HOMO-1 of \molocs{} (the right half of \autoref{fig:ocs-scanfield}(c)). This shows that the modifications of ionization rates in \molbrmet{} around orientations $\beta=\ang{0}$ cannot be caused by excited-state effects. 
These effects are likely to manifest around $\beta=\ang{180}$, so they might be responsible for the non-zero rate at this angle in the left half of \autoref{fig:ch3br}(a).
Lastly, we show the rates when the first-order correction is not included in the left half of \autoref{fig:ch3br}(c). As can be seen, it fails to reproduce the observed change in the orientation dependence even if the field has been increased from $0.0115$ to $0.055$. The full orientation-dependent plots of the rates at $F=0.0110$ from the two $\mathbf D$-diagonalizing HOMOs are shown in \autoref{fig:ch3br}(d) and \autoref{fig:ch3br}(e), where the position of the slices used for the OE-WFAT(1) data in Figs. \ref{fig:ch3br}(a)-(c) is also indicated. To produce \autoref{fig:ch3br}(d) and \autoref{fig:ch3br}(e), we run OE-WFAT(1) simulations on $18029$ $(\beta,\gamma)$ pairs. For this number of orientations, we observe a speed-up factor of about $1.5 \times 10^4$, emphasizing the efficiency of the partial-wave expansion method derived in the present work.

In \autoref{sec:degenerate}, we mentioned that \pink{we have used the approximation} in which the origins used for the $\mathbf{D}$-diagonalizing degenerate HOMOs are taken to be the same as those used for the $\mathbf{D}$-non-diagonalizing degenerate HOMOs. We show in \autoref{fig:ch3br}(f) that this approximation is well satisfied for \molbrmet{} by calculating $|\boldsymbol{\mu}_\text{MF} - \tilde{\boldsymbol{\mu}}_\text{MF}|$ averaged for the two HOMOs. 
The largest difference is ${\sim} \qty{0.04}{bohr}$, much smaller than \qty{1}{bohr}, the typical length scale for electronic wave functions. Such small differences are \pink{to be expected} since the off-diagonal elements of $\mathbf{D}$ in \molbrmet{} are much smaller than the diagonal elements.

\section{Conclusions and outlook \label{sec:conclusion}}

\pink{We have investigated two mechanisms that govern how the orientation dependence of tunneling ionization rates in several molecules change as the strength of the ionizing field increases: orbital distortion and parabolic channel contributions. While excited-state effects were expected to play an important role, we showed that, at least in \molco{}, \molocs{}, and \molbrmet{}, orbital distortion and parabolic channel effects were the dominant factors that determined how the orientation dependence of the tunneling ionization rates changed as a function of field strength. We found that: (1) first-order orbital distortion was necessary to reproduce experimental data in \molco{} and the RT-TDDFT results in \molocs{}, and (2) the contribution of higher parabolic channels was responsible for transforming a local minimum into a local maximum in the ionization rate from \molbrmet{} when the field pointed from C to Br as the field strength increased.}

\pink{We have found that first-order orbital distortion influences ionization rates by changing the ratio between rates at parallel orientations versus antiparallel orientations as field strength increases. While the positions of peaks and valleys in the orientation-dependent behavior typically remain unchanged, an important exception occurs when higher parabolic channels become significant. In these cases, the interaction between higher parabolic channels and first-order orbital distortion effects can reverse the orientation dependence, transforming what would normally be a minimum in the ionization rate into a maximum.}

Due to the profound effects orbital distortion and higher parabolic channel contributions can have on ionization rates at high field strengths, it is prudent to use a method for calculating rates that can account for these effects. In the present work, we have  shown that OE-WFAT(1) is up to the task, especially after reformulation using partial-wave expansion to greatly enhance its efficiency. We found, for example, a speed-up of about $1.5\times10^4$ times in calculations on \molbrmet{} as compared to the explicit method with virtually no loss of accuracy. 

\rladd{Taking into account our previous work in Ref. \cite{wahyutama-dft-mewfat-2022} in which adiabatic ionization was treated using WFAT, our present work demonstrated that OE-WFAT(1) could be a reliable and much more efficient alternative to explicit time-dependent methods to obtain molecular ionization rates due to a long-wavelength laser field with a correct orientation dependence.}
OE-WFAT(1) is most suitably used for molecules with low correlation effects so that the single-active electron approximation is valid. For correlated molecules, extending it to the many-electron treatment, resulting in ME-WFAT(1), is necessary.

\begin{acknowledgments}

This work was supported by the U.S. Department of Energy, Office of Science, Office of Basic Energy Sciences, under Award No. DE-SC0012462. Portions of this research were conducted with high performance computational resources provided by Louisiana State University and the Louisiana Optical Network Infrastructure.

\end{acknowledgments}

\appendix

\section{Definitions of some quantities \label{app:define_quantities}}

The improvement in OE-WFAT(1) from OE-WFAT(0) manifests in the appearance of first-order corrections to various quantities, some of them are $\psi^{(1)}(\mathbf{r})$, the static polarizability component $\alpha_{zz}$, \retaone{\nu}{\eta}, and \phione{\nu}{\xi}.
The zeroth- and first-order corrections to $\mathcal{R}_\nu(\eta)$ needed in \autoref{eq:Omega0_explicit} and \autoref{eq:Omega1_explicit} are given by \cite{ir_oewfat-2016}
\begin{align}
    \mathcal R_\nu^{(0)}(\eta) =
    &\,
    \frac{\Gamma(a)}{|m|!}
    (\varkappa \eta)^{b/2}
    e^{-\varkappa\eta/2}
    M(a,b,\varkappa\eta),
    \label{eq:define_r0_eta}
    \\
    \mathcal R_\nu^{(1)}(\eta) =
    &\,
    \frac{1}{\varkappa} \int_0^\eta d\eta'
    \left(
    \mathcal R_\nu^{(0)}(\eta') \mathcal O_\nu^{(0)}(\eta)
    - \mathcal R_\nu^{(0)}(\eta) \mathcal O_\nu^{(0)}(\eta')
    \right) \nonumber
    \\
    &\,
    \times
    \left(
    \frac{\beta_\nu^{(1)}}{\eta'}
    - \frac{\mu_z}{2}
    + \frac{\eta'}{4}
    \right)
    \mathcal R_\nu^{(0)}(\eta'),
    \label{eq:define_r1_eta}
\end{align}
where $a=(1+|m|)/2-\beta_\nu^{(0)}/\varkappa$, $b=1+|m|$, and $M(a,b,z)$ is the confluent hypergeometric function of the first kind \cite{math_handbook_abramowitz-1972}. 

The zeroth- and first-order corrections to the parabolic channel function $\phi_\nu(\xi)$ and its associated eigenvalue $\beta_\nu$ are given by \cite{wfat1_atom_ref-2013}
\begin{align}
    \phi_\nu^{(0)}(\xi) = 
    &\,
    \sqrt{\varkappa} \, (\varkappa \xi)^{|m|/2}
    e^{-\varkappa\xi/2}
    \sqrt{\frac{n_\xi!}{(n_\xi+|m|)!}}
    L_{n_\xi}^{(|m|)} (\varkappa\xi),
\end{align}
\begin{align}
    \phi_\nu^{(1)}(\xi) = 
    &\,
    \sum_{i=n_\xi-2, \, i \neq n_\xi}^{n_\xi+2}
    (C_1^{im} + \mu_z C_2^{im}) \phi_{im}^{(0)}(\xi),
    \label{eq:phi1_expand}
    \\
    \betazero{\nu} =
    &\,\,
    Z - \varkappa
    \left( n_\xi + \frac{|m|+1}{2} \right),
    \\
    \betaone{\nu} =
    &\,\,
    b_1^\nu + \mu_z b_2^\nu,
    \label{eq:define_beta1}
    \\
    b_1^\nu = 
    &\,
    -\frac{1}{4\varkappa^2}
    \left(
    6n_\xi(n_\xi+|m|+1) + m^2 + 3|m| + 2
    \right),
    \label{eq:define_b1}
    \\
    b_2^\nu = 
    &\,
    -\frac{1}{2\varkappa}
    (2n_\xi + |m| + 1).
    \label{eq:define_b2}
\end{align}
where $L_n^{(\alpha)}(x)$ are the generalized Laguerre polynomials \cite{math_handbook_abramowitz-1972}, and
\begin{subequations}
    \label{eq:phi1_expand_c}
    \begin{gather}
        C_r^{n_\xi-2,m} = \delta_{r1} c_{n_\xi-2} c_{n_\xi-1} / (8\varkappa^3),  
        \\
        C_r^{n_\xi-1,m} = -c_{n_\xi-1} \left[ \delta_{r2}\varkappa + \delta_{r1}(2n_\xi+|m|) \right] / (2\varkappa^3),
        \\
        C_r^{n_\xi+1,m} = c_{n_\xi} \left[ \delta_{r2}\varkappa + \delta_{r1} (2n_\xi+|m|+2) \right] / (2\varkappa^3),
        \\
        C_r^{n_\xi+2,m} = - \delta_{r1} c_{n_\xi} c_{n_\xi+1} / (8\varkappa^3),
        \\
        c_{n_\xi} = \sqrt{(n_\xi+1) (n_\xi+|m|+1)}.
    \end{gather}
\end{subequations}

The static polarizability component $\alpha_{zz}$ is the component of the polarizability tensor in the direction of the field in the LF, so it is given by
\begin{gather}
    \alpha_{zz} \equiv [\boldsymbol\alpha_\textrm{LF}]_{33} = \left[ R^T (\beta,\gamma) \, \boldsymbol \alpha_\textrm{MF} \, R (\beta,\gamma) \right]_{33},
    \label{eq:rotate_polarizability}
\end{gather}
with the components of static polarizability tensor in the MF given by
\begin{align}
    [\boldsymbol\alpha_\textrm{MF}]_{ss'} 
    =&\,\,
    -2 \, \big\langle \psi^{(0)} \big| x_s \big| \psi_{s'}^{(1)} \big\rangle 
    \nonumber \\
    =&\,\, 
    -2 \sum_{i; v_i \notin \mathcal D[n]} 
    \frac{\langle \psi^{(0)} |x_s| v_i \rangle \langle v_i |x_{s'}| \psi^{(0)} \rangle}{\varepsilon_n - \varepsilon_i},
    \label{eq:mf_polarizability}
\end{align}
where $s,s' = 1,2,3$. Note that $\boldsymbol\alpha_\textrm{MF}$ is a symmetric matrix.

The dependence of OE-WFAT(1) ionization rate on $\alpha_{zz}$ is encapsulated in $\tilde B_\nu$,
\begin{align}
    \tilde B_\nu = 
    &\,
    -\varkappa\alpha_{zz}
    - \frac{\mu_z^2}{\varkappa}
    + \frac{\mu_z}{\varkappa^2}
    + \frac{4\mu_z \beta_\nu^{(0)}}{\varkappa^3}
    - \frac{(9-6\gamma_m)\beta_\nu^{(0)}}{4\varkappa^4} \nonumber
    \\
    &\,
    -\frac{10+18\gamma_m+3\gamma_m^2}{24\varkappa^3}
    - \frac{(49+2\gamma_m)\beta_\nu^{(0)2}}{8\varkappa^5}
    + \frac{3\beta_\nu^{(0)3}}{2\varkappa^6} \nonumber
    \\
    &\,
    -\frac{\beta_\nu^{(0)4}}{8\varkappa^7},
    \label{eq:define_tbnu}
\end{align}
where $\gamma_m = (1-m^2)/4 - 2[\mathbf{d}_T^+]_z$. In our calculations, the origin is chosen so that $\mathbf{d}_T^+ = 0$, see Appendix \ref{app:origin}, therefore, we use $\gamma_m = (1-m^2)/4$.
While $A_\nu$ and $o_\nu$ do not depend on $\alpha_{zz}$, their orientation dependence only comes from $\mu_z$,
\begin{align}
    A_\nu = 
    &\,
    -\frac{2\beta_\nu^{(1)}}{\varkappa}
    - \frac{\gamma_m-2\mu_z \beta_\nu^{(0)}}{\varkappa^3}
    - \frac{3\beta_\nu^{(0)2}}{\varkappa^5},
    \label{eq:define_anu}
    \\
    o_\nu = &\, 
    o_1^\nu + \mu_z o_2^\nu,
    \label{eq:define_onu}
    \\
    o_1^\nu =
    &\, 
    \frac{1}{4\varkappa^3}
    \left(
    1 + \frac{\gamma_m^2}{2} - \gamma_m \left( \frac{4\beta_\nu^{(0)}}{\varkappa} - \frac{\beta_\nu^{(0)2}}{\varkappa^2} \right)
    + \frac{4\beta_\nu^{(0)}}{\varkappa} 
    \right. \nonumber
    \\
    &\,
    \left.
    + \, \frac{7\beta_\nu^{(0)2}}{2\varkappa^2}
    - \frac{4\beta_\nu^{(0)3}}{\varkappa^3}
    + \frac{\beta_\nu^{(0)4}}{2\varkappa^4}
    \right) \nonumber
    \\
    &\,
    + \frac{1}{\varkappa} \int_0^\infty d\eta
    \left(
    \left(
    \frac{b_1^\nu}{\eta} + \frac{\eta}{4}
    \right)
    \mathcal R_\nu^{(0)}(\eta) \mathcal O_\nu^{(0)}(\eta)
    - \frac{\eta}{4} \right. \nonumber
    \\
    &\,
    \left.
    - \, \frac{\beta_\nu^{(0)}}{2\varkappa^2}
    - \left(
    b_1^\nu + \frac{\gamma_m}{2\varkappa^2} + \frac{3\beta_\nu^{(0)2}}{2\varkappa^4}
    \right) \frac{1}{\eta+1}
    \right), 
    \\
    o_2^\nu =
    &\,
    -\frac{1}{2\varkappa^2}
    + \frac{1}{\varkappa}
    \int_0^\infty d\eta
    \left(
    \left(
    \frac{b_2^\nu}{\eta} - \frac{1}{2}
    \right)
    \mathcal R_\nu^{(0)}(\eta) \mathcal O_\nu^{(0)}(\eta) \right. \nonumber
    \\
    &\,
    \left.
    + \, \frac{1}{2}
    - \left(
    b_2^\nu - \frac{\beta_\nu^{(0)}}{\varkappa^2}
    \right)
    \frac{1}{\eta+1}
    \right),
\end{align}
where
\begin{gather}
    \mathcal O_\nu^{(0)}(\eta) =
    (\varkappa \eta)^{b/2}
    e^{-\varkappa\eta/2}
    U(a,b,\varkappa\eta),
    \label{eq:define_o0_eta}
\end{gather}
and $U(a,b,z)$ is the confluent hypergeometric function of the second kind \cite{math_handbook_abramowitz-1972}.

\section{Determination of the origin \label{app:origin}}
The WFAT formulation is origin independent; however, due to the inexact asymptotic behavior of numerical wave functions, the origin invariance is only approximately satisfied. Therefore, an optimal way to determine the origin is necessary; this is presented in Refs. \cite{wfat_partial_wave1-2017, ir_oewfat_grid-2018} where the origin is chosen to reduce the radial extent of $V_c(\mathbf{r})$. This is achieved by requiring the total dipole vector of the Koopman cation to be equal to zero \cite{wfat_partial_wave1-2017, ir_oewfat_grid-2018}. According to this prescription, the position vector of the optimal origin relative to an arbitrary origin in the MF is given by
\begin{equation}
    \mathbf{r}_c 
    = 
    \frac{\mathbf{d}_T^+}{Z}
    = 
    \frac{\mathbf{d}_T - \boldsymbol{\mu}_\text{MF}}{Z},
\end{equation}
where $\mathbf{d}_T^+ = \mathbf{d}_T - \boldsymbol{\mu}_\text{MF}$ and $\mathbf{d}_T$ are the total dipole vectors of the Koopman cation and neutral, respectively, and 
$\boldsymbol{\mu}_\text{MF} = - \big\langle \psi^{(0)} \big| \mathbf{r} \big| \psi^{(0)} \big\rangle$ is the dipole moment of the ionized orbital.
For single ionization starting from a neutral molecule, the denominator is equal to one, so, $\mathbf{r}_c = \mathbf{d}_T - \boldsymbol{\mu}_\text{MF}$.
Relative to the optimal origin, the dipole moment of the ionized orbital is given by
\begin{equation}
    \boldsymbol{\mu}'_\text{MF} = - \big\langle \psi^{(0)} \big| \mathbf{r} - \mathbf{r}_c \big| \psi^{(0)} \big\rangle
    =
    \boldsymbol{\mu}_\text{MF} + \mathbf{r}_c = \mathbf{d}_T.
\end{equation}
Therefore, the dipole moment of the ionized orbital in the optimal origin is equal to the total dipole of the neutral molecule, and is thus invariant of any unitary transform of the orbitals.

\section{Derivation of $Q_l^\nu(r)$
\label{app:derive_q}}

In order to obtain the expression of $Q_l^\nu(r)$, we must first determine an equation of which $\Omega_\nu^{(1)}(\mathbf r)$ is a solution. It is straightforward to verify that $\Omega_\nu^{(1)}(\mathbf r)$ given in \autoref{eq:Omega1_explicit} satisfies the following differential equation,
\begin{gather}
    \left(
    \frac{\nabla^2}{2} + \frac{Z}{r} + E^{(0)}
    \right)
    \Omega_\nu^{(1)} (\mathbf r)
    = 
    \left( z + \mu_z \right) \Omega_\nu^{(0)} (\mathbf r).
    \label{eq:pde_def_Omega1}
\end{gather}
We can now insert the partial wave expansion form of $\Omega_\nu^{(1)}(\mathbf r)$ given in Eq. \eqref{eq:Omega1_expand} into Eq. \eqref{eq:pde_def_Omega1} and project the resulting equation onto a particular spherical harmonics. These give
\begin{gather}
    \left(
    \odv*[fun]{r^2 \odv{}{r}}{r} 
    -
    l(l+1)
    +
    2 Z r
    +
    2 E^{(0)} r^2
    \right)
    Q_l^\nu(r)
    = \nonumber 
    \\
    2 r^2
    \left(
    k_{l|m|} r R_{l-1}^\nu(r)
    +
    k_{l+1,|m|} r R_{l+1}^\nu(r)
    +
    \mu_z R_l^\nu(r)
    \right).
    \label{eq:ode_def_q}
\end{gather}

The general solution of \autoref{eq:ode_def_q} may be written as
\begin{gather}
    Q_l^\nu(r) = 
    C_1 g_l^\nu(r) + C_2 h_l^\nu(r) + A_l^\nu(r),
    \label{eq:q_general_form}
\end{gather}
where $g_l^\nu(r)$ and $h_l^\nu(r)$ are the linearly independent solutions of a homogeneous differential equation obtained by setting the right-hand side of \autoref{eq:ode_def_q} to zero. We take the liberty of taking $h_l^\nu(r)$ as the solution that diverges at $r=0$, thus $g_l^\nu(0)$ is finite. These two functions are given by,
\begin{subequations}
    \begin{gather}
        g_l^\nu(r) = (\varkappa r)^l e^{-\varkappa r} M(l+1-Z/\varkappa, 2l+2, 2\varkappa r),
        \label{eq:def_g}
        \\
        h_l^\nu(r) = (\varkappa r)^l e^{-\varkappa r} U(l+1-Z/\varkappa, 2l+2, 2\varkappa r),
        \label{eq:def_h}
    \end{gather}
    \label{eq:def_gh}
\end{subequations}
With this convention, one may note that $g_l^\nu(r) \propto R_l^\nu(r)$.
Using the behavior of \retazero{\nu}{\eta} and \retaone{\nu}{\eta} near $\eta=0$, which may be found in Ref. \cite{ir_oewfat-2016}, we find the property $\omgone{\nu}{r=0,\theta,\varphi}=0$ (see \autoref{eq:Omega1_explicit}). This implies $Q_l^\nu(0)=0$, thus $C_2=0$.

To obtain $C_1$, we need the mathematical expression for $A_l^\nu(r)$, the particular solution of \autoref{eq:ode_def_q}. It can be shown that
\begin{align}
    A_l^\nu(r) =& 
    - \frac{1}{r^2 W(r)}
    \left[
    \int_0^r dr' \, h_l^\nu(r) g_l^\nu(r') Y(r') \,
    + \right. \nonumber \\
    &\left.
    \int_r^\infty dr' \, g_l^\nu(r) h_l^\nu(r') Y(r') 
    \right],
\end{align}
where $Y(r)$ is the right-hand side of \autoref{eq:ode_def_q} and $W(r)$ is the Wronskian of the homogeneous differential equation obtained by setting $Y(r)=0$ in \autoref{eq:ode_def_q}. By making use of the Wronskian of the defining differential equation for the Hypergeometric functions \cite{math_handbook_abramowitz-1972}, it can be shown that
\begin{gather}
    W(r) = -\frac{(2l+1)!}{\varkappa \, 2^{2l+1} \Gamma(l+1-Z/\varkappa) \, r^2}.
    \label{eq:wronskian}
\end{gather}
Therefore,
\begin{widetext}    
\begin{align}
    A_l^\nu(r) =
    \frac{\varkappa \, 2^{2l+1} \Gamma(l+1-Z/\varkappa)}{(2l+1)!}  
    \left[
    \int_0^r dr' \, h_l^\nu(r) g_l^\nu(r') Y(r') \,
    + 
    \int_r^\infty dr' \, g_l^\nu(r) h_l^\nu(r') Y(r') 
    \right].
    \label{eq:particular_solution}
\end{align}

Having obtained the expression for $A_l^\nu(r)$, we can now compute $C_1$ as
\begin{align}
    C_1 = \lim_{r \to 0} \frac {Q_l^\nu(r) - A_l^\nu(r)} {g_l^\nu(r)} 
    = 
    d_l^\nu 
    - \frac{\varkappa 2^{2l+1} \Gamma(l+1-Z/\varkappa)}{(2l+1)!}
    \int_0^\infty dr' \,
    h_l^\nu(r') Y(r')  
    \label{eq:c1}
\end{align}
where $d_l^\nu = \lim_{r \to 0} (Q_l^\nu(r)/g_l^\nu(r))$. We will now evaluate this limit. First, let us write $\delta_l^\nu(r) = Q_l^\nu(r)/g_l^\nu(r) = \delta_1^{\nu l}(r) + \delta_2^{\nu l}(r)$, and use $Q_l^\nu(r) = \int_0^{2\pi} d\varphi \int_0^\pi d\theta  \,\sin\theta \, Y_{lm}^*(\theta,\varphi) \, \omgone{\nu}{\mathbf{r}}$ (see \autoref{eq:Omega1_expand}) to derive
\begin{subequations}
    \begin{gather}
        \delta_1^{\nu l}(r) = 
        - \frac{2 \varkappa^{\betazero{\nu}/\varkappa-1}}{\sqrt{2\pi} g_l^\nu(r)} 
        \int_0^{2\pi} d\varphi \int_0^\pi \sin\theta \, d\theta \,
        Y_{lm}^*(\theta,\varphi)
        \frac{\retaone{\nu}{\eta}}{\sqrt{\eta}}
        \phizero{\nu}{\xi}
        e^{im\varphi},
        \label{eq:def_delta1}
    \\
        \delta_2^{\nu l}(r) = 
        - \frac{2 \varkappa^{\betazero{\nu}/\varkappa-1}}{\sqrt{2\pi} g_l^\nu(r)} 
        \int_0^{2\pi} d\varphi \int_0^\pi \sin\theta \, d\theta \,
        Y_{lm}^*(\theta,\varphi)
        \frac{\retazero{\nu}{\eta}}{\sqrt{\eta}}
        \phione{\nu}{\xi}
        e^{im\varphi}.
        \label{eq:def_delta2}
    \end{gather}
    \label{eq:def_delta}
\end{subequations}
We now have 
\begin{equation}
    d_l^\nu = \lim_{r \to 0} \left(\delta_1^{\nu l}(r) + \delta_2^{\nu l}(r) \right) .
    \label{eq:d_lim}
\end{equation}

In the following, we will first solve for $\lim_{r \to 0} \delta_1^{\nu l}(r)$ for $m \geq 0$. We start by expressing \retazero{\nu}{\eta}, \oetazero{\nu}{\eta}, and \retaone{\nu}{\eta} in infinite series form. The infinite series forms of \retazero{\nu}{\eta} and \oetazero{\nu}{\eta} can be obtained by using the infinite series representation of $M(a,b,\varkappa\eta)$, $U(a,b,\varkappa\eta)$, and $\exp(-\varkappa\eta/2)$ \cite{math_handbook_abramowitz-1972} in Eqs. \eqref{eq:define_r0_eta} and \eqref{eq:define_o0_eta}. Then using the so-obtained infinite series forms of \retazero{\nu}{\eta} and \oetazero{\nu}{\eta} in \autoref{eq:define_r1_eta} and performing the integration over $\eta'$, we arrive at the infinite series form of \retaone{\nu}{\eta},
\begin{align}
    \retaone{\nu}{\eta} =& \,\, 
    \left( \frac{\Gamma(a)}{\varkappa |m|!} \right)^2
    (\varkappa \eta)^{b/2} e^{-\varkappa \eta/2}
    \left[
    \frac{(-1)^{|m|+1}}{|m|! \Gamma(a-|m|)}
    \sum_{jikk'}^\infty
    (\varkappa \eta)^{|m|+S_1+1}
    \left(
    t_{ijkk'0}^\nu \betaone{\nu} \varkappa
    - 
    t_{ijkk'1}^\nu \frac{\mu_z}{2} \varkappa \eta
    + 
    t_{ijkk'2}^\nu \frac{(\varkappa \eta)^2}{4\varkappa}
    \right)
    \right.
    \nonumber \\
    &\,\,
    \left.
    - \frac{1}{\Gamma(a)}
    \sum_{ijk'}^\infty \sum_{k=1}^{|m|}
    (\varkappa \eta)^{|m|+S_2+1}
    \left(
    \tilde t_{ijkk'0}^\nu \betaone{\nu} \varkappa
    - 
    \tilde t_{ijkk'1}^\nu \frac{\mu_z}{2} \varkappa \eta
    +
    \tilde t_{ijkk'2}^\nu \frac{(\varkappa \eta)^2}{4\varkappa}
    \right)
    \right],
    \label{eq:retaone_series}
\end{align}
where
\begin{subequations}
        \begin{gather}
            t_{ijkk's}^\nu = 
            \frac{(-1)^i}{i!}
            \frac{(a)_j (a)_k (a)_{k'}}{(b)_j j! (b)_k k! (b)_{k'} k'!}
            \left(
            \frac{(k-k')(\psi(a+k) - \psi(1+k) - \psi(1+|m|+k))}{(\sigma_1+s+k')(\sigma_1+s+k)}
            +
            \frac{1}{(\sigma_1+s+k)^2}
            \right),
        \\
            \tilde t_{ijkk's}^\nu = 
            \frac{(-1)^i}{i!} 
            \frac{(a)_j (a)_{k'}}{(b)_j j! (b)_{k'} k'!}
            \frac{(k-1)! (1-a+k)_{|m|-k}}{(|m|-k)!}
            \frac{(k+k')}{(\sigma_1+s+k')(\sigma_1+s-k)},
        \end{gather}
\end{subequations}
$(x)_n$ is the rising factorial, $S_1 = i + j + k + k'$, $S_2 = i + j - k + k'$, and $\sigma_1 = |m|+1+i+j$.
By using \autoref{eq:retaone_series} for \retaone{\nu}{\eta}
in \autoref{eq:def_delta1}, and performing the angular integrations, we obtain
\begin{align}
    \delta_1^{\nu l} (r) =& \,\,
    \frac{-\sqrt{2} \varkappa^{\betazero{\nu}/\varkappa}}{M(l+1-Z/\varkappa, 2l+2, 2\varkappa r)}
    \left( \frac{\Gamma(a)}{\varkappa |m|!} \right)^2
    \sqrt{ \frac{n_\xi!(2l+1)(l-m)!}{(n_\xi+|m|)! (l+m)!} }
    \Bigg[
    \frac{(-1)^{|m|+1}}{|m|! \Gamma(a-|m|)}
    \sum_{ijkk'}^\infty \sum_{j'=0}^{n_\xi}
    (\varkappa r)^{2|m|+S_1+j'+1-l}  \nonumber\\
    &\,\, \left. \times
    \binom{n_\xi+|m|}{n_\xi-j'} \frac{(-1)^{j'}}{j'!}
    \left(
    t_{ijkk'0}^\nu I_{lm}^{|m|+S_1+1, j'} \betaone{\nu}\varkappa
    -
    t_{ijkk'1}^\nu I_{lm}^{|m|+S_1+2, j'} \frac{\mu_z}{2} \varkappa r
    +
    t_{ijkk'2}^\nu I_{lm}^{|m|+S_1+3, j'} \frac{(\varkappa r)^2}{4\varkappa}
    \right)
    -
    \right. \nonumber \\
    &\,\, \left.
    \frac{1}{\Gamma(a)} 
    \sum_{ijk'}^\infty \sum_{k=1}^{|m|} \sum_{j'=0}^{n_\xi}
    (\varkappa r)^{2|m|+S_2+j'+1-l}
    \binom{n_\xi+|m|}{n_\xi-j'} \frac{(-1)^{j'}}{j'!}
    \bigg(
    \tilde t_{ijkk'0}^\nu I_{lm}^{|m|+S_2+1,j'} \betaone{\nu}\varkappa
    -
    \tilde t_{ijkk'1}^\nu I_{lm}^{|m|+S_2+2,j'} \frac{\mu_z}{2} \varkappa r \right. \nonumber \\
    &\,\,
    + \,
    \tilde t_{ijkk'2}^\nu I_{lm}^{|m|+S_2+3,j'} \frac{(\varkappa r)^2}{4\varkappa}
    \bigg)
    \Bigg]
    \label{eq:def_delta1_series}
\end{align}
where
\begin{equation}
    I_{lm}^{pq} = \int_{-1}^1 du \, P_l^m(u) \, (1-u^2)^{|m|/2} (1-u)^p (1+u)^q
    =
    \left\{
        \begin{array}{ll}
            0,   & p+q < l-m \\
            & \\
            (-1)^{p+m} \frac{2(l+m)!}{(2l+1)!!},  & p+q = l-m
        \end{array}
    \right.
    \label{eq:ilmpq}
\end{equation}
with $P_l^m(u)$ being the associated Legendre polynomial and $m\geq0$. The value of $I_{lm}^{pq}$ for $p+q > l-m$ is not shown because we do not need them in our derivation. The combination of the limit operation, $\lim_{r \to 0}$, and the property of \autoref{eq:ilmpq} of $I_{lm}^{pq}$ can be used to reduce the infinite series in \autoref{eq:def_delta1_series} to a few surviving terms. For example, the terms involving the product $(\varkappa r)^{2|m| + S_1 + j' + 1 - l} I_{lm}^{|m|+S_1+1,j'}$ are equal to zero for $|m|+S_1+1+j' < l-m$ due to \autoref{eq:ilmpq} and for $|m|+S_1+1+j' > l-m$ because $\lim_{r \to 0} (\varkappa r)^{2|m| + S_1 + j' + 1 - l} = 0$. Therefore, we are left with the terms whose values of $S_1$ and $j'$ are such that $|m|+S_1+1+j' = l-m$. This allows us to arrive at the final expression for $\lim_{r \to 0} \delta_1^{\nu l}(r)$ with $m \geq 0$. To generalize the result to the case of $m < 0$, it can be shown that it suffices to include a multiplicative factor $(-1)^{(|m|-m)/2}$, thus we arrive at the following equation
\begin{align}
    \lim_{r \to 0} \delta_1^{\nu l} (r) =& \,\,
    \frac{(-1)^{l-(|m|-m)/2+1} \varkappa^{\betazero{\nu}/\varkappa-2}}{\Gamma(v + |m|)}
    \sqrt{(2l+1) (l+m)! (l-m)! (n_\xi+|m|)! n_\xi!} 
    \,
    \frac{2^{l+3/2} \, l!}{(2l+1)!} 
    \Bigg[
    \frac{(-1)^{|m|+1}}{\Gamma(v)} \nonumber \\
    &\,\, \times
    \left(
    \betaone{\nu} \varkappa H_1^{\nu l}
    -
    \frac{\mu_z}{2} H_2^{\nu l} 
    +
    \frac{1}{4 \varkappa} H_3^{\nu l}
    \right)
    -
    \Gamma(1-v)
    \left(
    \betaone{\nu} \varkappa \tilde H_1^{\nu l}
    -
    \frac{\mu_z}{2} \tilde H_2^{\nu l} 
    +
    \frac{1}{4 \varkappa} \tilde H_3^{\nu l}
    \right)
    \Bigg],
    \label{eq:lim_delta1}
\end{align}
where
\begin{align}
    H_p^{\nu l} =
    & \, 
    \sum_{(S_1 + j' = l - 2|m| - p)} 
    (-1)^i 
    \frac
    {\Gamma(v+|m|+j) \Gamma(v+|m|+k) \Gamma(v+|m|+k')} 
    {(|m|+j)! (|m|+j')! (|m|+k)! (|m|+k')! i! j! j'! k! k'! (n_\xi-j')!} \nonumber
    \\
    & \, \times
    \left(
    \frac
    {(k-k') (\psi(v+|m|+k) - \psi(1+k) - \psi(1+k+|m|))}
    {(l-|m|-k-j') (l-|m|-k'-j')}
    +
    \frac{1}{(l-|m|-k'-j')^2}
    \right),
    \label{eq:define_h_app}
\end{align}
\begin{align}
    \tilde H_p^{\nu l} =
    & \, 
    \sum_{(S_2 + j' = l - 2|m| - p)} 
    (-1)^i 
    \frac
    {\Gamma(v+|m|+j) \Gamma(v+|m|+k') (k-1)!}
    {(|m|+j)! (|m|+j')! (|m|-k)! (|m|+k')! i! j! j'! k'! \Gamma(1-v-|m|+k) (n_\xi-j')!} \nonumber
    \\
    & \, \times
    \frac{k+k'}{(l-|m|+k-j') (l-|m|-k'-j')}.
    \label{eq:define_ht_app}
\end{align}
The summation notation in \autoref{eq:define_h_app} is a shorthand for multiple sum $\sum_{i=0}^\infty \sum_{j=0}^\infty \sum_{k=0}^\infty \sum_{k'=0}^\infty \sum_{j'=0}^{n_\xi}$ such that $S_1 + j' = l - 2|m| - p$, whereas the one in \autoref{eq:define_ht_app} is a shorthand for $\sum_{i=0}^\infty \sum_{j=0}^\infty \sum_{k=1}^{|m|} \sum_{k'=0}^\infty \sum_{j'=0}^{n_\xi}$ such that $S_2 + j' = l - 2|m| - p$.

Next, we turn to solving for $\lim_{r \to 0} \delta_2^{\nu l}(r)$, we start by assuming $m \geq 0$. Inserting \autoref{eq:phi1_expand} into \autoref{eq:def_delta2}, we get
\begin{align}
    \delta_2^{\nu l}(r) = 
    -2 \varkappa^{\betazero{\nu}/\varkappa-1} 
    \sum_{i = n_\xi-2, \, i\neq n_\xi}^{n_\xi+2}
    C_{im} \, \tilde \delta_i^{\nu l}(r),
    \label{eq:def_delta2_series}
\end{align}
where $C_{im} = C_1^{im} + \mu_z C_2^{im}$ ($C_r^{im}$ for $r=1,2$ 
is given in \autoref{eq:phi1_expand_c}), and
\begin{equation}
    \tilde \delta_i^{\nu l}(r) = 
    \frac{1}{g_l^\nu(r)}
    \int_0^{2\pi} d\varphi \int_0^\pi \sin\theta \, d\theta \,
    Y_{lm}^*(\theta,\varphi) 
    \frac{\retazero{\nu}{\eta}}{\sqrt{2\pi\eta}}
    \phizero{im}{\xi}
    e^{im\varphi}.
    \label{eq:def_delta_i}
\end{equation}
By using the infinite series form of \retazero{\nu}{\eta} obtained previously in \autoref{eq:def_delta_i}, and performing the angular integrations, we obtain
\begin{align}
    \tilde \delta_i^{\nu l}(r) =& \,\,
    \frac{\Gamma(a) \varkappa^{|m|+1-l}}{M(l+1-Z/\varkappa,2l+2,2\varkappa r)}
    \frac{1}{|m|!}
    \sqrt{ \frac{i! (2l+1) (l-m)!}{2 (i+|m|)! (l+m)!} }
    \sum_{j=0}^\infty \sum_{k=0}^i
    (-1)^k \frac{\varkappa^{j+k}}{j!k!} \frac{(a)_j}{(b)_j} \binom{i+|m|}{i-k} I_{lm}^{jk} \, r^{|m|+j+k-l}.
    \label{eq:delta_i_series}
\end{align}
We then insert \autoref{eq:delta_i_series} for $\tilde \delta_i^{\nu l}(r)$ in \autoref{eq:def_delta2_series}, and then apply $\lim_{r \to 0}$ and the property \autoref{eq:ilmpq} to reduce the resulting infinite series to only a few surviving terms. To generalize the final result to the case of $m<0$, we only need to multiply it with $(-1)^{(|m|-m)/2}$. This yields
\begin{align}
    \lim_{r \to 0} \delta_2^{\nu l}(r) =& \,\,
    (-1)^{l-(|m|-m)/2+1} \,
    \varkappa^{\betazero{\nu}/\varkappa}
    \sqrt{(2l+1) (l+m)! (l-m)!}
    \frac{2^{l+3/2} l!}{(2l+1)!}
    \sum_{i = n_\xi-2, \, i\neq n_\xi}^{n_\xi+2}
    \sqrt{(i+|m|)! i!} \, C_{im} \nonumber \\
    &\,\, \times
    \sum_{k=0}^{\min(i,l-|m|)}
    \frac{\Gamma(l+v-k)}{k! (l-k)! (|m|+k)! (l-|m|-k)! (i-k)!}.
    \label{eq:lim_delta2}
\end{align}
Equations \eqref{eq:d_lim}, \eqref{eq:lim_delta1}, and \eqref{eq:lim_delta2} allow us to arrive at the sought expression for $d_l^\nu$,
\begin{align}
    d_l^\nu
    =
    & \,
    (-1)^{l+(m-|m|)/2+1} \,
    \varkappa^{\beta_\nu^{(0)}/\varkappa}
    \sqrt{(2l+1) (l+m)! (l-m)!} \frac{2^{l+3/2} l!}{(2l+1)!} \nonumber
    \\
    & \,
    \left[
    \sum_{i=n_\xi-2; i \neq n_\xi}^{n_\xi+2} 
    \sqrt{(i+|m|)! \, i!} \, C_{im} \sum_{k=0}^{\textrm{min}(i,l-|m|)}
    \frac{\Gamma(l+v-k)}{k!(l-k)!(|m|+k)!(l-|m|-k)!(i-k)!}\, 
    + \frac{\sqrt{(n_\xi+|m|)! n_\xi!}}{\varkappa^2 \Gamma(v+|m|)}
    \right. \nonumber
    \\
    & \,
    \left.
    \left\{
    \frac{(-1)^{|m|+1}}{\Gamma(v)}
    \left(
    \beta_\nu^{(1)} \varkappa H_1^{\nu l}
    -
    \frac{\mu_z}{2} H_2^{\nu l}
    +
    \frac{1}{4\varkappa} H_3^{\nu l}
    \right)
    - 
    \Gamma(1-v)
    \left(
    \beta_\nu^{(1)} \varkappa \tilde H_1^{\nu l}
    -
    \frac{\mu_z}{2} \tilde H_2^{\nu l}
    +
    \frac{1}{4\varkappa} \tilde H_3^{\nu l}
    \right)
    \right\}
    \right].
    \label{eq:define_d_app}
\end{align}
Finally, by substituting $d_l^\nu$ in \autoref{eq:c1} using \autoref{eq:define_d_app}, we obtain an explicit expression for $C_1$.

We now have everything to compute $Q_l^\nu(r)$, namely,
\autoref{eq:q_general_form};
the expression for $C_1$ obtained in the previous step;
the result $C_2=0$;
the expressions for $g_l^\nu(r)$ and $h_l^\nu(r)$ in \autoref{eq:def_gh};
and the expression for $A_l^\nu(r)$ in \autoref{eq:particular_solution}.
These yield
\begin{align}
    Q_l^\nu(r) 
    =& \,\,
    d_l^\nu g_l^\nu(r)
    + 
    \frac{2^{2l+2} \varkappa \, \Gamma(l+1-Z/\varkappa)}{(2l+1)!} 
    \int_0^r dr' r'^2 
    \left(
    g_l^\nu(r) h_l^\nu(r')
    -
    g_l^\nu(r') h_l^\nu(r)
    \right)  
    \left(
    k_{l|m|} r' R_{l-1}^\nu(r') \right. \nonumber
    \\
    & \,
    \left. 
    + \,
    k_{l+1,|m|} r' R_{l+1}^\nu(r')
    + 
    \mu_z R_l^\nu(r')
    \right).
    \label{eq:q_final}
\end{align}
Equations \eqref{eq:q_split}, \eqref{eq:define_q1}, \eqref{eq:define_q2}, and \eqref{eq:define_d} then result from splitting \autoref{eq:q_final} into terms that depend on $\mu_z$ and those that do not (see \autoref{eq:q_split}).

\end{widetext}

\section{Basis functions for noble gases and \molhhc{} \label{app:dvr_basis}}

For our calculations on the noble gas atoms and \molhhc{}, we employ a single-center basis based on a discrete variable representation (DVR) to represent \wfnzero{\mathbf{r}} and \wfnone{\mathbf{r}}. Mathematically, this basis factorizes into functions that depend on $r$, $\theta$, and $\varphi$,
\begin{equation}
    f_{Iij\pm}^m (\mathbf{r}) = r^{-1} \rho_i^I(r) \tau_j^m(\theta) \Phi_{\pm}^m(\varphi).
    \label{eq:define_dvr_basis}
\end{equation}
We employ finite-element (FE)DVR for $\rho_i^I(r)$ in which the radial coordinate is segmented into finite elements (indexed with $I$), and each finite element contains several DVR bases (indexed by $i$) \cite{fedvr-grid-scatter-2000, fedvr-three-body-2004, fedvr-parallel-solver-2006, fedvr-quant_dyn-2011}. Each DVR basis in a finite element is characterized by a radial quadrature point, $r_i^I$.

Due to the axial symmetry of noble gases and \molhhc{}, the eigenfunctions of their respective Hamiltonian can have a definite magnetic quantum number. So, we choose the azimuthal part such that $\Phi_+^m(\varphi) = ((1+\delta_{0m})\pi)^{-1/2} \cos m\varphi$ and $\Phi_-^m(\varphi) = \pi^{-1/2} \sin m\varphi$, where $m \geq 0$ is the magnetic quantum number.

For the polar angular part, we employ associated-Legendre (AL)DVR basis. The set of underlying orthogonal polynomials for the associated quadrature rule for this basis is given by
\begin{equation}
    t_l^m(u) = \frac{(-1)^{m}}{2^l l!} 
    \odv*[order=l+m]{}{u}
    (u^2-1)^l
\end{equation}
where $m$ defines the polynomials set, $l \geq m$ enumerates the polynomials within the set, and  $u = \cos \theta$. The orthogonality property of $t_l^m(u)$ is given by $\int_{-1}^1 du \, (1-u^2)^m t_l^m(u) t_{l'}^m(u) = \delta_{ll'} N_{lm}$, where $N_{lm} = 
2(l+m)!
\left(
(2l+1)(l-m)!
\right)^{-1}$. The ALDVR basis is then constructed through
\begin{equation}
    \tau_i^{m} (\theta) = 
    \sum_{j=1}^{N_\theta} 
    \sqrt{w_i^m} \left( 1 - (u_i^m)^2 \right)^{-m/2} T_j^m(u_i^m) \, T_j^m(u)
\end{equation}
for $i = 1, \ldots, N_\theta$, where $N_\theta$ is the number of $\theta$ quadrature grids and
\begin{equation}
    T_j^m(u) = N_{lm}^{-1/2} (1-u^2)^{m/2} \, t_{m+j-1}^m(u) ,
\end{equation}
with $j = 1, \ldots, N_\theta$, and $w_i^m$ and $u_i^m$ are the quadrature weights and nodes. $\{u_i^m ; i = 1, \ldots, N_\theta \}$ are the roots of $t_{m+N_\theta}^m(u)$ and $u_i^m \in [-1,1]$. Note that $(1-u^2)^{m/2} t_l^m(u)$ is the associated Legendre function.

The matrix representation of an axially symmetric local operator $V(r,\theta)$ in our DVR-based basis of \eqref{eq:define_dvr_basis} is approximately diagonal, with the diagonal elements given by $V(r_i^I,\theta_i^m)$ where $\cos \theta_i^m = u_i^m$.

\nocite{wahyutama_2025_15844886}
\bibliography{apssamp}

\providecommand{\noopsort}[1]{}\providecommand{\singleletter}[1]{#1}%
\begin{thebibliography}{71}%
\makeatletter
\providecommand \@ifxundefined [1]{%
 \@ifx{#1\undefined}
}%
\providecommand \@ifnum [1]{%
 \ifnum #1\expandafter \@firstoftwo
 \else \expandafter \@secondoftwo
 \fi
}%
\providecommand \@ifx [1]{%
 \ifx #1\expandafter \@firstoftwo
 \else \expandafter \@secondoftwo
 \fi
}%
\providecommand \natexlab [1]{#1}%
\providecommand \enquote  [1]{``#1''}%
\providecommand \bibnamefont  [1]{#1}%
\providecommand \bibfnamefont [1]{#1}%
\providecommand \citenamefont [1]{#1}%
\providecommand \href@noop [0]{\@secondoftwo}%
\providecommand \href [0]{\begingroup \@sanitize@url \@href}%
\providecommand \@href[1]{\@@startlink{#1}\@@href}%
\providecommand \@@href[1]{\endgroup#1\@@endlink}%
\providecommand \@sanitize@url [0]{\catcode `\\12\catcode `\$12\catcode `\&12\catcode `\#12\catcode `\^12\catcode `\_12\catcode `\%12\relax}%
\providecommand \@@startlink[1]{}%
\providecommand \@@endlink[0]{}%
\providecommand \url  [0]{\begingroup\@sanitize@url \@url }%
\providecommand \@url [1]{\endgroup\@href {#1}{\urlprefix }}%
\providecommand \urlprefix  [0]{URL }%
\providecommand \Eprint [0]{\href }%
\providecommand \doibase [0]{https://doi.org/}%
\providecommand \selectlanguage [0]{\@gobble}%
\providecommand \bibinfo  [0]{\@secondoftwo}%
\providecommand \bibfield  [0]{\@secondoftwo}%
\providecommand \translation [1]{[#1]}%
\providecommand \BibitemOpen [0]{}%
\providecommand \bibitemStop [0]{}%
\providecommand \bibitemNoStop [0]{.\EOS\space}%
\providecommand \EOS [0]{\spacefactor3000\relax}%
\providecommand \BibitemShut  [1]{\csname bibitem#1\endcsname}%
\let\auto@bib@innerbib\@empty
\bibitem [{\citenamefont {Itatani}\ \emph {et~al.}(2004)\citenamefont {Itatani}, \citenamefont {Levesque}, \citenamefont {Zeidler}, \citenamefont {Niikura}, \citenamefont {P{\'e}pin}, \citenamefont {Kieffer}, \citenamefont {Corkum},\ and\ \citenamefont {Villeneuve}}]{mot-2004}%
  \BibitemOpen
  \bibfield  {author} {\bibinfo {author} {\bibfnamefont {J.}~\bibnamefont {Itatani}}, \bibinfo {author} {\bibfnamefont {J.}~\bibnamefont {Levesque}}, \bibinfo {author} {\bibfnamefont {D.}~\bibnamefont {Zeidler}}, \bibinfo {author} {\bibfnamefont {H.}~\bibnamefont {Niikura}}, \bibinfo {author} {\bibfnamefont {H.}~\bibnamefont {P{\'e}pin}}, \bibinfo {author} {\bibfnamefont {J.-C.}\ \bibnamefont {Kieffer}}, \bibinfo {author} {\bibfnamefont {P.~B.}\ \bibnamefont {Corkum}},\ and\ \bibinfo {author} {\bibfnamefont {D.~M.}\ \bibnamefont {Villeneuve}},\ }\bibfield  {title} {\bibinfo {title} {Tomographic imaging of molecular orbitals},\ }\href {https://doi.org/10.1038/nature03183} {\bibfield  {journal} {\bibinfo  {journal} {Nature}\ }\textbf {\bibinfo {volume} {432}},\ \bibinfo {pages} {867} (\bibinfo {year} {2004})}\BibitemShut {NoStop}%
\bibitem [{\citenamefont {Patchkovskii}\ \emph {et~al.}(2006)\citenamefont {Patchkovskii}, \citenamefont {Zhao}, \citenamefont {Brabec},\ and\ \citenamefont {Villeneuve}}]{mot-multiel-2006}%
  \BibitemOpen
  \bibfield  {author} {\bibinfo {author} {\bibfnamefont {S.}~\bibnamefont {Patchkovskii}}, \bibinfo {author} {\bibfnamefont {Z.}~\bibnamefont {Zhao}}, \bibinfo {author} {\bibfnamefont {T.}~\bibnamefont {Brabec}},\ and\ \bibinfo {author} {\bibfnamefont {D.~M.}\ \bibnamefont {Villeneuve}},\ }\bibfield  {title} {\bibinfo {title} {High harmonic generation and molecular orbital tomography in multielectron systems: Beyond the single active electron approximation},\ }\href {https://doi.org/10.1103/PhysRevLett.97.123003} {\bibfield  {journal} {\bibinfo  {journal} {Phys. Rev. Lett.}\ }\textbf {\bibinfo {volume} {97}},\ \bibinfo {pages} {123003} (\bibinfo {year} {2006})}\BibitemShut {NoStop}%
\bibitem [{\citenamefont {Haessler}\ \emph {et~al.}(2010)\citenamefont {Haessler}, \citenamefont {Caillat}, \citenamefont {Boutu}, \citenamefont {Giovanetti-Teixeira}, \citenamefont {Ruchon}, \citenamefont {Auguste}, \citenamefont {Diveki}, \citenamefont {Breger}, \citenamefont {Maquet}, \citenamefont {Carr{\'e}} \emph {et~al.}}]{mot-attoimaging-2010}%
  \BibitemOpen
  \bibfield  {author} {\bibinfo {author} {\bibfnamefont {S.}~\bibnamefont {Haessler}}, \bibinfo {author} {\bibfnamefont {J.}~\bibnamefont {Caillat}}, \bibinfo {author} {\bibfnamefont {W.}~\bibnamefont {Boutu}}, \bibinfo {author} {\bibfnamefont {C.}~\bibnamefont {Giovanetti-Teixeira}}, \bibinfo {author} {\bibfnamefont {T.}~\bibnamefont {Ruchon}}, \bibinfo {author} {\bibfnamefont {T.}~\bibnamefont {Auguste}}, \bibinfo {author} {\bibfnamefont {Z.}~\bibnamefont {Diveki}}, \bibinfo {author} {\bibfnamefont {P.}~\bibnamefont {Breger}}, \bibinfo {author} {\bibfnamefont {A.}~\bibnamefont {Maquet}}, \bibinfo {author} {\bibfnamefont {B.}~\bibnamefont {Carr{\'e}}}, \emph {et~al.},\ }\bibfield  {title} {\bibinfo {title} {Attosecond imaging of molecular electronic wavepackets},\ }\href {https://doi.org/10.1038/nphys1511} {\bibfield  {journal} {\bibinfo  {journal} {Nat. Phys.}\ }\textbf {\bibinfo {volume} {6}},\ \bibinfo {pages} {200} (\bibinfo {year} {2010})}\BibitemShut {NoStop}%
\bibitem [{\citenamefont {Vozzi}\ \emph {et~al.}(2011)\citenamefont {Vozzi}, \citenamefont {Negro}, \citenamefont {Calegari}, \citenamefont {Sansone}, \citenamefont {Nisoli}, \citenamefont {De~Silvestri},\ and\ \citenamefont {Stagira}}]{mot-generalized-2011}%
  \BibitemOpen
  \bibfield  {author} {\bibinfo {author} {\bibfnamefont {C.}~\bibnamefont {Vozzi}}, \bibinfo {author} {\bibfnamefont {M.}~\bibnamefont {Negro}}, \bibinfo {author} {\bibfnamefont {F.}~\bibnamefont {Calegari}}, \bibinfo {author} {\bibfnamefont {G.}~\bibnamefont {Sansone}}, \bibinfo {author} {\bibfnamefont {M.}~\bibnamefont {Nisoli}}, \bibinfo {author} {\bibfnamefont {S.}~\bibnamefont {De~Silvestri}},\ and\ \bibinfo {author} {\bibfnamefont {S.}~\bibnamefont {Stagira}},\ }\bibfield  {title} {\bibinfo {title} {Generalized molecular orbital tomography},\ }\href {https://doi.org/10.1038/nphys2029} {\bibfield  {journal} {\bibinfo  {journal} {Nat. Phys.}\ }\textbf {\bibinfo {volume} {7}},\ \bibinfo {pages} {822} (\bibinfo {year} {2011})}\BibitemShut {NoStop}%
\bibitem [{\citenamefont {Negro}\ \emph {et~al.}(2014)\citenamefont {Negro}, \citenamefont {Devetta}, \citenamefont {Faccialá}, \citenamefont {De~Silvestri}, \citenamefont {Vozzi},\ and\ \citenamefont {Stagira}}]{mot-hhs-molimage-2014}%
  \BibitemOpen
  \bibfield  {author} {\bibinfo {author} {\bibfnamefont {M.}~\bibnamefont {Negro}}, \bibinfo {author} {\bibfnamefont {M.}~\bibnamefont {Devetta}}, \bibinfo {author} {\bibfnamefont {D.}~\bibnamefont {Faccialá}}, \bibinfo {author} {\bibfnamefont {S.}~\bibnamefont {De~Silvestri}}, \bibinfo {author} {\bibfnamefont {C.}~\bibnamefont {Vozzi}},\ and\ \bibinfo {author} {\bibfnamefont {S.}~\bibnamefont {Stagira}},\ }\bibfield  {title} {\bibinfo {title} {High-order harmonic spectroscopy for molecular imaging of polyatomic molecules},\ }\href {https://doi.org/10.1039/C4FD00033A} {\bibfield  {journal} {\bibinfo  {journal} {Faraday Discuss.}\ }\textbf {\bibinfo {volume} {171}},\ \bibinfo {pages} {133} (\bibinfo {year} {2014})}\BibitemShut {NoStop}%
\bibitem [{\citenamefont {Ren}\ \emph {et~al.}(2023)\citenamefont {Ren}, \citenamefont {Zhang}, \citenamefont {Yang}, \citenamefont {Zhu}, \citenamefont {Zhao},\ and\ \citenamefont {Zhao}}]{mot-h2o-2023}%
  \BibitemOpen
  \bibfield  {author} {\bibinfo {author} {\bibfnamefont {Z.}~\bibnamefont {Ren}}, \bibinfo {author} {\bibfnamefont {B.}~\bibnamefont {Zhang}}, \bibinfo {author} {\bibfnamefont {Y.}~\bibnamefont {Yang}}, \bibinfo {author} {\bibfnamefont {Y.}~\bibnamefont {Zhu}}, \bibinfo {author} {\bibfnamefont {J.}~\bibnamefont {Zhao}},\ and\ \bibinfo {author} {\bibfnamefont {Z.}~\bibnamefont {Zhao}},\ }\bibfield  {title} {\bibinfo {title} {{Molecular orbital tomography of HOMO and HOMO-1 of the nonlinear molecule H$_2$O: The study on multi-orbital effects}},\ }\href {https://doi.org/https://doi.org/10.1016/j.rinp.2023.107181} {\bibfield  {journal} {\bibinfo  {journal} {Results Phys.}\ }\textbf {\bibinfo {volume} {55}},\ \bibinfo {pages} {107181} (\bibinfo {year} {2023})}\BibitemShut {NoStop}%
\bibitem [{\citenamefont {Ren}\ \emph {et~al.}(2024)\citenamefont {Ren}, \citenamefont {Zhang}, \citenamefont {Yang}, \citenamefont {Zhu}, \citenamefont {Bai}, \citenamefont {Liu}, \citenamefont {Zhao},\ and\ \citenamefont {Zhao}}]{slimp2-2024}%
  \BibitemOpen
  \bibfield  {author} {\bibinfo {author} {\bibfnamefont {Z.}~\bibnamefont {Ren}}, \bibinfo {author} {\bibfnamefont {B.}~\bibnamefont {Zhang}}, \bibinfo {author} {\bibfnamefont {Y.}~\bibnamefont {Yang}}, \bibinfo {author} {\bibfnamefont {Y.}~\bibnamefont {Zhu}}, \bibinfo {author} {\bibfnamefont {G.}~\bibnamefont {Bai}}, \bibinfo {author} {\bibfnamefont {J.}~\bibnamefont {Liu}}, \bibinfo {author} {\bibfnamefont {J.}~\bibnamefont {Zhao}},\ and\ \bibinfo {author} {\bibfnamefont {Z.}~\bibnamefont {Zhao}},\ }\bibfield  {title} {\bibinfo {title} {{SLIMP 2.0: A new version of strong laser interaction model package for atoms and molecules, now with molecular orbital tomography based on high-order harmonic spectra}},\ }\href {https://doi.org/https://doi.org/10.1016/j.cpc.2024.109213} {\bibfield  {journal} {\bibinfo  {journal} {Comput. Phys. Commun.}\ }\textbf {\bibinfo {volume} {301}},\ \bibinfo {pages} {109213} (\bibinfo {year} {2024})}\BibitemShut {NoStop}%
\bibitem [{\citenamefont {Meckel}\ \emph {et~al.}(2008)\citenamefont {Meckel}, \citenamefont {Comtois}, \citenamefont {Zeidler}, \citenamefont {Staudte}, \citenamefont {Pavičić}, \citenamefont {Bandulet}, \citenamefont {Pépin}, \citenamefont {Kieffer}, \citenamefont {Dörner}, \citenamefont {Villeneuve},\ and\ \citenamefont {Corkum}}]{lied-science-2008}%
  \BibitemOpen
  \bibfield  {author} {\bibinfo {author} {\bibfnamefont {M.}~\bibnamefont {Meckel}}, \bibinfo {author} {\bibfnamefont {D.}~\bibnamefont {Comtois}}, \bibinfo {author} {\bibfnamefont {D.}~\bibnamefont {Zeidler}}, \bibinfo {author} {\bibfnamefont {A.}~\bibnamefont {Staudte}}, \bibinfo {author} {\bibfnamefont {D.}~\bibnamefont {Pavičić}}, \bibinfo {author} {\bibfnamefont {H.~C.}\ \bibnamefont {Bandulet}}, \bibinfo {author} {\bibfnamefont {H.}~\bibnamefont {Pépin}}, \bibinfo {author} {\bibfnamefont {J.~C.}\ \bibnamefont {Kieffer}}, \bibinfo {author} {\bibfnamefont {R.}~\bibnamefont {Dörner}}, \bibinfo {author} {\bibfnamefont {D.~M.}\ \bibnamefont {Villeneuve}},\ and\ \bibinfo {author} {\bibfnamefont {P.~B.}\ \bibnamefont {Corkum}},\ }\bibfield  {title} {\bibinfo {title} {Laser-induced electron tunneling and diffraction},\ }\href {https://doi.org/10.1126/science.1157980} {\bibfield  {journal} {\bibinfo  {journal} {Science}\ }\textbf {\bibinfo {volume} {320}},\ \bibinfo {pages} {1478} (\bibinfo {year}
  {2008})}\BibitemShut {NoStop}%
\bibitem [{\citenamefont {Lin}\ \emph {et~al.}(2010)\citenamefont {Lin}, \citenamefont {Le}, \citenamefont {Chen}, \citenamefont {Morishita},\ and\ \citenamefont {Lucchese}}]{lied-review-2010}%
  \BibitemOpen
  \bibfield  {author} {\bibinfo {author} {\bibfnamefont {C.~D.}\ \bibnamefont {Lin}}, \bibinfo {author} {\bibfnamefont {A.-T.}\ \bibnamefont {Le}}, \bibinfo {author} {\bibfnamefont {Z.}~\bibnamefont {Chen}}, \bibinfo {author} {\bibfnamefont {T.}~\bibnamefont {Morishita}},\ and\ \bibinfo {author} {\bibfnamefont {R.}~\bibnamefont {Lucchese}},\ }\bibfield  {title} {\bibinfo {title} {Strong-field rescattering physics—self-imaging of a molecule by its own electrons},\ }\href {https://doi.org/10.1088/0953-4075/43/12/122001} {\bibfield  {journal} {\bibinfo  {journal} {J. Phys. B}\ }\textbf {\bibinfo {volume} {43}},\ \bibinfo {pages} {122001} (\bibinfo {year} {2010})}\BibitemShut {NoStop}%
\bibitem [{\citenamefont {Xu}\ \emph {et~al.}(2010)\citenamefont {Xu}, \citenamefont {Chen}, \citenamefont {Le},\ and\ \citenamefont {Lin}}]{lied-self-image-2010}%
  \BibitemOpen
  \bibfield  {author} {\bibinfo {author} {\bibfnamefont {J.}~\bibnamefont {Xu}}, \bibinfo {author} {\bibfnamefont {Z.}~\bibnamefont {Chen}}, \bibinfo {author} {\bibfnamefont {A.-T.}\ \bibnamefont {Le}},\ and\ \bibinfo {author} {\bibfnamefont {C.~D.}\ \bibnamefont {Lin}},\ }\bibfield  {title} {\bibinfo {title} {Self-imaging of molecules from diffraction spectra by laser-induced rescattering electrons},\ }\href {https://doi.org/10.1103/PhysRevA.82.033403} {\bibfield  {journal} {\bibinfo  {journal} {Phys. Rev. A}\ }\textbf {\bibinfo {volume} {82}},\ \bibinfo {pages} {033403} (\bibinfo {year} {2010})}\BibitemShut {NoStop}%
\bibitem [{\citenamefont {Belsa}\ \emph {et~al.}(2021)\citenamefont {Belsa}, \citenamefont {Amini}, \citenamefont {Liu}, \citenamefont {Sanchez}, \citenamefont {Steinle}, \citenamefont {Steinmetzer}, \citenamefont {Le}, \citenamefont {Moshammer}, \citenamefont {Pfeifer}, \citenamefont {Ullrich}, \citenamefont {Moszynski}, \citenamefont {Lin}, \citenamefont {Gräfe},\ and\ \citenamefont {Biegert}}]{lied-nh3-umbrella-2021}%
  \BibitemOpen
  \bibfield  {author} {\bibinfo {author} {\bibfnamefont {B.}~\bibnamefont {Belsa}}, \bibinfo {author} {\bibfnamefont {K.}~\bibnamefont {Amini}}, \bibinfo {author} {\bibfnamefont {X.}~\bibnamefont {Liu}}, \bibinfo {author} {\bibfnamefont {A.}~\bibnamefont {Sanchez}}, \bibinfo {author} {\bibfnamefont {T.}~\bibnamefont {Steinle}}, \bibinfo {author} {\bibfnamefont {J.}~\bibnamefont {Steinmetzer}}, \bibinfo {author} {\bibfnamefont {A.~T.}\ \bibnamefont {Le}}, \bibinfo {author} {\bibfnamefont {R.}~\bibnamefont {Moshammer}}, \bibinfo {author} {\bibfnamefont {T.}~\bibnamefont {Pfeifer}}, \bibinfo {author} {\bibfnamefont {J.}~\bibnamefont {Ullrich}}, \bibinfo {author} {\bibfnamefont {R.}~\bibnamefont {Moszynski}}, \bibinfo {author} {\bibfnamefont {C.~D.}\ \bibnamefont {Lin}}, \bibinfo {author} {\bibfnamefont {S.}~\bibnamefont {Gräfe}},\ and\ \bibinfo {author} {\bibfnamefont {J.}~\bibnamefont {Biegert}},\ }\bibfield  {title} {\bibinfo {title} {Laser-induced electron diffraction of the ultrafast umbrella motion in
  ammonia},\ }\href {https://doi.org/10.1063/4.0000046} {\bibfield  {journal} {\bibinfo  {journal} {Struct. Dyn.}\ }\textbf {\bibinfo {volume} {8}},\ \bibinfo {pages} {014301} (\bibinfo {year} {2021})}\BibitemShut {NoStop}%
\bibitem [{\citenamefont {Corkum}(1993)}]{sfi-plasma-perspective-1993}%
  \BibitemOpen
  \bibfield  {author} {\bibinfo {author} {\bibfnamefont {P.~B.}\ \bibnamefont {Corkum}},\ }\bibfield  {title} {\bibinfo {title} {Plasma perspective on strong field multiphoton ionization},\ }\href {https://doi.org/10.1103/PhysRevLett.71.1994} {\bibfield  {journal} {\bibinfo  {journal} {Phys. Rev. Lett.}\ }\textbf {\bibinfo {volume} {71}},\ \bibinfo {pages} {1994} (\bibinfo {year} {1993})}\BibitemShut {NoStop}%
\bibitem [{\citenamefont {Kulander}\ \emph {et~al.}(1993)\citenamefont {Kulander}, \citenamefont {Schafer},\ and\ \citenamefont {Krause}}]{super-intense-1993}%
  \BibitemOpen
  \bibfield  {author} {\bibinfo {author} {\bibfnamefont {K.~C.}\ \bibnamefont {Kulander}}, \bibinfo {author} {\bibfnamefont {K.~J.}\ \bibnamefont {Schafer}},\ and\ \bibinfo {author} {\bibfnamefont {J.~L.}\ \bibnamefont {Krause}},\ }\bibfield  {title} {\bibinfo {title} {Dynamics of short-pulse excitation, ionization and harmonic conversion},\ }in\ \href {https://doi.org/10.1007/978-1-4615-7963-2} {\emph {\bibinfo {booktitle} {Super-Intense Laser-Atom Physics}}},\ \bibinfo {editor} {edited by\ \bibinfo {editor} {\bibfnamefont {A.}~\bibnamefont {L'Huillier}}, \bibinfo {editor} {\bibfnamefont {B.}~\bibnamefont {Piraux}},\ and\ \bibinfo {editor} {\bibfnamefont {K.}~\bibnamefont {Rzazewski}}}\ (\bibinfo  {publisher} {Springer},\ \bibinfo {address} {New York},\ \bibinfo {year} {1993})\ p.~\bibinfo {pages} {95}\BibitemShut {NoStop}%
\bibitem [{\citenamefont {Schafer}\ \emph {et~al.}(1993)\citenamefont {Schafer}, \citenamefont {Yang}, \citenamefont {DiMauro},\ and\ \citenamefont {Kulander}}]{ati-1993}%
  \BibitemOpen
  \bibfield  {author} {\bibinfo {author} {\bibfnamefont {K.~J.}\ \bibnamefont {Schafer}}, \bibinfo {author} {\bibfnamefont {B.}~\bibnamefont {Yang}}, \bibinfo {author} {\bibfnamefont {L.~F.}\ \bibnamefont {DiMauro}},\ and\ \bibinfo {author} {\bibfnamefont {K.~C.}\ \bibnamefont {Kulander}},\ }\bibfield  {title} {\bibinfo {title} {Above threshold ionization beyond the high harmonic cutoff},\ }\href {https://doi.org/10.1103/PhysRevLett.70.1599} {\bibfield  {journal} {\bibinfo  {journal} {Phys. Rev. Lett.}\ }\textbf {\bibinfo {volume} {70}},\ \bibinfo {pages} {1599} (\bibinfo {year} {1993})}\BibitemShut {NoStop}%
\bibitem [{\citenamefont {Ammosov}\ \emph {et~al.}(1986)\citenamefont {Ammosov}, \citenamefont {Delone},\ and\ \citenamefont {Krainov}}]{adk-1986}%
  \BibitemOpen
  \bibfield  {author} {\bibinfo {author} {\bibfnamefont {M.~V.}\ \bibnamefont {Ammosov}}, \bibinfo {author} {\bibfnamefont {N.~B.}\ \bibnamefont {Delone}},\ and\ \bibinfo {author} {\bibfnamefont {V.~P.}\ \bibnamefont {Krainov}},\ }\bibfield  {title} {\bibinfo {title} {{Tunnel Ionization of Complex Atoms and Atomic Ions in Electromagnetic Field}},\ }in\ \href {https://doi.org/10.1117/12.938695} {\emph {\bibinfo {booktitle} {High Intensity Laser Processes}}},\ Vol.\ \bibinfo {volume} {0664},\ \bibinfo {editor} {edited by\ \bibinfo {editor} {\bibfnamefont {J.~A.}\ \bibnamefont {Alcock}}},\ \bibinfo {organization} {International Society for Optics and Photonics}\ (\bibinfo  {publisher} {SPIE},\ \bibinfo {year} {1986})\ pp.\ \bibinfo {pages} {138 -- 141}\BibitemShut {NoStop}%
\bibitem [{\citenamefont {Tong}\ \emph {et~al.}(2002)\citenamefont {Tong}, \citenamefont {Zhao},\ and\ \citenamefont {Lin}}]{moadk-2002}%
  \BibitemOpen
  \bibfield  {author} {\bibinfo {author} {\bibfnamefont {X.~M.}\ \bibnamefont {Tong}}, \bibinfo {author} {\bibfnamefont {Z.~X.}\ \bibnamefont {Zhao}},\ and\ \bibinfo {author} {\bibfnamefont {C.~D.}\ \bibnamefont {Lin}},\ }\bibfield  {title} {\bibinfo {title} {Theory of molecular tunneling ionization},\ }\href {https://doi.org/10.1103/PhysRevA.66.033402} {\bibfield  {journal} {\bibinfo  {journal} {Phys. Rev. A}\ }\textbf {\bibinfo {volume} {66}},\ \bibinfo {pages} {033402} (\bibinfo {year} {2002})}\BibitemShut {NoStop}%
\bibitem [{\citenamefont {Kjeldsen}\ \emph {et~al.}(2005)\citenamefont {Kjeldsen}, \citenamefont {Bisgaard}, \citenamefont {Madsen},\ and\ \citenamefont {Stapelfeldt}}]{moadk-kjeldsen-2005}%
  \BibitemOpen
  \bibfield  {author} {\bibinfo {author} {\bibfnamefont {T.~K.}\ \bibnamefont {Kjeldsen}}, \bibinfo {author} {\bibfnamefont {C.~Z.}\ \bibnamefont {Bisgaard}}, \bibinfo {author} {\bibfnamefont {L.~B.}\ \bibnamefont {Madsen}},\ and\ \bibinfo {author} {\bibfnamefont {H.}~\bibnamefont {Stapelfeldt}},\ }\bibfield  {title} {\bibinfo {title} {Influence of molecular symmetry on strong-field ionization: Studies on ethylene, benzene, fluorobenzene, and chlorofluorobenzene},\ }\href {https://doi.org/10.1103/PhysRevA.71.013418} {\bibfield  {journal} {\bibinfo  {journal} {Phys. Rev. A}\ }\textbf {\bibinfo {volume} {71}},\ \bibinfo {pages} {013418} (\bibinfo {year} {2005})}\BibitemShut {NoStop}%
\bibitem [{\citenamefont {Murray}\ \emph {et~al.}(2010)\citenamefont {Murray}, \citenamefont {Liu},\ and\ \citenamefont {Ivanov}}]{ft-tunnel-atom-2010}%
  \BibitemOpen
  \bibfield  {author} {\bibinfo {author} {\bibfnamefont {R.}~\bibnamefont {Murray}}, \bibinfo {author} {\bibfnamefont {W.-K.}\ \bibnamefont {Liu}},\ and\ \bibinfo {author} {\bibfnamefont {M.~Y.}\ \bibnamefont {Ivanov}},\ }\bibfield  {title} {\bibinfo {title} {Partial fourier-transform approach to tunnel ionization: Atomic systems},\ }\href {https://doi.org/10.1103/PhysRevA.81.023413} {\bibfield  {journal} {\bibinfo  {journal} {Phys. Rev. A}\ }\textbf {\bibinfo {volume} {81}},\ \bibinfo {pages} {023413} (\bibinfo {year} {2010})}\BibitemShut {NoStop}%
\bibitem [{\citenamefont {Tolstikhin}\ \emph {et~al.}(2011)\citenamefont {Tolstikhin}, \citenamefont {Morishita},\ and\ \citenamefont {Madsen}}]{wfat_theory1-2011}%
  \BibitemOpen
  \bibfield  {author} {\bibinfo {author} {\bibfnamefont {O.~I.}\ \bibnamefont {Tolstikhin}}, \bibinfo {author} {\bibfnamefont {T.}~\bibnamefont {Morishita}},\ and\ \bibinfo {author} {\bibfnamefont {L.~B.}\ \bibnamefont {Madsen}},\ }\bibfield  {title} {\bibinfo {title} {Theory of tunneling ionization of molecules: Weak-field asymptotics including dipole effects},\ }\href {https://doi.org/10.1103/PhysRevA.84.053423} {\bibfield  {journal} {\bibinfo  {journal} {Phys. Rev. A}\ }\textbf {\bibinfo {volume} {84}},\ \bibinfo {pages} {053423} (\bibinfo {year} {2011})}\BibitemShut {NoStop}%
\bibitem [{\citenamefont {Madsen}\ \emph {et~al.}(2013)\citenamefont {Madsen}, \citenamefont {Jensen}, \citenamefont {Tolstikhin},\ and\ \citenamefont {Morishita}}]{wfat-structurefactors-2013}%
  \BibitemOpen
  \bibfield  {author} {\bibinfo {author} {\bibfnamefont {L.~B.}\ \bibnamefont {Madsen}}, \bibinfo {author} {\bibfnamefont {F.}~\bibnamefont {Jensen}}, \bibinfo {author} {\bibfnamefont {O.~I.}\ \bibnamefont {Tolstikhin}},\ and\ \bibinfo {author} {\bibfnamefont {T.}~\bibnamefont {Morishita}},\ }\bibfield  {title} {\bibinfo {title} {Structure factors for tunneling ionization rates of molecules},\ }\href {https://doi.org/10.1103/PhysRevA.87.013406} {\bibfield  {journal} {\bibinfo  {journal} {Phys. Rev. A}\ }\textbf {\bibinfo {volume} {87}},\ \bibinfo {pages} {013406} (\bibinfo {year} {2013})}\BibitemShut {NoStop}%
\bibitem [{\citenamefont {Dnestryan}\ and\ \citenamefont {Tolstikhin}(2016)}]{ir_oewfat-2016}%
  \BibitemOpen
  \bibfield  {author} {\bibinfo {author} {\bibfnamefont {A.~I.}\ \bibnamefont {Dnestryan}}\ and\ \bibinfo {author} {\bibfnamefont {O.~I.}\ \bibnamefont {Tolstikhin}},\ }\bibfield  {title} {\bibinfo {title} {Integral-equation approach to the weak-field asymptotic theory of tunneling ionization},\ }\href {https://doi.org/10.1103/PhysRevA.93.033412} {\bibfield  {journal} {\bibinfo  {journal} {Phys. Rev. A}\ }\textbf {\bibinfo {volume} {93}},\ \bibinfo {pages} {033412} (\bibinfo {year} {2016})}\BibitemShut {NoStop}%
\bibitem [{\citenamefont {Liu}\ and\ \citenamefont {Liu}(2016)}]{ft-tunnel-molecule-2016}%
  \BibitemOpen
  \bibfield  {author} {\bibinfo {author} {\bibfnamefont {M.}~\bibnamefont {Liu}}\ and\ \bibinfo {author} {\bibfnamefont {Y.}~\bibnamefont {Liu}},\ }\bibfield  {title} {\bibinfo {title} {Application of the partial-fourier-transform approach for tunnel ionization of molecules},\ }\href {https://doi.org/10.1103/PhysRevA.93.043426} {\bibfield  {journal} {\bibinfo  {journal} {Phys. Rev. A}\ }\textbf {\bibinfo {volume} {93}},\ \bibinfo {pages} {043426} (\bibinfo {year} {2016})}\BibitemShut {NoStop}%
\bibitem [{\citenamefont {Tolstikhin}\ \emph {et~al.}(2014)\citenamefont {Tolstikhin}, \citenamefont {Madsen},\ and\ \citenamefont {Morishita}}]{tr_mewfat-2014}%
  \BibitemOpen
  \bibfield  {author} {\bibinfo {author} {\bibfnamefont {O.~I.}\ \bibnamefont {Tolstikhin}}, \bibinfo {author} {\bibfnamefont {L.~B.}\ \bibnamefont {Madsen}},\ and\ \bibinfo {author} {\bibfnamefont {T.}~\bibnamefont {Morishita}},\ }\bibfield  {title} {\bibinfo {title} {Weak-field asymptotic theory of tunneling ionization in many-electron atomic and molecular systems},\ }\href {https://doi.org/10.1103/PhysRevA.89.013421} {\bibfield  {journal} {\bibinfo  {journal} {Phys. Rev. A}\ }\textbf {\bibinfo {volume} {89}},\ \bibinfo {pages} {013421} (\bibinfo {year} {2014})}\BibitemShut {NoStop}%
\bibitem [{\citenamefont {Tolstikhina}\ \emph {et~al.}(2014)\citenamefont {Tolstikhina}, \citenamefont {Morishita},\ and\ \citenamefont {Tolstikhin}}]{tr_mewfat_app2-2014}%
  \BibitemOpen
  \bibfield  {author} {\bibinfo {author} {\bibfnamefont {I.~Y.}\ \bibnamefont {Tolstikhina}}, \bibinfo {author} {\bibfnamefont {T.}~\bibnamefont {Morishita}},\ and\ \bibinfo {author} {\bibfnamefont {O.~I.}\ \bibnamefont {Tolstikhin}},\ }\bibfield  {title} {\bibinfo {title} {Application of the many-electron weak-field asymptotic theory of tunneling ionization to atoms},\ }\href {https://doi.org/10.1103/PhysRevA.90.053413} {\bibfield  {journal} {\bibinfo  {journal} {Phys. Rev. A}\ }\textbf {\bibinfo {volume} {90}},\ \bibinfo {pages} {053413} (\bibinfo {year} {2014})}\BibitemShut {NoStop}%
\bibitem [{\citenamefont {Yue}\ \emph {et~al.}(2017)\citenamefont {Yue}, \citenamefont {Bauch},\ and\ \citenamefont {Madsen}}]{tr_mewfat_app-2017}%
  \BibitemOpen
  \bibfield  {author} {\bibinfo {author} {\bibfnamefont {L.}~\bibnamefont {Yue}}, \bibinfo {author} {\bibfnamefont {S.}~\bibnamefont {Bauch}},\ and\ \bibinfo {author} {\bibfnamefont {L.~B.}\ \bibnamefont {Madsen}},\ }\bibfield  {title} {\bibinfo {title} {Electron correlation in tunneling ionization of diatomic molecules: An application of the many-electron weak-field asymptotic theory with a generalized-active-space partition scheme},\ }\href {https://doi.org/10.1103/PhysRevA.96.043408} {\bibfield  {journal} {\bibinfo  {journal} {Phys. Rev. A}\ }\textbf {\bibinfo {volume} {96}},\ \bibinfo {pages} {043408} (\bibinfo {year} {2017})}\BibitemShut {NoStop}%
\bibitem [{\citenamefont {Wahyutama}\ \emph {et~al.}(2022)\citenamefont {Wahyutama}, \citenamefont {Jayasinghe}, \citenamefont {Mauger}, \citenamefont {Lopata}, \citenamefont {Gaarde},\ and\ \citenamefont {Schafer}}]{wahyutama-dft-mewfat-2022}%
  \BibitemOpen
  \bibfield  {author} {\bibinfo {author} {\bibfnamefont {I.~S.}\ \bibnamefont {Wahyutama}}, \bibinfo {author} {\bibfnamefont {D.~D.}\ \bibnamefont {Jayasinghe}}, \bibinfo {author} {\bibfnamefont {F.}~\bibnamefont {Mauger}}, \bibinfo {author} {\bibfnamefont {K.}~\bibnamefont {Lopata}}, \bibinfo {author} {\bibfnamefont {M.~B.}\ \bibnamefont {Gaarde}},\ and\ \bibinfo {author} {\bibfnamefont {K.~J.}\ \bibnamefont {Schafer}},\ }\bibfield  {title} {\bibinfo {title} {All-electron, density-functional-based method for angle-resolved tunneling ionization in the adiabatic regime},\ }\href {https://doi.org/10.1103/PhysRevA.106.052211} {\bibfield  {journal} {\bibinfo  {journal} {Phys. Rev. A}\ }\textbf {\bibinfo {volume} {106}},\ \bibinfo {pages} {052211} (\bibinfo {year} {2022})}\BibitemShut {NoStop}%
\bibitem [{\citenamefont {Wahyutama}\ \emph {et~al.}(2025{\natexlab{a}})\citenamefont {Wahyutama}, \citenamefont {Jayasinghe}, \citenamefont {Mauger}, \citenamefont {Lopata},\ and\ \citenamefont {Schafer}}]{wahyutama-mewfat-fundamental-2025}%
  \BibitemOpen
  \bibfield  {author} {\bibinfo {author} {\bibfnamefont {I.~S.}\ \bibnamefont {Wahyutama}}, \bibinfo {author} {\bibfnamefont {D.~D.}\ \bibnamefont {Jayasinghe}}, \bibinfo {author} {\bibfnamefont {F.}~\bibnamefont {Mauger}}, \bibinfo {author} {\bibfnamefont {K.}~\bibnamefont {Lopata}},\ and\ \bibinfo {author} {\bibfnamefont {K.~J.}\ \bibnamefont {Schafer}},\ }\bibfield  {title} {\bibinfo {title} {All-electron molecular tunnel ionization based on the weak-field asymptotic theory in the integral representation},\ }\href {https://doi.org/10.1103/PhysRevA.111.013117} {\bibfield  {journal} {\bibinfo  {journal} {Phys. Rev. A}\ }\textbf {\bibinfo {volume} {111}},\ \bibinfo {pages} {013117} (\bibinfo {year} {2025}{\natexlab{a}})}\BibitemShut {NoStop}%
\bibitem [{\citenamefont {Mikosch}\ \emph {et~al.}(2013)\citenamefont {Mikosch}, \citenamefont {Boguslavskiy}, \citenamefont {Wilkinson}, \citenamefont {Spanner}, \citenamefont {Patchkovskii},\ and\ \citenamefont {Stolow}}]{channel-ati-2013}%
  \BibitemOpen
  \bibfield  {author} {\bibinfo {author} {\bibfnamefont {J.}~\bibnamefont {Mikosch}}, \bibinfo {author} {\bibfnamefont {A.~E.}\ \bibnamefont {Boguslavskiy}}, \bibinfo {author} {\bibfnamefont {I.}~\bibnamefont {Wilkinson}}, \bibinfo {author} {\bibfnamefont {M.}~\bibnamefont {Spanner}}, \bibinfo {author} {\bibfnamefont {S.}~\bibnamefont {Patchkovskii}},\ and\ \bibinfo {author} {\bibfnamefont {A.}~\bibnamefont {Stolow}},\ }\bibfield  {title} {\bibinfo {title} {Channel- and angle-resolved above threshold ionization in the molecular frame},\ }\href {https://doi.org/10.1103/PhysRevLett.110.023004} {\bibfield  {journal} {\bibinfo  {journal} {Phys. Rev. Lett.}\ }\textbf {\bibinfo {volume} {110}},\ \bibinfo {pages} {023004} (\bibinfo {year} {2013})}\BibitemShut {NoStop}%
\bibitem [{\citenamefont {\ifmmode~\acute{S}\else \'{S}\fi{}piewanowski}\ and\ \citenamefont {Madsen}(2015)}]{CO-orbdistort-2015}%
  \BibitemOpen
  \bibfield  {author} {\bibinfo {author} {\bibfnamefont {M.~D.}\ \bibnamefont {\ifmmode~\acute{S}\else \'{S}\fi{}piewanowski}}\ and\ \bibinfo {author} {\bibfnamefont {L.~B.}\ \bibnamefont {Madsen}},\ }\bibfield  {title} {\bibinfo {title} {Alignment- and orientation-dependent strong-field ionization of molecules: Field-induced orbital distortion effects},\ }\href {https://doi.org/10.1103/PhysRevA.91.043406} {\bibfield  {journal} {\bibinfo  {journal} {Phys. Rev. A}\ }\textbf {\bibinfo {volume} {91}},\ \bibinfo {pages} {043406} (\bibinfo {year} {2015})}\BibitemShut {NoStop}%
\bibitem [{\citenamefont {Hoang}\ \emph {et~al.}(2017)\citenamefont {Hoang}, \citenamefont {Zhao}, \citenamefont {Le},\ and\ \citenamefont {Le}}]{CO-permdipole-2017}%
  \BibitemOpen
  \bibfield  {author} {\bibinfo {author} {\bibfnamefont {V.-H.}\ \bibnamefont {Hoang}}, \bibinfo {author} {\bibfnamefont {S.-F.}\ \bibnamefont {Zhao}}, \bibinfo {author} {\bibfnamefont {V.-H.}\ \bibnamefont {Le}},\ and\ \bibinfo {author} {\bibfnamefont {A.-T.}\ \bibnamefont {Le}},\ }\bibfield  {title} {\bibinfo {title} {Influence of permanent dipole and dynamic core-electron polarization on tunneling ionization of polar molecules},\ }\href {https://doi.org/10.1103/PhysRevA.95.023407} {\bibfield  {journal} {\bibinfo  {journal} {Phys. Rev. A}\ }\textbf {\bibinfo {volume} {95}},\ \bibinfo {pages} {023407} (\bibinfo {year} {2017})}\BibitemShut {NoStop}%
\bibitem [{\citenamefont {Penka~Fowe}\ and\ \citenamefont {Bandrauk}(2011)}]{OCS-CS2-tddft-2011}%
  \BibitemOpen
  \bibfield  {author} {\bibinfo {author} {\bibfnamefont {E.}~\bibnamefont {Penka~Fowe}}\ and\ \bibinfo {author} {\bibfnamefont {A.~D.}\ \bibnamefont {Bandrauk}},\ }\bibfield  {title} {\bibinfo {title} {Nonperturbative time-dependent density-functional theory of ionization and harmonic generation in {OCS} and {CS}${}_{2}$ molecules with ultrashort intense laser pulses: Intensity and orientational effects},\ }\href {https://doi.org/10.1103/PhysRevA.84.035402} {\bibfield  {journal} {\bibinfo  {journal} {Phys. Rev. A}\ }\textbf {\bibinfo {volume} {84}},\ \bibinfo {pages} {035402} (\bibinfo {year} {2011})}\BibitemShut {NoStop}%
\bibitem [{\citenamefont {Penka}\ \emph {et~al.}(2014)\citenamefont {Penka}, \citenamefont {Couture-Bienvenue},\ and\ \citenamefont {Bandrauk}}]{CO-OCS-tddft-2014}%
  \BibitemOpen
  \bibfield  {author} {\bibinfo {author} {\bibfnamefont {E.~F.}\ \bibnamefont {Penka}}, \bibinfo {author} {\bibfnamefont {E.}~\bibnamefont {Couture-Bienvenue}},\ and\ \bibinfo {author} {\bibfnamefont {A.~D.}\ \bibnamefont {Bandrauk}},\ }\bibfield  {title} {\bibinfo {title} {Ionization and harmonic generation in {CO} and {$\text{H}_{2}\text{CO}$} and their cations with ultrashort intense laser pulses with time-dependent density-functional theory},\ }\href {https://doi.org/10.1103/PhysRevA.89.023414} {\bibfield  {journal} {\bibinfo  {journal} {Phys. Rev. A}\ }\textbf {\bibinfo {volume} {89}},\ \bibinfo {pages} {023414} (\bibinfo {year} {2014})}\BibitemShut {NoStop}%
\bibitem [{\citenamefont {\ifmmode~\acute{S}\else \'{S}\fi{}piewanowski}\ and\ \citenamefont {Madsen}(2014)}]{orbdistort-hhg-2014}%
  \BibitemOpen
  \bibfield  {author} {\bibinfo {author} {\bibfnamefont {M.~D.}\ \bibnamefont {\ifmmode~\acute{S}\else \'{S}\fi{}piewanowski}}\ and\ \bibinfo {author} {\bibfnamefont {L.~B.}\ \bibnamefont {Madsen}},\ }\bibfield  {title} {\bibinfo {title} {Field-induced orbital distortion in high-order-harmonic generation from aligned and oriented molecules within adiabatic strong-field approximation},\ }\href {https://doi.org/10.1103/PhysRevA.89.043407} {\bibfield  {journal} {\bibinfo  {journal} {Phys. Rev. A}\ }\textbf {\bibinfo {volume} {89}},\ \bibinfo {pages} {043407} (\bibinfo {year} {2014})}\BibitemShut {NoStop}%
\bibitem [{\citenamefont {Trinh}\ \emph {et~al.}(2015)\citenamefont {Trinh}, \citenamefont {Pham}, \citenamefont {Tolstikhin},\ and\ \citenamefont {Morishita}}]{oewfat1-h2_c-2015}%
  \BibitemOpen
  \bibfield  {author} {\bibinfo {author} {\bibfnamefont {V.~H.}\ \bibnamefont {Trinh}}, \bibinfo {author} {\bibfnamefont {V.~N.~T.}\ \bibnamefont {Pham}}, \bibinfo {author} {\bibfnamefont {O.~I.}\ \bibnamefont {Tolstikhin}},\ and\ \bibinfo {author} {\bibfnamefont {T.}~\bibnamefont {Morishita}},\ }\bibfield  {title} {\bibinfo {title} {Weak-field asymptotic theory of tunneling ionization including the first-order correction terms: Application to molecules},\ }\href {https://doi.org/10.1103/PhysRevA.91.063410} {\bibfield  {journal} {\bibinfo  {journal} {Phys. Rev. A}\ }\textbf {\bibinfo {volume} {91}},\ \bibinfo {pages} {063410} (\bibinfo {year} {2015})}\BibitemShut {NoStop}%
\bibitem [{\citenamefont {Silverstone}\ \emph {et~al.}(1981)\citenamefont {Silverstone}, \citenamefont {Harrell},\ and\ \citenamefont {Grot}}]{high-order-stark-1981}%
  \BibitemOpen
  \bibfield  {author} {\bibinfo {author} {\bibfnamefont {H.~J.}\ \bibnamefont {Silverstone}}, \bibinfo {author} {\bibfnamefont {E.}~\bibnamefont {Harrell}},\ and\ \bibinfo {author} {\bibfnamefont {C.}~\bibnamefont {Grot}},\ }\bibfield  {title} {\bibinfo {title} {{High-order perturbation theory of the imaginary part of the resonance eigenvalues of the Stark effect in hydrogen and of the anharmonic oscillator with negative anharmonicity}},\ }\href {https://doi.org/10.1103/PhysRevA.24.1925} {\bibfield  {journal} {\bibinfo  {journal} {Phys. Rev. A}\ }\textbf {\bibinfo {volume} {24}},\ \bibinfo {pages} {1925} (\bibinfo {year} {1981})}\BibitemShut {NoStop}%
\bibitem [{\citenamefont {Trinh}\ \emph {et~al.}(2013)\citenamefont {Trinh}, \citenamefont {Tolstikhin}, \citenamefont {Madsen},\ and\ \citenamefont {Morishita}}]{wfat1_atom_ref-2013}%
  \BibitemOpen
  \bibfield  {author} {\bibinfo {author} {\bibfnamefont {V.~H.}\ \bibnamefont {Trinh}}, \bibinfo {author} {\bibfnamefont {O.~I.}\ \bibnamefont {Tolstikhin}}, \bibinfo {author} {\bibfnamefont {L.~B.}\ \bibnamefont {Madsen}},\ and\ \bibinfo {author} {\bibfnamefont {T.}~\bibnamefont {Morishita}},\ }\bibfield  {title} {\bibinfo {title} {First-order correction terms in the weak-field asymptotic theory of tunneling ionization},\ }\href {https://doi.org/10.1103/PhysRevA.87.043426} {\bibfield  {journal} {\bibinfo  {journal} {Phys. Rev. A}\ }\textbf {\bibinfo {volume} {87}},\ \bibinfo {pages} {043426} (\bibinfo {year} {2013})}\BibitemShut {NoStop}%
\bibitem [{\citenamefont {Madsen}\ \emph {et~al.}(2017)\citenamefont {Madsen}, \citenamefont {Jensen}, \citenamefont {Dnestryan},\ and\ \citenamefont {Tolstikhin}}]{wfat_partial_wave1-2017}%
  \BibitemOpen
  \bibfield  {author} {\bibinfo {author} {\bibfnamefont {L.~B.}\ \bibnamefont {Madsen}}, \bibinfo {author} {\bibfnamefont {F.}~\bibnamefont {Jensen}}, \bibinfo {author} {\bibfnamefont {A.~I.}\ \bibnamefont {Dnestryan}},\ and\ \bibinfo {author} {\bibfnamefont {O.~I.}\ \bibnamefont {Tolstikhin}},\ }\bibfield  {title} {\bibinfo {title} {{Structure factors for tunneling ionization rates of molecules: General Hartree-Fock-based integral representation}},\ }\href {https://doi.org/10.1103/PhysRevA.96.013423} {\bibfield  {journal} {\bibinfo  {journal} {Phys. Rev. A}\ }\textbf {\bibinfo {volume} {96}},\ \bibinfo {pages} {013423} (\bibinfo {year} {2017})}\BibitemShut {NoStop}%
\bibitem [{\citenamefont {Dnestryan}\ \emph {et~al.}(2018)\citenamefont {Dnestryan}, \citenamefont {Tolstikhin}, \citenamefont {Madsen},\ and\ \citenamefont {Jensen}}]{ir_oewfat_grid-2018}%
  \BibitemOpen
  \bibfield  {author} {\bibinfo {author} {\bibfnamefont {A.~I.}\ \bibnamefont {Dnestryan}}, \bibinfo {author} {\bibfnamefont {O.~I.}\ \bibnamefont {Tolstikhin}}, \bibinfo {author} {\bibfnamefont {L.~B.}\ \bibnamefont {Madsen}},\ and\ \bibinfo {author} {\bibfnamefont {F.}~\bibnamefont {Jensen}},\ }\bibfield  {title} {\bibinfo {title} {{Structure factors for tunneling ionization rates of molecules: General grid-based methodology and convergence studies}},\ }\href {https://doi.org/10.1063/1.5046902} {\bibfield  {journal} {\bibinfo  {journal} {J. Chem. Phys.}\ }\textbf {\bibinfo {volume} {149}},\ \bibinfo {pages} {164107} (\bibinfo {year} {2018})}\BibitemShut {NoStop}%
\bibitem [{nwc(2025)}]{nwchem-github-dev-2025}%
  \BibitemOpen
  \href@noop {} {\bibinfo {title} {{NWChem (forked from upstream)}}},\ \bibinfo {howpublished} {\url{https://github.com/iswhy/nwchem/tree/FEATURE-WFAT}} (\bibinfo {year} {2025})\BibitemShut {NoStop}%
\bibitem [{\citenamefont {Song}\ \emph {et~al.}(2023)\citenamefont {Song}, \citenamefont {Zhu}, \citenamefont {Ni},\ and\ \citenamefont {Wu}}]{pystructurefactor-2023}%
  \BibitemOpen
  \bibfield  {author} {\bibinfo {author} {\bibfnamefont {S.}~\bibnamefont {Song}}, \bibinfo {author} {\bibfnamefont {M.}~\bibnamefont {Zhu}}, \bibinfo {author} {\bibfnamefont {H.}~\bibnamefont {Ni}},\ and\ \bibinfo {author} {\bibfnamefont {J.}~\bibnamefont {Wu}},\ }\bibfield  {title} {\bibinfo {title} {{PyStructureFactor: A Python code for the molecular structure factor in tunneling ionization rates}},\ }\href {https://doi.org/10.1016/j.cpc.2023.108882} {\bibfield  {journal} {\bibinfo  {journal} {Comput. Phys. Commun.}\ }\textbf {\bibinfo {volume} {292}},\ \bibinfo {pages} {108882} (\bibinfo {year} {2023})}\BibitemShut {NoStop}%
\bibitem [{\citenamefont {Kraus}\ \emph {et~al.}(2015{\natexlab{a}})\citenamefont {Kraus}, \citenamefont {Tolstikhin}, \citenamefont {Baykusheva}, \citenamefont {Rupenyan}, \citenamefont {Schneider}, \citenamefont {Bisgaard}, \citenamefont {Morishita}, \citenamefont {Jensen}, \citenamefont {Madsen},\ and\ \citenamefont {W{\"o}rner}}]{wfat-application-laser-elstru-2015}%
  \BibitemOpen
  \bibfield  {author} {\bibinfo {author} {\bibfnamefont {P.}~\bibnamefont {Kraus}}, \bibinfo {author} {\bibfnamefont {O.~I.}\ \bibnamefont {Tolstikhin}}, \bibinfo {author} {\bibfnamefont {D.}~\bibnamefont {Baykusheva}}, \bibinfo {author} {\bibfnamefont {A.}~\bibnamefont {Rupenyan}}, \bibinfo {author} {\bibfnamefont {J.}~\bibnamefont {Schneider}}, \bibinfo {author} {\bibfnamefont {C.~Z.}\ \bibnamefont {Bisgaard}}, \bibinfo {author} {\bibfnamefont {T.}~\bibnamefont {Morishita}}, \bibinfo {author} {\bibfnamefont {F.}~\bibnamefont {Jensen}}, \bibinfo {author} {\bibfnamefont {L.~B.}\ \bibnamefont {Madsen}},\ and\ \bibinfo {author} {\bibfnamefont {H.~J.}\ \bibnamefont {W{\"o}rner}},\ }\bibfield  {title} {\bibinfo {title} {Observation of laser-induced electronic structure in oriented polyatomic molecules},\ }\href {https://doi.org/10.1038/ncomms8039} {\bibfield  {journal} {\bibinfo  {journal} {Nat. Commun.}\ }\textbf {\bibinfo {volume} {6}},\ \bibinfo {pages} {7039} (\bibinfo {year}
  {2015}{\natexlab{a}})}\BibitemShut {NoStop}%
\bibitem [{\citenamefont {Kraus}\ \emph {et~al.}(2015{\natexlab{b}})\citenamefont {Kraus}, \citenamefont {Mignolet}, \citenamefont {Baykusheva}, \citenamefont {Rupenyan}, \citenamefont {Horný}, \citenamefont {Penka}, \citenamefont {Grassi}, \citenamefont {Tolstikhin}, \citenamefont {Schneider}, \citenamefont {Jensen}, \citenamefont {Madsen}, \citenamefont {Bandrauk}, \citenamefont {Remacle},\ and\ \citenamefont {Wörner}}]{wfat-application-atto-cm-2015}%
  \BibitemOpen
  \bibfield  {author} {\bibinfo {author} {\bibfnamefont {P.~M.}\ \bibnamefont {Kraus}}, \bibinfo {author} {\bibfnamefont {B.}~\bibnamefont {Mignolet}}, \bibinfo {author} {\bibfnamefont {D.}~\bibnamefont {Baykusheva}}, \bibinfo {author} {\bibfnamefont {A.}~\bibnamefont {Rupenyan}}, \bibinfo {author} {\bibfnamefont {L.}~\bibnamefont {Horný}}, \bibinfo {author} {\bibfnamefont {E.~F.}\ \bibnamefont {Penka}}, \bibinfo {author} {\bibfnamefont {G.}~\bibnamefont {Grassi}}, \bibinfo {author} {\bibfnamefont {O.~I.}\ \bibnamefont {Tolstikhin}}, \bibinfo {author} {\bibfnamefont {J.}~\bibnamefont {Schneider}}, \bibinfo {author} {\bibfnamefont {F.}~\bibnamefont {Jensen}}, \bibinfo {author} {\bibfnamefont {L.~B.}\ \bibnamefont {Madsen}}, \bibinfo {author} {\bibfnamefont {A.~D.}\ \bibnamefont {Bandrauk}}, \bibinfo {author} {\bibfnamefont {F.}~\bibnamefont {Remacle}},\ and\ \bibinfo {author} {\bibfnamefont {H.~J.}\ \bibnamefont {Wörner}},\ }\bibfield  {title} {\bibinfo {title} {Measurement and laser control of attosecond
  charge migration in ionized iodoacetylene},\ }\href {https://doi.org/10.1126/science.aab2160} {\bibfield  {journal} {\bibinfo  {journal} {Science}\ }\textbf {\bibinfo {volume} {350}},\ \bibinfo {pages} {790} (\bibinfo {year} {2015}{\natexlab{b}})}\BibitemShut {NoStop}%
\bibitem [{\citenamefont {Endo}\ \emph {et~al.}(2019)\citenamefont {Endo}, \citenamefont {Fujise}, \citenamefont {Hasegawa}, \citenamefont {Matsuda}, \citenamefont {Fushitani}, \citenamefont {Tolstikhin}, \citenamefont {Morishita},\ and\ \citenamefont {Hishikawa}}]{dissoc-tunnel-no-2019}%
  \BibitemOpen
  \bibfield  {author} {\bibinfo {author} {\bibfnamefont {T.}~\bibnamefont {Endo}}, \bibinfo {author} {\bibfnamefont {H.}~\bibnamefont {Fujise}}, \bibinfo {author} {\bibfnamefont {H.}~\bibnamefont {Hasegawa}}, \bibinfo {author} {\bibfnamefont {A.}~\bibnamefont {Matsuda}}, \bibinfo {author} {\bibfnamefont {M.}~\bibnamefont {Fushitani}}, \bibinfo {author} {\bibfnamefont {O.~I.}\ \bibnamefont {Tolstikhin}}, \bibinfo {author} {\bibfnamefont {T.}~\bibnamefont {Morishita}},\ and\ \bibinfo {author} {\bibfnamefont {A.}~\bibnamefont {Hishikawa}},\ }\bibfield  {title} {\bibinfo {title} {Angle dependence of dissociative tunneling ionization of {NO} in asymmetric two-color intense laser fields},\ }\href {https://doi.org/10.1103/PhysRevA.100.053422} {\bibfield  {journal} {\bibinfo  {journal} {Phys. Rev. A}\ }\textbf {\bibinfo {volume} {100}},\ \bibinfo {pages} {053422} (\bibinfo {year} {2019})}\BibitemShut {NoStop}%
\bibitem [{\citenamefont {Tehlar}\ \emph {et~al.}(2024)\citenamefont {Tehlar}, \citenamefont {Casanova}, \citenamefont {Dnestryan}, \citenamefont {Jensen}, \citenamefont {Madsen}, \citenamefont {Tolstikhin},\ and\ \citenamefont {Wörner}}]{wfat-application-hhs-2024}%
  \BibitemOpen
  \bibfield  {author} {\bibinfo {author} {\bibfnamefont {A.}~\bibnamefont {Tehlar}}, \bibinfo {author} {\bibfnamefont {J.~T.}\ \bibnamefont {Casanova}}, \bibinfo {author} {\bibfnamefont {A.}~\bibnamefont {Dnestryan}}, \bibinfo {author} {\bibfnamefont {F.}~\bibnamefont {Jensen}}, \bibinfo {author} {\bibfnamefont {L.~B.}\ \bibnamefont {Madsen}}, \bibinfo {author} {\bibfnamefont {O.~I.}\ \bibnamefont {Tolstikhin}},\ and\ \bibinfo {author} {\bibfnamefont {H.~J.}\ \bibnamefont {Wörner}},\ }\bibfield  {title} {\bibinfo {title} {{High-harmonic spectroscopy of impulsively aligned 1,3-cyclohexadiene: Signatures of attosecond charge migration}},\ }\href {https://doi.org/10.1063/4.0000227} {\bibfield  {journal} {\bibinfo  {journal} {Struct. Dyn.}\ }\textbf {\bibinfo {volume} {11}},\ \bibinfo {pages} {014304} (\bibinfo {year} {2024})}\BibitemShut {NoStop}%
\bibitem [{\citenamefont {Becke}(1988)}]{becke_cell-1988}%
  \BibitemOpen
  \bibfield  {author} {\bibinfo {author} {\bibfnamefont {A.~D.}\ \bibnamefont {Becke}},\ }\bibfield  {title} {\bibinfo {title} {A multicenter numerical integration scheme for polyatomic molecules},\ }\href {https://doi.org/10.1063/1.454033} {\bibfield  {journal} {\bibinfo  {journal} {J. Chem. Phys.}\ }\textbf {\bibinfo {volume} {88}},\ \bibinfo {pages} {2547} (\bibinfo {year} {1988})}\BibitemShut {NoStop}%
\bibitem [{\citenamefont {Madsen}\ \emph {et~al.}(2012)\citenamefont {Madsen}, \citenamefont {Tolstikhin},\ and\ \citenamefont {Morishita}}]{oewfat0-linearmol-2012}%
  \BibitemOpen
  \bibfield  {author} {\bibinfo {author} {\bibfnamefont {L.~B.}\ \bibnamefont {Madsen}}, \bibinfo {author} {\bibfnamefont {O.~I.}\ \bibnamefont {Tolstikhin}},\ and\ \bibinfo {author} {\bibfnamefont {T.}~\bibnamefont {Morishita}},\ }\bibfield  {title} {\bibinfo {title} {Application of the weak-field asymptotic theory to the analysis of tunneling ionization of linear molecules},\ }\href {https://doi.org/10.1103/PhysRevA.85.053404} {\bibfield  {journal} {\bibinfo  {journal} {Phys. Rev. A}\ }\textbf {\bibinfo {volume} {85}},\ \bibinfo {pages} {053404} (\bibinfo {year} {2012})}\BibitemShut {NoStop}%
\bibitem [{\citenamefont {Varshalovich}\ \emph {et~al.}(1988)\citenamefont {Varshalovich}, \citenamefont {Moskalev},\ and\ \citenamefont {Khersonskii}}]{varshalovich_angular_momentum-1988}%
  \BibitemOpen
  \bibfield  {author} {\bibinfo {author} {\bibfnamefont {D.~A.}\ \bibnamefont {Varshalovich}}, \bibinfo {author} {\bibfnamefont {A.~N.}\ \bibnamefont {Moskalev}},\ and\ \bibinfo {author} {\bibfnamefont {V.~K.}\ \bibnamefont {Khersonskii}},\ }\href {https://doi.org/10.1142/0270} {\emph {\bibinfo {title} {Quantum Theory of Angular Momentum}}}\ (\bibinfo  {publisher} {World Scientific},\ \bibinfo {year} {1988})\BibitemShut {NoStop}%
\bibitem [{\citenamefont {Rescigno}\ and\ \citenamefont {McCurdy}(2000)}]{fedvr-grid-scatter-2000}%
  \BibitemOpen
  \bibfield  {author} {\bibinfo {author} {\bibfnamefont {T.~N.}\ \bibnamefont {Rescigno}}\ and\ \bibinfo {author} {\bibfnamefont {C.~W.}\ \bibnamefont {McCurdy}},\ }\bibfield  {title} {\bibinfo {title} {Numerical grid methods for quantum-mechanical scattering problems},\ }\href {https://doi.org/10.1103/PhysRevA.62.032706} {\bibfield  {journal} {\bibinfo  {journal} {Phys. Rev. A}\ }\textbf {\bibinfo {volume} {62}},\ \bibinfo {pages} {032706} (\bibinfo {year} {2000})}\BibitemShut {NoStop}%
\bibitem [{\citenamefont {McCurdy}\ \emph {et~al.}(2004)\citenamefont {McCurdy}, \citenamefont {Baertschy},\ and\ \citenamefont {Rescigno}}]{fedvr-three-body-2004}%
  \BibitemOpen
  \bibfield  {author} {\bibinfo {author} {\bibfnamefont {C.~W.}\ \bibnamefont {McCurdy}}, \bibinfo {author} {\bibfnamefont {M.}~\bibnamefont {Baertschy}},\ and\ \bibinfo {author} {\bibfnamefont {T.~N.}\ \bibnamefont {Rescigno}},\ }\bibfield  {title} {\bibinfo {title} {Solving the three-body {Coulomb} breakup problem using exterior complex scaling},\ }\href {https://doi.org/10.1088/0953-4075/37/17/R01} {\bibfield  {journal} {\bibinfo  {journal} {J. Phys. B}\ }\textbf {\bibinfo {volume} {37}},\ \bibinfo {pages} {R137} (\bibinfo {year} {2004})}\BibitemShut {NoStop}%
\bibitem [{\citenamefont {Schneider}\ \emph {et~al.}(2006)\citenamefont {Schneider}, \citenamefont {Collins},\ and\ \citenamefont {Hu}}]{fedvr-parallel-solver-2006}%
  \BibitemOpen
  \bibfield  {author} {\bibinfo {author} {\bibfnamefont {B.~I.}\ \bibnamefont {Schneider}}, \bibinfo {author} {\bibfnamefont {L.~A.}\ \bibnamefont {Collins}},\ and\ \bibinfo {author} {\bibfnamefont {S.~X.}\ \bibnamefont {Hu}},\ }\bibfield  {title} {\bibinfo {title} {Parallel solver for the time-dependent linear and nonlinear {Schr\"odinger} equation},\ }\href {https://doi.org/10.1103/PhysRevE.73.036708} {\bibfield  {journal} {\bibinfo  {journal} {Phys. Rev. E}\ }\textbf {\bibinfo {volume} {73}},\ \bibinfo {pages} {036708} (\bibinfo {year} {2006})}\BibitemShut {NoStop}%
\bibitem [{\citenamefont {Schneider}\ \emph {et~al.}(2011)\citenamefont {Schneider}, \citenamefont {Feist}, \citenamefont {Nagele}, \citenamefont {Pazourek}, \citenamefont {Hu}, \citenamefont {Collins},\ and\ \citenamefont {Burgd\"orfer}}]{fedvr-quant_dyn-2011}%
  \BibitemOpen
  \bibfield  {author} {\bibinfo {author} {\bibfnamefont {B.~I.}\ \bibnamefont {Schneider}}, \bibinfo {author} {\bibfnamefont {J.}~\bibnamefont {Feist}}, \bibinfo {author} {\bibfnamefont {S.}~\bibnamefont {Nagele}}, \bibinfo {author} {\bibfnamefont {R.}~\bibnamefont {Pazourek}}, \bibinfo {author} {\bibfnamefont {S.~X.}\ \bibnamefont {Hu}}, \bibinfo {author} {\bibfnamefont {L.~A.}\ \bibnamefont {Collins}},\ and\ \bibinfo {author} {\bibfnamefont {J.}~\bibnamefont {Burgd\"orfer}},\ }\bibfield  {title} {\bibinfo {title} {{Recent Advances in Computational Methods for the Solution of the Time-Dependent Schr\"odinger Equation for the Interaction of Short, Intense Radiation with One and Two Electron Systems}},\ }in\ \href {https://doi.org/10.1007/978-1-4419-9491-2} {\emph {\bibinfo {booktitle} {Quantum Dynamic Imaging}}},\ \bibinfo {editor} {edited by\ \bibinfo {editor} {\bibfnamefont {A.~D.}\ \bibnamefont {Bandrauk}}\ and\ \bibinfo {editor} {\bibfnamefont {M.}~\bibnamefont {Ivanov}}}\ (\bibinfo  {publisher}
  {Springer},\ \bibinfo {address} {New York},\ \bibinfo {year} {2011})\ p.\ \bibinfo {pages} {149}\BibitemShut {NoStop}%
\bibitem [{Note1()}]{Note1}%
  \BibitemOpen
  \bibinfo {note} {We choose to use FEDVR instead of Laguerre-based DVR for the radial basis because the latter is sensitive to the choice of the scaling factor in the argument of the exponential prefactor appearing in Laguerre functions, which, if improperly chosen, can make the WFAT integrals (Eq. \protect \eqref {eq:define-kr}, \protect \eqref {eq:define-js}, and \protect \eqref {eq:define-i}) diverge.}\BibitemShut {Stop}%
\bibitem [{\citenamefont {Li}\ \emph {et~al.}(2011)\citenamefont {Li}, \citenamefont {Ray}, \citenamefont {De}, \citenamefont {Znakovskaya}, \citenamefont {Cao}, \citenamefont {Laurent}, \citenamefont {Wang}, \citenamefont {Kling}, \citenamefont {Le},\ and\ \citenamefont {Cocke}}]{CO-NO-twocolor-2011}%
  \BibitemOpen
  \bibfield  {author} {\bibinfo {author} {\bibfnamefont {H.}~\bibnamefont {Li}}, \bibinfo {author} {\bibfnamefont {D.}~\bibnamefont {Ray}}, \bibinfo {author} {\bibfnamefont {S.}~\bibnamefont {De}}, \bibinfo {author} {\bibfnamefont {I.}~\bibnamefont {Znakovskaya}}, \bibinfo {author} {\bibfnamefont {W.}~\bibnamefont {Cao}}, \bibinfo {author} {\bibfnamefont {G.}~\bibnamefont {Laurent}}, \bibinfo {author} {\bibfnamefont {Z.}~\bibnamefont {Wang}}, \bibinfo {author} {\bibfnamefont {M.~F.}\ \bibnamefont {Kling}}, \bibinfo {author} {\bibfnamefont {A.~T.}\ \bibnamefont {Le}},\ and\ \bibinfo {author} {\bibfnamefont {C.~L.}\ \bibnamefont {Cocke}},\ }\bibfield  {title} {\bibinfo {title} {Orientation dependence of the ionization of {CO} and {NO} in an intense femtosecond two-color laser field},\ }\href {https://doi.org/10.1103/PhysRevA.84.043429} {\bibfield  {journal} {\bibinfo  {journal} {Phys. Rev. A}\ }\textbf {\bibinfo {volume} {84}},\ \bibinfo {pages} {043429} (\bibinfo {year} {2011})}\BibitemShut {NoStop}%
\bibitem [{\citenamefont {Wu}\ \emph {et~al.}(2012)\citenamefont {Wu}, \citenamefont {Schmidt}, \citenamefont {Kunitski}, \citenamefont {Meckel}, \citenamefont {Voss}, \citenamefont {Sann}, \citenamefont {Kim}, \citenamefont {Jahnke}, \citenamefont {Czasch},\ and\ \citenamefont {D\"orner}}]{CO-ionrate1-2012}%
  \BibitemOpen
  \bibfield  {author} {\bibinfo {author} {\bibfnamefont {J.}~\bibnamefont {Wu}}, \bibinfo {author} {\bibfnamefont {L.~P.~H.}\ \bibnamefont {Schmidt}}, \bibinfo {author} {\bibfnamefont {M.}~\bibnamefont {Kunitski}}, \bibinfo {author} {\bibfnamefont {M.}~\bibnamefont {Meckel}}, \bibinfo {author} {\bibfnamefont {S.}~\bibnamefont {Voss}}, \bibinfo {author} {\bibfnamefont {H.}~\bibnamefont {Sann}}, \bibinfo {author} {\bibfnamefont {H.}~\bibnamefont {Kim}}, \bibinfo {author} {\bibfnamefont {T.}~\bibnamefont {Jahnke}}, \bibinfo {author} {\bibfnamefont {A.}~\bibnamefont {Czasch}},\ and\ \bibinfo {author} {\bibfnamefont {R.}~\bibnamefont {D\"orner}},\ }\bibfield  {title} {\bibinfo {title} {{Multiorbital Tunneling Ionization of the CO Molecule}},\ }\href {https://doi.org/10.1103/PhysRevLett.108.183001} {\bibfield  {journal} {\bibinfo  {journal} {Phys. Rev. Lett.}\ }\textbf {\bibinfo {volume} {108}},\ \bibinfo {pages} {183001} (\bibinfo {year} {2012})}\BibitemShut {NoStop}%
\bibitem [{\citenamefont {Zhang}\ \emph {et~al.}(2013)\citenamefont {Zhang}, \citenamefont {Yuan},\ and\ \citenamefont {Zhao}}]{CO-dyncore-2013}%
  \BibitemOpen
  \bibfield  {author} {\bibinfo {author} {\bibfnamefont {B.}~\bibnamefont {Zhang}}, \bibinfo {author} {\bibfnamefont {J.}~\bibnamefont {Yuan}},\ and\ \bibinfo {author} {\bibfnamefont {Z.}~\bibnamefont {Zhao}},\ }\bibfield  {title} {\bibinfo {title} {Dynamic core polarization in strong-field ionization of {CO} molecules},\ }\href {https://doi.org/10.1103/PhysRevLett.111.163001} {\bibfield  {journal} {\bibinfo  {journal} {Phys. Rev. Lett.}\ }\textbf {\bibinfo {volume} {111}},\ \bibinfo {pages} {163001} (\bibinfo {year} {2013})}\BibitemShut {NoStop}%
\bibitem [{\citenamefont {Saito}\ \emph {et~al.}(2015)\citenamefont {Saito}, \citenamefont {Tolstikhin}, \citenamefont {Madsen},\ and\ \citenamefont {Morishita}}]{wfat-structurefactors-atomic-data-2015}%
  \BibitemOpen
  \bibfield  {author} {\bibinfo {author} {\bibfnamefont {R.}~\bibnamefont {Saito}}, \bibinfo {author} {\bibfnamefont {O.~I.}\ \bibnamefont {Tolstikhin}}, \bibinfo {author} {\bibfnamefont {L.~B.}\ \bibnamefont {Madsen}},\ and\ \bibinfo {author} {\bibfnamefont {T.}~\bibnamefont {Morishita}},\ }\bibfield  {title} {\bibinfo {title} {Structure factors for tunneling ionization rates of diatomic molecules},\ }\href {https://doi.org/https://doi.org/10.1016/j.adt.2015.02.001} {\bibfield  {journal} {\bibinfo  {journal} {At. Data Nucl. Data Tables}\ }\textbf {\bibinfo {volume} {103-104}},\ \bibinfo {pages} {4} (\bibinfo {year} {2015})}\BibitemShut {NoStop}%
\bibitem [{\citenamefont {Majety}\ and\ \citenamefont {Scrinzi}(2015)}]{CO_hacc-2015}%
  \BibitemOpen
  \bibfield  {author} {\bibinfo {author} {\bibfnamefont {V.~P.}\ \bibnamefont {Majety}}\ and\ \bibinfo {author} {\bibfnamefont {A.}~\bibnamefont {Scrinzi}},\ }\bibfield  {title} {\bibinfo {title} {Static field ionization rates for multi-electron atoms and small molecules},\ }\href {https://doi.org/10.1088/0953-4075/48/24/245603} {\bibfield  {journal} {\bibinfo  {journal} {J. Phys. B}\ }\textbf {\bibinfo {volume} {48}},\ \bibinfo {pages} {245603} (\bibinfo {year} {2015})}\BibitemShut {NoStop}%
\bibitem [{\citenamefont {Abu-samha}\ and\ \citenamefont {Madsen}(2020{\natexlab{a}})}]{CO_tdse_sae-2020}%
  \BibitemOpen
  \bibfield  {author} {\bibinfo {author} {\bibfnamefont {M.}~\bibnamefont {Abu-samha}}\ and\ \bibinfo {author} {\bibfnamefont {L.~B.}\ \bibnamefont {Madsen}},\ }\bibfield  {title} {\bibinfo {title} {Effect of multielectron polarization in the strong-field ionization of the oriented {CO} molecule},\ }\href {https://doi.org/10.1103/PhysRevA.101.013433} {\bibfield  {journal} {\bibinfo  {journal} {Phys. Rev. A}\ }\textbf {\bibinfo {volume} {101}},\ \bibinfo {pages} {013433} (\bibinfo {year} {2020}{\natexlab{a}})}\BibitemShut {NoStop}%
\bibitem [{\citenamefont {Sissay}\ \emph {et~al.}(2016)\citenamefont {Sissay}, \citenamefont {Abanador}, \citenamefont {Mauger}, \citenamefont {Gaarde}, \citenamefont {Schafer},\ and\ \citenamefont {Lopata}}]{angle-sfi-tuned-rs-2016}%
  \BibitemOpen
  \bibfield  {author} {\bibinfo {author} {\bibfnamefont {A.}~\bibnamefont {Sissay}}, \bibinfo {author} {\bibfnamefont {P.}~\bibnamefont {Abanador}}, \bibinfo {author} {\bibfnamefont {F.}~\bibnamefont {Mauger}}, \bibinfo {author} {\bibfnamefont {M.}~\bibnamefont {Gaarde}}, \bibinfo {author} {\bibfnamefont {K.~J.}\ \bibnamefont {Schafer}},\ and\ \bibinfo {author} {\bibfnamefont {K.}~\bibnamefont {Lopata}},\ }\bibfield  {title} {\bibinfo {title} {Angle-dependent strong-field molecular ionization rates with tuned range-separated time-dependent density functional theory},\ }\href {https://doi.org/10.1063/1.4961731} {\bibfield  {journal} {\bibinfo  {journal} {The Journal of Chemical Physics}\ }\textbf {\bibinfo {volume} {145}},\ \bibinfo {pages} {094105} (\bibinfo {year} {2016})}\BibitemShut {NoStop}%
\bibitem [{\citenamefont {Hoerner}\ and\ \citenamefont {Schlegel}(2017)}]{ch3x_ionization-2017}%
  \BibitemOpen
  \bibfield  {author} {\bibinfo {author} {\bibfnamefont {P.}~\bibnamefont {Hoerner}}\ and\ \bibinfo {author} {\bibfnamefont {H.~B.}\ \bibnamefont {Schlegel}},\ }\bibfield  {title} {\bibinfo {title} {{Angular Dependence of Strong Field Ionization of CH3X (X = F, Cl, Br, or I) Using Time-Dependent Configuration Interaction with an Absorbing Potential}},\ }\href {https://doi.org/10.1021/acs.jpca.7b06108} {\bibfield  {journal} {\bibinfo  {journal} {J. Phys. Chem. A}\ }\textbf {\bibinfo {volume} {121}},\ \bibinfo {pages} {5940} (\bibinfo {year} {2017})}\BibitemShut {NoStop}%
\bibitem [{\citenamefont {Valiev}\ \emph {et~al.}(2010)\citenamefont {Valiev}, \citenamefont {Bylaska}, \citenamefont {Govind}, \citenamefont {Kowalski}, \citenamefont {Straatsma}, \citenamefont {{Van Dam}}, \citenamefont {Wang}, \citenamefont {Nieplocha}, \citenamefont {Apra}, \citenamefont {Windus},\ and\ \citenamefont {{de Jong}}}]{nwchem-2010}%
  \BibitemOpen
  \bibfield  {author} {\bibinfo {author} {\bibfnamefont {M.}~\bibnamefont {Valiev}}, \bibinfo {author} {\bibfnamefont {E.}~\bibnamefont {Bylaska}}, \bibinfo {author} {\bibfnamefont {N.}~\bibnamefont {Govind}}, \bibinfo {author} {\bibfnamefont {K.}~\bibnamefont {Kowalski}}, \bibinfo {author} {\bibfnamefont {T.}~\bibnamefont {Straatsma}}, \bibinfo {author} {\bibfnamefont {H.}~\bibnamefont {{Van Dam}}}, \bibinfo {author} {\bibfnamefont {D.}~\bibnamefont {Wang}}, \bibinfo {author} {\bibfnamefont {J.}~\bibnamefont {Nieplocha}}, \bibinfo {author} {\bibfnamefont {E.}~\bibnamefont {Apra}}, \bibinfo {author} {\bibfnamefont {T.}~\bibnamefont {Windus}},\ and\ \bibinfo {author} {\bibfnamefont {W.}~\bibnamefont {{de Jong}}},\ }\bibfield  {title} {\bibinfo {title} {Nwchem: A comprehensive and scalable open-source solution for large scale molecular simulations},\ }\href {https://doi.org/https://doi.org/10.1016/j.cpc.2010.04.018} {\bibfield  {journal} {\bibinfo  {journal} {Comput. Phys. Commun.}\ }\textbf {\bibinfo {volume}
  {181}},\ \bibinfo {pages} {1477} (\bibinfo {year} {2010})}\BibitemShut {NoStop}%
\bibitem [{\citenamefont {Apr\`a}\ \emph {et~al.}(2020)\citenamefont {Apr\`a}, \citenamefont {Bylaska}, \citenamefont {de~Jong}, \citenamefont {Govind}, \citenamefont {Kowalski}, \citenamefont {Straatsma}, \citenamefont {Valiev}, \citenamefont {van Dam}, \citenamefont {Alexeev}, \citenamefont {Anchell}, \citenamefont {Anisimov}, \citenamefont {Aquino}, \citenamefont {Atta-Fynn}, \citenamefont {Autschbach}, \citenamefont {Bauman}, \citenamefont {Becca}, \citenamefont {Bernholdt}, \citenamefont {Bhaskaran-Nair}, \citenamefont {Bogatko}, \citenamefont {Borowski}, \citenamefont {Boschen}, \citenamefont {Brabec}, \citenamefont {Bruner}, \citenamefont {Cau\"et}, \citenamefont {Chen}, \citenamefont {Chuev}, \citenamefont {Cramer}, \citenamefont {Daily}, \citenamefont {Deegan}, \citenamefont {Dunning}, \citenamefont {Dupuis}, \citenamefont {Dyall}, \citenamefont {Fann}, \citenamefont {Fischer}, \citenamefont {Fonari}, \citenamefont {Fr\"uchtl}, \citenamefont {Gagliardi}, \citenamefont {Garza}, \citenamefont
  {Gawande}, \citenamefont {Ghosh}, \citenamefont {Glaesemann}, \citenamefont {G\"otz}, \citenamefont {Hammond}, \citenamefont {Helms}, \citenamefont {Hermes}, \citenamefont {Hirao}, \citenamefont {Hirata}, \citenamefont {Jacquelin}, \citenamefont {Jensen}, \citenamefont {Johnson}, \citenamefont {J\'onsson}, \citenamefont {Kendall}, \citenamefont {Klemm}, \citenamefont {Kobayashi}, \citenamefont {Konkov}, \citenamefont {Krishnamoorthy}, \citenamefont {Krishnan}, \citenamefont {Lin}, \citenamefont {Lins}, \citenamefont {Littlefield}, \citenamefont {Logsdail}, \citenamefont {Lopata}, \citenamefont {Ma}, \citenamefont {Marenich}, \citenamefont {Martin~del Campo}, \citenamefont {Mejia-Rodriguez}, \citenamefont {Moore}, \citenamefont {Mullin}, \citenamefont {Nakajima}, \citenamefont {Nascimento}, \citenamefont {Nichols}, \citenamefont {Nichols}, \citenamefont {Nieplocha}, \citenamefont {Otero-de-la Roza}, \citenamefont {Palmer}, \citenamefont {Panyala}, \citenamefont {Pirojsirikul}, \citenamefont {Peng},
  \citenamefont {Peverati}, \citenamefont {Pittner}, \citenamefont {Pollack}, \citenamefont {Richard}, \citenamefont {Sadayappan}, \citenamefont {Schatz}, \citenamefont {Shelton}, \citenamefont {Silverstein}, \citenamefont {Smith}, \citenamefont {Soares}, \citenamefont {Song}, \citenamefont {Swart}, \citenamefont {Taylor}, \citenamefont {Thomas}, \citenamefont {Tipparaju}, \citenamefont {Truhlar}, \citenamefont {Tsemekhman}, \citenamefont {Van~Voorhis}, \citenamefont {V\'azquez-Mayagoitia}, \citenamefont {Verma}, \citenamefont {Villa}, \citenamefont {Vishnu}, \citenamefont {Vogiatzis}, \citenamefont {Wang}, \citenamefont {Weare}, \citenamefont {Williamson}, \citenamefont {Windus}, \citenamefont {Woli\'nski}, \citenamefont {Wong}, \citenamefont {Wu}, \citenamefont {Yang}, \citenamefont {Yu}, \citenamefont {Zacharias}, \citenamefont {Zhang}, \citenamefont {Zhao},\ and\ \citenamefont {Harrison}}]{nwchem2-2020}%
  \BibitemOpen
  \bibfield  {author} {\bibinfo {author} {\bibfnamefont {E.}~\bibnamefont {Apr\`a}}, \bibinfo {author} {\bibfnamefont {E.~J.}\ \bibnamefont {Bylaska}}, \bibinfo {author} {\bibfnamefont {W.~A.}\ \bibnamefont {de~Jong}}, \bibinfo {author} {\bibfnamefont {N.}~\bibnamefont {Govind}}, \bibinfo {author} {\bibfnamefont {K.}~\bibnamefont {Kowalski}}, \bibinfo {author} {\bibfnamefont {T.~P.}\ \bibnamefont {Straatsma}}, \bibinfo {author} {\bibfnamefont {M.}~\bibnamefont {Valiev}}, \bibinfo {author} {\bibfnamefont {H.~J.~J.}\ \bibnamefont {van Dam}}, \bibinfo {author} {\bibfnamefont {Y.}~\bibnamefont {Alexeev}}, \bibinfo {author} {\bibfnamefont {J.}~\bibnamefont {Anchell}}, \bibinfo {author} {\bibfnamefont {V.}~\bibnamefont {Anisimov}}, \bibinfo {author} {\bibfnamefont {F.~W.}\ \bibnamefont {Aquino}}, \bibinfo {author} {\bibfnamefont {R.}~\bibnamefont {Atta-Fynn}}, \bibinfo {author} {\bibfnamefont {J.}~\bibnamefont {Autschbach}}, \bibinfo {author} {\bibfnamefont {N.~P.}\ \bibnamefont {Bauman}}, \bibinfo {author}
  {\bibfnamefont {J.~C.}\ \bibnamefont {Becca}}, \bibinfo {author} {\bibfnamefont {D.~E.}\ \bibnamefont {Bernholdt}}, \bibinfo {author} {\bibfnamefont {K.}~\bibnamefont {Bhaskaran-Nair}}, \bibinfo {author} {\bibfnamefont {S.}~\bibnamefont {Bogatko}}, \bibinfo {author} {\bibfnamefont {P.}~\bibnamefont {Borowski}}, \bibinfo {author} {\bibfnamefont {J.}~\bibnamefont {Boschen}}, \bibinfo {author} {\bibfnamefont {J.}~\bibnamefont {Brabec}}, \bibinfo {author} {\bibfnamefont {A.}~\bibnamefont {Bruner}}, \bibinfo {author} {\bibfnamefont {E.}~\bibnamefont {Cau\"et}}, \bibinfo {author} {\bibfnamefont {Y.}~\bibnamefont {Chen}}, \bibinfo {author} {\bibfnamefont {G.~N.}\ \bibnamefont {Chuev}}, \bibinfo {author} {\bibfnamefont {C.~J.}\ \bibnamefont {Cramer}}, \bibinfo {author} {\bibfnamefont {J.}~\bibnamefont {Daily}}, \bibinfo {author} {\bibfnamefont {M.~J.~O.}\ \bibnamefont {Deegan}}, \bibinfo {author} {\bibfnamefont {J.}~\bibnamefont {Dunning}, \bibfnamefont {T.~H.}}, \bibinfo {author} {\bibfnamefont {M.}~\bibnamefont
  {Dupuis}}, \bibinfo {author} {\bibfnamefont {K.~G.}\ \bibnamefont {Dyall}}, \bibinfo {author} {\bibfnamefont {G.~I.}\ \bibnamefont {Fann}}, \bibinfo {author} {\bibfnamefont {S.~A.}\ \bibnamefont {Fischer}}, \bibinfo {author} {\bibfnamefont {A.}~\bibnamefont {Fonari}}, \bibinfo {author} {\bibfnamefont {H.}~\bibnamefont {Fr\"uchtl}}, \bibinfo {author} {\bibfnamefont {L.}~\bibnamefont {Gagliardi}}, \bibinfo {author} {\bibfnamefont {J.}~\bibnamefont {Garza}}, \bibinfo {author} {\bibfnamefont {N.}~\bibnamefont {Gawande}}, \bibinfo {author} {\bibfnamefont {S.}~\bibnamefont {Ghosh}}, \bibinfo {author} {\bibfnamefont {K.}~\bibnamefont {Glaesemann}}, \bibinfo {author} {\bibfnamefont {A.~W.}\ \bibnamefont {G\"otz}}, \bibinfo {author} {\bibfnamefont {J.}~\bibnamefont {Hammond}}, \bibinfo {author} {\bibfnamefont {V.}~\bibnamefont {Helms}}, \bibinfo {author} {\bibfnamefont {E.~D.}\ \bibnamefont {Hermes}}, \bibinfo {author} {\bibfnamefont {K.}~\bibnamefont {Hirao}}, \bibinfo {author} {\bibfnamefont {S.}~\bibnamefont
  {Hirata}}, \bibinfo {author} {\bibfnamefont {M.}~\bibnamefont {Jacquelin}}, \bibinfo {author} {\bibfnamefont {L.}~\bibnamefont {Jensen}}, \bibinfo {author} {\bibfnamefont {B.~G.}\ \bibnamefont {Johnson}}, \bibinfo {author} {\bibfnamefont {H.}~\bibnamefont {J\'onsson}}, \bibinfo {author} {\bibfnamefont {R.~A.}\ \bibnamefont {Kendall}}, \bibinfo {author} {\bibfnamefont {M.}~\bibnamefont {Klemm}}, \bibinfo {author} {\bibfnamefont {R.}~\bibnamefont {Kobayashi}}, \bibinfo {author} {\bibfnamefont {V.}~\bibnamefont {Konkov}}, \bibinfo {author} {\bibfnamefont {S.}~\bibnamefont {Krishnamoorthy}}, \bibinfo {author} {\bibfnamefont {M.}~\bibnamefont {Krishnan}}, \bibinfo {author} {\bibfnamefont {Z.}~\bibnamefont {Lin}}, \bibinfo {author} {\bibfnamefont {R.~D.}\ \bibnamefont {Lins}}, \bibinfo {author} {\bibfnamefont {R.~J.}\ \bibnamefont {Littlefield}}, \bibinfo {author} {\bibfnamefont {A.~J.}\ \bibnamefont {Logsdail}}, \bibinfo {author} {\bibfnamefont {K.}~\bibnamefont {Lopata}}, \bibinfo {author} {\bibfnamefont
  {W.}~\bibnamefont {Ma}}, \bibinfo {author} {\bibfnamefont {A.~V.}\ \bibnamefont {Marenich}}, \bibinfo {author} {\bibfnamefont {J.}~\bibnamefont {Martin~del Campo}}, \bibinfo {author} {\bibfnamefont {D.}~\bibnamefont {Mejia-Rodriguez}}, \bibinfo {author} {\bibfnamefont {J.~E.}\ \bibnamefont {Moore}}, \bibinfo {author} {\bibfnamefont {J.~M.}\ \bibnamefont {Mullin}}, \bibinfo {author} {\bibfnamefont {T.}~\bibnamefont {Nakajima}}, \bibinfo {author} {\bibfnamefont {D.~R.}\ \bibnamefont {Nascimento}}, \bibinfo {author} {\bibfnamefont {J.~A.}\ \bibnamefont {Nichols}}, \bibinfo {author} {\bibfnamefont {P.~J.}\ \bibnamefont {Nichols}}, \bibinfo {author} {\bibfnamefont {J.}~\bibnamefont {Nieplocha}}, \bibinfo {author} {\bibfnamefont {A.}~\bibnamefont {Otero-de-la Roza}}, \bibinfo {author} {\bibfnamefont {B.}~\bibnamefont {Palmer}}, \bibinfo {author} {\bibfnamefont {A.}~\bibnamefont {Panyala}}, \bibinfo {author} {\bibfnamefont {T.}~\bibnamefont {Pirojsirikul}}, \bibinfo {author} {\bibfnamefont {B.}~\bibnamefont
  {Peng}}, \bibinfo {author} {\bibfnamefont {R.}~\bibnamefont {Peverati}}, \bibinfo {author} {\bibfnamefont {J.}~\bibnamefont {Pittner}}, \bibinfo {author} {\bibfnamefont {L.}~\bibnamefont {Pollack}}, \bibinfo {author} {\bibfnamefont {R.~M.}\ \bibnamefont {Richard}}, \bibinfo {author} {\bibfnamefont {P.}~\bibnamefont {Sadayappan}}, \bibinfo {author} {\bibfnamefont {G.~C.}\ \bibnamefont {Schatz}}, \bibinfo {author} {\bibfnamefont {W.~A.}\ \bibnamefont {Shelton}}, \bibinfo {author} {\bibfnamefont {D.~W.}\ \bibnamefont {Silverstein}}, \bibinfo {author} {\bibfnamefont {D.~M.~A.}\ \bibnamefont {Smith}}, \bibinfo {author} {\bibfnamefont {T.~A.}\ \bibnamefont {Soares}}, \bibinfo {author} {\bibfnamefont {D.}~\bibnamefont {Song}}, \bibinfo {author} {\bibfnamefont {M.}~\bibnamefont {Swart}}, \bibinfo {author} {\bibfnamefont {H.~L.}\ \bibnamefont {Taylor}}, \bibinfo {author} {\bibfnamefont {G.~S.}\ \bibnamefont {Thomas}}, \bibinfo {author} {\bibfnamefont {V.}~\bibnamefont {Tipparaju}}, \bibinfo {author} {\bibfnamefont
  {D.~G.}\ \bibnamefont {Truhlar}}, \bibinfo {author} {\bibfnamefont {K.}~\bibnamefont {Tsemekhman}}, \bibinfo {author} {\bibfnamefont {T.}~\bibnamefont {Van~Voorhis}}, \bibinfo {author} {\bibfnamefont {A.}~\bibnamefont {V\'azquez-Mayagoitia}}, \bibinfo {author} {\bibfnamefont {P.}~\bibnamefont {Verma}}, \bibinfo {author} {\bibfnamefont {O.}~\bibnamefont {Villa}}, \bibinfo {author} {\bibfnamefont {A.}~\bibnamefont {Vishnu}}, \bibinfo {author} {\bibfnamefont {K.~D.}\ \bibnamefont {Vogiatzis}}, \bibinfo {author} {\bibfnamefont {D.}~\bibnamefont {Wang}}, \bibinfo {author} {\bibfnamefont {J.~H.}\ \bibnamefont {Weare}}, \bibinfo {author} {\bibfnamefont {M.~J.}\ \bibnamefont {Williamson}}, \bibinfo {author} {\bibfnamefont {T.~L.}\ \bibnamefont {Windus}}, \bibinfo {author} {\bibfnamefont {K.}~\bibnamefont {Woli\'nski}}, \bibinfo {author} {\bibfnamefont {A.~T.}\ \bibnamefont {Wong}}, \bibinfo {author} {\bibfnamefont {Q.}~\bibnamefont {Wu}}, \bibinfo {author} {\bibfnamefont {C.}~\bibnamefont {Yang}}, \bibinfo {author}
  {\bibfnamefont {Q.}~\bibnamefont {Yu}}, \bibinfo {author} {\bibfnamefont {M.}~\bibnamefont {Zacharias}}, \bibinfo {author} {\bibfnamefont {Z.}~\bibnamefont {Zhang}}, \bibinfo {author} {\bibfnamefont {Y.}~\bibnamefont {Zhao}},\ and\ \bibinfo {author} {\bibfnamefont {R.~J.}\ \bibnamefont {Harrison}},\ }\bibfield  {title} {\bibinfo {title} {{NWChem: Past, present, and future}},\ }\href {https://doi.org/10.1063/5.0004997} {\bibfield  {journal} {\bibinfo  {journal} {J. Chem. Phys.}\ }\textbf {\bibinfo {volume} {152}},\ \bibinfo {pages} {184102} (\bibinfo {year} {2020})}\BibitemShut {NoStop}%
\bibitem [{\citenamefont {Dimitrovski}\ \emph {et~al.}(2011)\citenamefont {Dimitrovski}, \citenamefont {Abu-samha}, \citenamefont {Madsen}, \citenamefont {Filsinger}, \citenamefont {Meijer}, \citenamefont {K\"upper}, \citenamefont {Holmegaard}, \citenamefont {Kalh\o{}j}, \citenamefont {Nielsen},\ and\ \citenamefont {Stapelfeldt}}]{OCS-circpol-2011}%
  \BibitemOpen
  \bibfield  {author} {\bibinfo {author} {\bibfnamefont {D.}~\bibnamefont {Dimitrovski}}, \bibinfo {author} {\bibfnamefont {M.}~\bibnamefont {Abu-samha}}, \bibinfo {author} {\bibfnamefont {L.~B.}\ \bibnamefont {Madsen}}, \bibinfo {author} {\bibfnamefont {F.}~\bibnamefont {Filsinger}}, \bibinfo {author} {\bibfnamefont {G.}~\bibnamefont {Meijer}}, \bibinfo {author} {\bibfnamefont {J.}~\bibnamefont {K\"upper}}, \bibinfo {author} {\bibfnamefont {L.}~\bibnamefont {Holmegaard}}, \bibinfo {author} {\bibfnamefont {L.}~\bibnamefont {Kalh\o{}j}}, \bibinfo {author} {\bibfnamefont {J.~H.}\ \bibnamefont {Nielsen}},\ and\ \bibinfo {author} {\bibfnamefont {H.}~\bibnamefont {Stapelfeldt}},\ }\bibfield  {title} {\bibinfo {title} {Ionization of oriented carbonyl sulfide molecules by intense circularly polarized laser pulses},\ }\href {https://doi.org/10.1103/PhysRevA.83.023405} {\bibfield  {journal} {\bibinfo  {journal} {Phys. Rev. A}\ }\textbf {\bibinfo {volume} {83}},\ \bibinfo {pages} {023405} (\bibinfo {year}
  {2011})}\BibitemShut {NoStop}%
\bibitem [{\citenamefont {Hansen}\ \emph {et~al.}(2011)\citenamefont {Hansen}, \citenamefont {Holmegaard}, \citenamefont {Nielsen}, \citenamefont {Stapelfeldt}, \citenamefont {Dimitrovski},\ and\ \citenamefont {Madsen}}]{OCS-asymtop-2011}%
  \BibitemOpen
  \bibfield  {author} {\bibinfo {author} {\bibfnamefont {J.~L.}\ \bibnamefont {Hansen}}, \bibinfo {author} {\bibfnamefont {L.}~\bibnamefont {Holmegaard}}, \bibinfo {author} {\bibfnamefont {J.~H.}\ \bibnamefont {Nielsen}}, \bibinfo {author} {\bibfnamefont {H.}~\bibnamefont {Stapelfeldt}}, \bibinfo {author} {\bibfnamefont {D.}~\bibnamefont {Dimitrovski}},\ and\ \bibinfo {author} {\bibfnamefont {L.~B.}\ \bibnamefont {Madsen}},\ }\bibfield  {title} {\bibinfo {title} {Orientation-dependent ionization yields from strong-field ionization of fixed-in-space linear and asymmetric top molecules},\ }\href {https://doi.org/10.1088/0953-4075/45/1/015101} {\bibfield  {journal} {\bibinfo  {journal} {J. Phys. B}\ }\textbf {\bibinfo {volume} {45}},\ \bibinfo {pages} {015101} (\bibinfo {year} {2011})}\BibitemShut {NoStop}%
\bibitem [{\citenamefont {Johansen}\ \emph {et~al.}(2016)\citenamefont {Johansen}, \citenamefont {Bay}, \citenamefont {Christensen}, \citenamefont {Thøgersen}, \citenamefont {Dimitrovski}, \citenamefont {Madsen},\ and\ \citenamefont {Stapelfeldt}}]{OCS-align-sfi-2016}%
  \BibitemOpen
  \bibfield  {author} {\bibinfo {author} {\bibfnamefont {R.}~\bibnamefont {Johansen}}, \bibinfo {author} {\bibfnamefont {K.~G.}\ \bibnamefont {Bay}}, \bibinfo {author} {\bibfnamefont {L.}~\bibnamefont {Christensen}}, \bibinfo {author} {\bibfnamefont {J.}~\bibnamefont {Thøgersen}}, \bibinfo {author} {\bibfnamefont {D.}~\bibnamefont {Dimitrovski}}, \bibinfo {author} {\bibfnamefont {L.~B.}\ \bibnamefont {Madsen}},\ and\ \bibinfo {author} {\bibfnamefont {H.}~\bibnamefont {Stapelfeldt}},\ }\bibfield  {title} {\bibinfo {title} {Alignment-dependent strong-field ionization yields of carbonyl sulfide molecules induced by mid-infrared laser pulses},\ }\href {https://doi.org/10.1088/0953-4075/49/20/205601} {\bibfield  {journal} {\bibinfo  {journal} {J. Phys. B}\ }\textbf {\bibinfo {volume} {49}},\ \bibinfo {pages} {205601} (\bibinfo {year} {2016})}\BibitemShut {NoStop}%
\bibitem [{\citenamefont {S\'andor}\ \emph {et~al.}(2018)\citenamefont {S\'andor}, \citenamefont {Sissay}, \citenamefont {Mauger}, \citenamefont {Abanador}, \citenamefont {Gorman}, \citenamefont {Scarborough}, \citenamefont {Gaarde}, \citenamefont {Lopata}, \citenamefont {Schafer},\ and\ \citenamefont {Jones}}]{OCS-single-double-2018}%
  \BibitemOpen
  \bibfield  {author} {\bibinfo {author} {\bibfnamefont {P.}~\bibnamefont {S\'andor}}, \bibinfo {author} {\bibfnamefont {A.}~\bibnamefont {Sissay}}, \bibinfo {author} {\bibfnamefont {F.}~\bibnamefont {Mauger}}, \bibinfo {author} {\bibfnamefont {P.~M.}\ \bibnamefont {Abanador}}, \bibinfo {author} {\bibfnamefont {T.~T.}\ \bibnamefont {Gorman}}, \bibinfo {author} {\bibfnamefont {T.~D.}\ \bibnamefont {Scarborough}}, \bibinfo {author} {\bibfnamefont {M.~B.}\ \bibnamefont {Gaarde}}, \bibinfo {author} {\bibfnamefont {K.}~\bibnamefont {Lopata}}, \bibinfo {author} {\bibfnamefont {K.~J.}\ \bibnamefont {Schafer}},\ and\ \bibinfo {author} {\bibfnamefont {R.~R.}\ \bibnamefont {Jones}},\ }\bibfield  {title} {\bibinfo {title} {Angle dependence of strong-field single and double ionization of carbonyl sulfide},\ }\href {https://doi.org/10.1103/PhysRevA.98.043425} {\bibfield  {journal} {\bibinfo  {journal} {Phys. Rev. A}\ }\textbf {\bibinfo {volume} {98}},\ \bibinfo {pages} {043425} (\bibinfo {year} {2018})}\BibitemShut
  {NoStop}%
\bibitem [{\citenamefont {Abu-samha}\ and\ \citenamefont {Madsen}(2020{\natexlab{b}})}]{OCS-multiel-2020}%
  \BibitemOpen
  \bibfield  {author} {\bibinfo {author} {\bibfnamefont {M.}~\bibnamefont {Abu-samha}}\ and\ \bibinfo {author} {\bibfnamefont {L.~B.}\ \bibnamefont {Madsen}},\ }\bibfield  {title} {\bibinfo {title} {Multielectron effects in strong-field ionization of the oriented {OCS} molecule},\ }\href {https://doi.org/10.1103/PhysRevA.102.063111} {\bibfield  {journal} {\bibinfo  {journal} {Phys. Rev. A}\ }\textbf {\bibinfo {volume} {102}},\ \bibinfo {pages} {063111} (\bibinfo {year} {2020}{\natexlab{b}})}\BibitemShut {NoStop}%
\bibitem [{\citenamefont {Durden}\ and\ \citenamefont {Schlegel}(2024)}]{cap-reduce-cost-2024}%
  \BibitemOpen
  \bibfield  {author} {\bibinfo {author} {\bibfnamefont {A.~S.}\ \bibnamefont {Durden}}\ and\ \bibinfo {author} {\bibfnamefont {H.~B.}\ \bibnamefont {Schlegel}},\ }\bibfield  {title} {\bibinfo {title} {{Reducing the Cost of TD-CI Simulations of Strong Field Ionization}},\ }\href {https://doi.org/10.1021/acs.jpca.4c01732} {\bibfield  {journal} {\bibinfo  {journal} {J. Phys. Chem. A}\ }\textbf {\bibinfo {volume} {128}},\ \bibinfo {pages} {7440} (\bibinfo {year} {2024})}\BibitemShut {NoStop}%
\bibitem [{\citenamefont {Sándor}\ \emph {et~al.}(2019)\citenamefont {Sándor}, \citenamefont {Sissay}, \citenamefont {Mauger}, \citenamefont {Gordon}, \citenamefont {Gorman}, \citenamefont {Scarborough}, \citenamefont {Gaarde}, \citenamefont {Lopata}, \citenamefont {Schafer},\ and\ \citenamefont {Jones}}]{halomethane_ionization-2019}%
  \BibitemOpen
  \bibfield  {author} {\bibinfo {author} {\bibfnamefont {P.}~\bibnamefont {Sándor}}, \bibinfo {author} {\bibfnamefont {A.}~\bibnamefont {Sissay}}, \bibinfo {author} {\bibfnamefont {F.}~\bibnamefont {Mauger}}, \bibinfo {author} {\bibfnamefont {M.~W.}\ \bibnamefont {Gordon}}, \bibinfo {author} {\bibfnamefont {T.~T.}\ \bibnamefont {Gorman}}, \bibinfo {author} {\bibfnamefont {T.~D.}\ \bibnamefont {Scarborough}}, \bibinfo {author} {\bibfnamefont {M.~B.}\ \bibnamefont {Gaarde}}, \bibinfo {author} {\bibfnamefont {K.}~\bibnamefont {Lopata}}, \bibinfo {author} {\bibfnamefont {K.~J.}\ \bibnamefont {Schafer}},\ and\ \bibinfo {author} {\bibfnamefont {R.~R.}\ \bibnamefont {Jones}},\ }\bibfield  {title} {\bibinfo {title} {Angle-dependent strong-field ionization of halomethanes},\ }\href {https://doi.org/10.1063/1.5121711} {\bibfield  {journal} {\bibinfo  {journal} {J. Chem. Phys.}\ }\textbf {\bibinfo {volume} {151}},\ \bibinfo {pages} {194308} (\bibinfo {year} {2019})}\BibitemShut {NoStop}%
\bibitem [{\citenamefont {Slater}(1972)}]{math_handbook_abramowitz-1972}%
  \BibitemOpen
  \bibfield  {author} {\bibinfo {author} {\bibfnamefont {L.~J.}\ \bibnamefont {Slater}},\ }in\ \href@noop {} {\emph {\bibinfo {booktitle} {Handbook of Mathematical Functions with Formulas, Graphs, and Mathematical Tables}}},\ \bibinfo {editor} {edited by\ \bibinfo {editor} {\bibfnamefont {M.}~\bibnamefont {Abramowitz}}\ and\ \bibinfo {editor} {\bibfnamefont {I.~A.}\ \bibnamefont {Stegun}}}\ (\bibinfo  {publisher} {U.S. Dept. of Commerce},\ \bibinfo {address} {New York},\ \bibinfo {year} {1972})\ Chap.~\bibinfo {chapter} {13}\BibitemShut {NoStop}%
\bibitem [{\citenamefont {Wahyutama}\ \emph {et~al.}(2025{\natexlab{b}})\citenamefont {Wahyutama}, \citenamefont {Jayasinghe}, \citenamefont {Mauger}, \citenamefont {Lopata},\ and\ \citenamefont {Schafer}}]{wahyutama_2025_15844886}%
  \BibitemOpen
  \bibfield  {author} {\bibinfo {author} {\bibfnamefont {I.}~\bibnamefont {Wahyutama}}, \bibinfo {author} {\bibfnamefont {D.}~\bibnamefont {Jayasinghe}}, \bibinfo {author} {\bibfnamefont {F.}~\bibnamefont {Mauger}}, \bibinfo {author} {\bibfnamefont {K.}~\bibnamefont {Lopata}},\ and\ \bibinfo {author} {\bibfnamefont {K.}~\bibnamefont {Schafer}},\ }\bibfield  {title} {\bibinfo {title} {{Dataset for article ``Orbital distortion and parabolic channel effects transform minima in molecular ionization probabilities into maxima''}},\ }\href {https://doi.org/10.5281/zenodo.15844886} {10.5281/zenodo.15844886} (\bibinfo {year} {2025}{\natexlab{b}})\BibitemShut {NoStop}%
\end{thebibliography}%

\end{document}